\documentclass[twocolumn,showpacs,nofootinbib,preprintnumbers,amsmath,amssymb,
showkeys,superscriptaddress]{revtex4}

\usepackage[pdftex]{graphicx}  
\usepackage{color}
\usepackage{slashed}
\usepackage[normalem]{ulem}
\usepackage{soul}

\newcommand \beq{\begin{eqnarray}}
\newcommand \eeq{\end{eqnarray}}

\begin{document}

\title{The ghost-antighost-gluon vertex from the Curci-Ferrari model: Two-loop corrections}

\author{Nahuel Barrios\vspace{.4cm}}%
\affiliation{%
Instituto de F\'{\i}sica, Facultad de Ingenier\'{\i}a, Universidad de
la Rep\'ublica, J. H. y Reissig 565, 11000 Montevideo, Uruguay.
\vspace{.1cm}}%
\affiliation{%
Centre de Physique Th\'eorique (CPHT), CNRS, Ecole Polytechnique,\\ Institut Polytechnique de Paris,  Route de Saclay, F-91128 Palaiseau, France.
\vspace{.1cm}}%

\author{Marcela Pel\'aez\vspace{.4cm}}%
\affiliation{%
Instituto de F\'{\i}sica, Facultad de Ingenier\'{\i}a, Universidad de
la Rep\'ublica, J. H. y Reissig 565, 11000 Montevideo, Uruguay.
\vspace{.1cm}}%

\author{Urko Reinosa}%
\affiliation{%
Centre de Physique Th\'eorique (CPHT), CNRS, Ecole Polytechnique,\\ Institut Polytechnique de Paris,  Route de Saclay, F-91128 Palaiseau, France.
\vspace{.1cm}}%

\author{Nicol\'as Wschebor}%
\affiliation{%
 Instituto de F\'{\i}sica, Facultad de Ingenier\'{\i}a, Universidad de
 la Rep\'ublica, J. H. y Reissig 565, 11000 Montevideo, Uruguay.
\vspace{.1cm}}%

\date{\today}

\begin{abstract}
The Curci-Ferrari model has been shown to provide a good grasp on pure Yang-Mills correlation functions in the
Landau gauge, already at one-loop order. In a recent work, the robustness of these results
has been tested by evaluating the two-loop corrections to the gluon and ghost propagators. We pursue this systematic
investigation by computing the ghost-antighost-gluon vertex to the same accuracy in a particular kinematic configuration
that makes the calculations simpler. Because both the parameters of the model and the normalizations of the fields have already
been fixed in a previous work, the present calculation represents both a pure prediction and a stringent test of the approach. 
We find that the two-loop results systematically improve the comparison to Monte-Carlo simulations as compared to earlier
one-loop results. The improvement is particularly significative in the SU($3$) case where the predicted ghost-antighost-gluon
vertex is in very good agreement with the data. The same comparison in the SU($2$) case is not as good, however. This may be
due to the presence of a larger coupling constant in the infrared in that case although we note that a similar mismatch has been
quoted in non-perturbative continuum approaches. Despite these features of the SU($2$) case, it is possible to find sets of
parameters fitting both the propagators and the ghost-antighost-gluon vertex to a reasonable accuracy.

\end{abstract}

\maketitle

\section{Introduction}
Many years after the formulation of Quantum Chromodynamics (QCD), the theoretical description of the infrared behavior of
strong interactions remains largely an open problem. Questions such as the confinement of colored partons or the dynamics of
spontaneous chiral symmetry breaking count as some of the biggest challenges in the field. Of course, extensive numerical lattice
studies have allowed for the first principle extraction of many hadronic properties \cite{Alexandrou:2008tn,Carrasco:2014cwa}.
However, these Monte-Carlo simulations are extremely costly from a numerical point of view, and, some central questions, such
as the study of the QCD phase diagram at finite baryonic density \cite{Philipsen:2010gj},
remain so far out of reach. As a consequence, any analytical or semi-analytical approach that is able to describe
at least some aspects of the infrared behavior of strong interactions is welcome. Of course, standard perturbative
approaches (the most straightforward analytical procedure in field theory) do not work in the infrared regime of QCD, which is
why this regime is usually referred to as ``non-perturbative''.

Among the semi-analytical methods that aim at going beyond the standard perturbative QCD paradigm, one can identify essentially two types. The vast majority of approaches put their focus in constructing non-perturbative approximation schemes in the continuum. These include truncations of the hierarchy of Dyson-Schwinger (DSE)
\cite{Alkofer00,vonSmekal97,Atkinson:1998zc,Zwanziger:2001kw,Lerche:2002ep,Fischer:2002hna,Maas:2004se,Boucaud06,Huber:2007kc,Aguilar07,Aguilar08,Boucaud:2008ji,Boucaud08,Dall'Olio:2012zw,Huber:2012zj,Huber:2016tvc,Huber:2020keu}
or functional renormalization group (FRG) \cite{Ellwanger96,Pawlowski:2003hq,Fischer:2004uk,Fischer:2006vf,Fischer08,Cyrol:2016tym,Dupuis:2020fhh}
equations as well as variational ans\"atze in the Hamiltonian formalism (HF) \cite{Schleifenbaum:2006bq,Quandt:2013wna,Quandt:2015aaa}. A generic feature of all these approaches is that they are not formulated in terms of gauge-invariant observables but rely, instead, on the evaluation of correlation functions for the partonic degrees of freedom. Of course, physics is determined by hadronic gauge-invariant observables and an effort has been made in order to reconstruct physical observables from correlation functions (see, for instance, \cite{Roberts-Hadron,Fu:2019hdw}).
Practical calculations require, however, the use of a gauge-fixed action. In order to keep Lorentz invariance manifest, covariant gauges are usually preferred and for many reasons to be discussed below, most studies, by far, are done in
the Landau gauge.
 
Covariant gauge fixing  in a nonperturbative setting is not a trivial problem, however, and the standard Faddeev-Popov (FP) prescription, well justified in the ultraviolet, together with its underlying local Becchi-Rouet-Stora-Tyutin (BRST) symmetry, cannot be applied in a straightforward way in the infrared \cite{Neuberger:1986vv}. This problem is intimately related to the ambiguity that exists when one tries to fix the gauge in a covariant manner, the so-called Gribov problem \cite{Gribov77}. It has lead some groups to try to tackle the infrared properties of non-abelian theories from a different perspective, with a focus on first extending the gauge fixing procedure beyond its ultraviolet FP realization, before any prejudice on the type of method to be used in the determination of the correlation functions in the infrared. The most known of these approaches is certainly the Gribov-Zwanziger framework where the Gribov ambiguity is partially lifted by removing so-called infinitesimal Gribov copies \cite{Gribov77,Zwanziger89,Vandersickel:2012tz,Dudal:2008sp}.\footnote{Interestingly, a generalization of the BRST symmetry has been recently
discovered in this context  \cite{Capri:2015ixa,Capri:2016gut}.} Ideally, one would like to eliminate any type of copy but
this remains, to date, an arduous task. 

Let us mention that this second type of approach is not totally disconnected from the previous one. For instance, it is known that
DS equations are formally the same for a theory with or without infinitesimal copies (with the exception of
ghost correlation functions) \cite{Zwanziger:2003cf}. What changes are the boundary
conditions to be applied on these equations (because the underlying actions are different of course).
In nonperturbative approaches, one particular handle on the boundary
conditions\footnote{Another such handle is related to the value of the ghost dressing function at zero momentum.}
is provided by the fact that the (necessary) regularization
breaks the BRST symmetry explicitly, placing inevitably the model within a larger class of models with less symmetry, and, therefore,
with more operators/couplings. How these extra couplings should be fixed in order to retrieve a
BRST invariant theory and whether or not BRST should be retrieved at all in the IR are questions that are still open to
debate.\footnote{In a recent work \cite{Huber:2020keu}, while it is acknowledged that there is an arbitrariness related to the
removal of quadratic divergences that impacted previous implementations, it is claimed that, within a new
implementation of the truncation of DS equations, this arbitrariness has almost no impact on the gluon propagator, once expressed
in physical units. This is certainly an interesting claim that deserves attention. Whether the full arbitrariness that the subtraction
of quadratic divergences entails has been tested in \cite{Huber:2020keu} as well as how the observed insensivity to this subtraction
depends on the specifics of the truncation and how it can be implemented in other non-perturbative continuum approaches remain open
questions.} 

This large activity around the semi-analytical evaluation of Landau gauge correlation functions has
motivated numerous gauge-fixed lattice simulations. In fact, one of the main reasons explaining the focus on the Landau gauge is that the gauge fixing can be formulated as the extremization of the functional $W_A[U]\equiv\int_x {\rm tr}\,A^U_\mu(x) A^U_\mu(x)$. For a given gauge field configuration $A_\mu$, the latter admits many extrema $U_i$ along the gauge orbit $A_\mu^U$, corresponding to the Gribov copies mentioned above. Since gauge fixing amounts to choosing one copy per orbit one can restrict to copies that minimize the functional $W_A[U]$, turning the gauge fixing into a minimization problem well suited for numerical simulations. Various ways of choosing these minimizing Gribov copies have been considered \cite{Maas:2016frv}, the simplest of which consists in
randomly picking one copy on each orbit, defining the so-called minimal Landau gauge,
which explicitly breaks the BRST symmetry of the FP Lagrangian.\footnote{The possibility that the FP construction in the
Landau gauge is correct at a nonperturbative level, despite the presence of Gribov copies has been suggested
in \cite{Hirschfeld:1978yq,vonSmekal:2013cla,vonSmekal:2008en} but remains unproven so far.} 

Rather independently of the precise choice of copy, lattice studies in $\smash{d=3}$ and $\smash{d=4}$ dimensions\footnote{The case $d=2$ requires a separate discussion, see \cite{Maas:2007uv,Cucchieri:2011um,Cucchieri:2011ig}.} have clearly demonstrated that the gluon propagator saturates to a finite non-zero value at vanishing
momentum, corresponding to a massive-like
behavior \cite{Bonnet:2000kw,Bonnet:2001uh,Cucchieri_08b,Bogolubsky09,Bornyakov09,Iritani:2009mp,Maas:2011se,Oliveira:2012eh}.
At the same time, this behavior is a non-standard one for it features a violation of positivity. The ghost dressing function (the corresponding propagator times the momentum square) has also been found to saturate at a finite non-zero value for vanishing momentum.
Finally, the gauge coupling extracted from the ghost-antighost-gluon vertex stays finite for all momenta and even
becomes small in the deep infrared \cite{Bogolubsky09,Boucaud:2011ug}.

All these results are clearly at odds with standard perturbation theory based on the FP procedure, which features an infrared
Landau pole in the running of the coupling constant. The non-perturbative approaches referred to above typically find two classes of solutions, known as {\it scaling} and {\it decoupling}, depending on how the boundary conditions are chosen. The class of decoupling solutions allows for a very good comparison to lattice data. As for the Gribov-Zwanziger approach, in its simplest form, it leads instead to a scaling type solution, at odds with the lattice results. A refinement based on the dynamical generation of condensates could reconcile the approach
with the lattice results at tree-level \cite{Dudal10}. It remains to see how this survives the inclusion of higher order
corrections.\footnote{There exist examples where tree level masses generated by condensates are cancelled by one-loop
corrections \cite{Meerleer2020}.}

Next to these two main approaches and their myriad of results, a third way has been put forward and has proven quite successful in determining many infrared properties of Yang-Mills (YM) theories. It belongs to the class of approaches that aim at extending the gauge fixing beyond its ultraviolet FP prescription but it is more phenomenological in spirit than the Gribov-Zwanziger approach: rather than trying to infer the complete gauge-fixed action by eliminating as many copies as possible, one exploits the lattice results in the Landau gauge in order to guess the main ingredients that would compose such an action. As proposed initially in \cite{Tissier:2010ts,Tissier:2011ey},
one considers a massive deformation of the FP Lagrangian in the Landau gauge. This deformation is rather minimalistic since the
only modification to the Landau gauge FP Feynman rules is that the gluon propagator, while remaining transverse, becomes massive.

The model corresponds to the Landau limit of the Curci-Ferrari (CF) model \cite{Curci76} which has a long history. It was
proven to be renormalizable long time ago \cite{deBoer95,Delduc89,Tissier:2008nw} but was discarded due to violations of
positivity \cite{Curci:1976kh,deBoer95}. Indeed, the model possess a BRST-like symmetry but it is not nilpotent and it turns out not to
be sufficient to prove that the standard
definition of the perturbative physical space \cite{Kugo:1977mk,Kugo:1977zq} only contains positive norm states.
A Hilbert space with only positive norm states is a necessary step in defining a physical space on which to verify unitarity.
It is to be stressed, however, that perturbative unitarity is not sufficient in general to prove the true unitarity of a given model.
Even in QED, unitarity requires one to consider the S-matrix between any possible scattering state, including scattering between
the elementary constituents of the model and possible bound states.\footnote{This is pretty clear in non-relativistic scattering
where the scattering operator is unitary on the ``asymptotic space'' that labels all these possible scattering states including
the scattering of bound states \cite{JRTaylor}.} This is even more true in YM theory or QCD where confinement forbids the presence
of elementary constituents among the asymptotic states.  In such models the true physical space is certainly not one that includes
quarks or transverse gluons (as it is the case for the standard perturbative physical space) but rather glueballs and hadrons and
it is on such a physical space that the question of unitarity needs to be addressed. The possibility to construct such a version of
the physical space in the CF model
or in any model where the Gribov problem is taken into account in one way or another remains of course an open question.
We stress, however, that first-principle lattice simulations \cite{Cucchieri:2004mf,Bowman07} have shown an unambiguous violation
of reflection-positivity in the (transverse) gluon propagator, which are well reproduced by the CF
model \cite{Tissier:2010ts,Siringo:2017ide,Kondo:2019ywt}. This observed positivity violation raises serious doubts on the
applicability of the standard definition of the perturbative physical space for both YM theories and QCD and, as a consequence,
the criticisms regarding the CF model must be reconsidered.
  
Letting aside these interesting but to date unsolved questions, the CF model has been used to evaluate many correlation
functions of YM theory in the Landau gauge and yields unexpectedly good results within a simple perturbative expansion.
With appropriate renormalization conditions
\cite{Tissier:2011ey,Weber:2011nw} the model is infrared-safe in the sense that there is a family of renormalization-group
trajectories without Landau-pole. The corresponding correlation functions are then regular for any Euclidean momentum, down to zero momentum. Moreover,
the trajectories that actually reproduce lattice data correspond to moderate couplings,
allowing for a reasonable control of perturbation theory \cite{Tissier:2011ey,Reinosa:2017qtf}.\footnote{We refer
to \cite{Siringo:2015jea} for a similar approach based on a massive modification of perturbation theory which also leads to very good
results at one-loop order. The premises of this approach are however quite different from those of the CF approach, since it is
assumed from the beginning that the Faddeev-Popov action is a good starting point to study the infrared properties.}

Lattice two-point YM vertex functions are very well reproduced at one-loop order \cite{Tissier:2010ts,Tissier:2011ey,Reinosa:2017qtf,Pelaez:2013cpa} and
recently the corresponding two-loop perturbative corrections have been evaluated. They are found to be tiny and tend to improve the one-loop results, confirming the validity of perturbation theory in the CF model, at least as far as YM two-point functions are concerned \cite{Gracey:2019xom}. The same analysis has been performed for the YM three-point functions, but so far only at one-loop order \cite{Pelaez:2013cpa}. The comparison to lattice data remains good, although not as good as with the two-point
functions. It is the purpose of the present paper to extend the systematic evaluation of two-loop corrections to the
three-point YM vertices, in view of further testing the validity of the perturbative CF picture. In this work, we consider the ghost-antighost-gluon vertex in a particular momentum configuration that makes the calculation of the same level of difficulty than that of the two-point functions.

Before closing this Introduction, let us mention that the CF model has also been used to investigate many other properties of YM theory and QCD. It has been extended to include quarks, yielding a reasonable agreement
with lattice data  but also showing that the relevant coupling in the quark-gluon sector is
significantly larger than in the YM sector \cite{Pelaez:2014mxa,Pelaez:2015tba}. This observation is particularly important in order to study the spontaneous breaking of chiral symmetry,
which, as expected, can not be obtained by purely perturbative means. Nevertheless, it was shown that the smallness of
the YM coupling allows for the formulation of controlled approximations that reproduce the lattice data accurately and also explain the spontaneous chiral symmetry breaking in QCD in a controlled manner \cite{Pelaez:2017bhh}.
The model has also been extended to finite temperature and density. In the case of YM theory as well as QCD in the limit of heavy quarks, it has allowed to
successfully capture various features of the phase diagram, in particular the confinement-deconfinement transition and its associated order parameter
\cite{Reinosa:2014ooa,Reinosa:2014zta,Reinosa:2015gxn,Reinosa:2015oua,Reinosa:2016iml,Siringo:2017svp,Maelger:2017amh,Maelger:2018vow}. More recently, the phase diagram at finite temperature and
finite chemical potential began to be studied within the CF model in the presence of chiral quarks \cite{Maelger:2019cbk}.

The article is organized as follows. In Sec.~\ref{sec:CF}, we briefly review the CF model together with its renormalization, and we summarize some of the results relevant to this work. In Sec.~\ref{sec:vertex}, we describe the main properties of the ghost-antighost-gluon vertex and discuss its perturbative contributions
at one- and two-loop order which we reduce to master integrals and split into UV divergent and finite parts.
In Sec.~\ref{sec:checks},
we present the various crosschecks that we employed in order to verify the large and tedious output of the two-loop calculation. 
We present our results in Sec.~\ref{sec:results} together with a comparison with lattice results. We conclude in Sec.~\ref{sec:conclusion} and gather some technical details in the appendices.

\section{The Curci-Ferrari model}\label{sec:CF}
In what follows, we work with the Euclidean Lagrangian density
\beq
  \label{eq_lagrang}
  \mathcal L=\frac 14 (F_{\mu\nu}^a)^2+\partial _\mu\overline c^a(D_\mu
  c)^a+ih^a\partial_\mu A_\mu^a+\frac {m^2}2 (A_\mu^a)^2\,,
\eeq
where Latin indices label the generators of the $SU(N)$ color group. The covariant derivative in the adjoint representation is given by 
\beq
(D_\mu c)^a\equiv\partial_\mu c^a+g f^{abc}A_\mu^b c^c\,,
\eeq 
and the corresponding field-strength tensor reads
\beq
F_{\mu\nu}^a\equiv\partial_\mu A_\nu^a-\partial_\nu A_\mu^a+gf^{abc}A_\mu^bA_\nu^c\,,
\eeq 
with $g$ the coupling constant. 

As already mentioned in the Introduction, the Lagrangian density (\ref{eq_lagrang}) corresponds to a particular case of the CF model~\cite{Curci76}, obtained in the limit of vanishing gauge parameter (i.e. the Landau gauge). At tree level, the gluon propagator is massive and transverse in
momentum space, which ensures that the  model is renormalizable. We refer the reader to Refs. \cite{Tissier:2008nw,Tissier:2011ey}
for a more detailed
account of the model, including its many symmetries.

\subsection{Infrared-safe renormalization scheme}\label{ssec:IS}
The model is regularized in $\smash{d=4-2\epsilon}$ dimensions. It is renormalized  as usual by rescaling both the bare fields,
\beq 
A_B^{a,\mu}=\sqrt{Z_A}\,A^{a,\mu}\,, \quad c_B^{a}=\sqrt{Z_c}\,c^{a}\,, \quad \bar c_B^{a}=\sqrt{Z_c}\,\bar c^{a}\,,
\eeq
and the bare parameters,
\beq\label{eq:ren_param}
g_B=Z_g\,g\,, \quad m_B^2=Z_{m^2}\,m^2\,,
\eeq
where we have denoted bare quantities with a subscript ``$B$''. 

One interesting feature of the model (\ref{eq_lagrang}) is that the renormalization factors $Z_X$ are constrained by two non-renormalization
theorems \cite{Taylor71,jag12,jag13,Dudal02,Wschebor07,Tissier08}. First, owing both to the particular form of the ghost-antighost-gluon
interaction and to the transversality of the
gluon propagator, the ghost-antighost-gluon vertex receives no corrections beyond tree-level in the limit of vanishing ghost
momentum \cite{Taylor71}.\footnote{Indeed, in this limit, and for any diagrammatic contribution beyond tree level, one finds a
factor $\smash{P^\perp_{\rho\sigma}(q)\,q_\sigma=0}$, where the transverse
projector $\smash{P^\perp_{\rho\sigma}(q)\equiv\delta_{\rho\sigma}-q_\rho q_\sigma/q^2}$ originates from the gluon propagator
attached to the same vertex than the ghost leg, while $q_\sigma$ is the antighost momentum leaving that same
vertex (which equals the gluon momentum in the limit where the external ghost momentum is taken to zero).} A direct consequence of
this observation is that the combination $Z_g\sqrt{Z_A}Z_c$ is finite. Similarly, owing to various symmetries enjoyed by the model,
one can argue that $Z_{m^2}Z_AZ_c$ is finite as well \cite{jag12,jag13,Dudal02,Wschebor07,Tissier08}. 

In particular, this means that one can choose renormalization schemes where
\beq\label{eq:nonrenorm}
Z_g\sqrt{Z_A}Z_c=Z_{m^2}Z_AZ_c=1\,,
\eeq
and, therefore, such that all renormalization factors can be obtained from the sole knowledge of the gluon and ghost propagators, denoted $G$ and $D$ respectively. In what follows, we work within this set-up and fix the remaining renormalization factors by imposing the following conditions on the gluon and ghost propagators at the running scale $\mu$:
\beq
G^{-1}(k=\mu)=\mu^2+m^2(\mu)\,, \hspace{.4cm}D^{-1}(k=\mu)=\mu^2\,.
\eeq
These constraints, together with those in Eq.~(\ref{eq:nonrenorm}), define the so-called Infrared Safe (IS) scheme \cite{Tissier:2011ey}. 

We recall for completeness that, in dimensional regularization, the bare coupling that appears in the Lagrangian density
has mass dimension $\epsilon$ and is usually written $g_B\mu^\epsilon$. Despite this explicit dependence on $\mu$, the dimensionful bare coupling $g_B\mu^\epsilon$ should be considered $\mu$-independent. This is particularly important when deriving the beta functions that govern the $\mu$-evolution of the renormalized parameters $g$ and $m$.

\subsection{Summary of results}
At one- and two-loop order of perturbation theory, the CF model in the IS scheme displays two classes of
renormalization group (RG) trajectories in the space of dimensionless parameters $(m^2/\mu^2,g^2)$, separated by a particular trajectory
connecting an UV and an IR fixed point \cite{Reinosa:2017qtf}. On one side of this separatrix, the renormalization group flow
becomes singular at a finite scale $\mu_{\rm Landau}$, which generalizes the Landau pole of the Faddeev-Popov
model (corresponding to the limit $m\to 0$). In contrast, on the other side of the separatrix, the RG trajectories are defined
for all values of the renormalization scale and are characterized by a bounded coupling that approaches zero both in the UV limit and in the IR limit.

Strictly speaking, only the IS trajectories for which the coupling remains perturbative should be taken seriously within this perturbative
determination of the RG flow. Luckily enough, these are the trajectories that best describe the lattice data for the
Landau gauge YM correlation
functions \cite{Tissier:2010ts,Tissier:2011ey,Reinosa:2017qtf}. 

\begin{figure}
\hglue-6mm\includegraphics[width=0.45\textwidth]{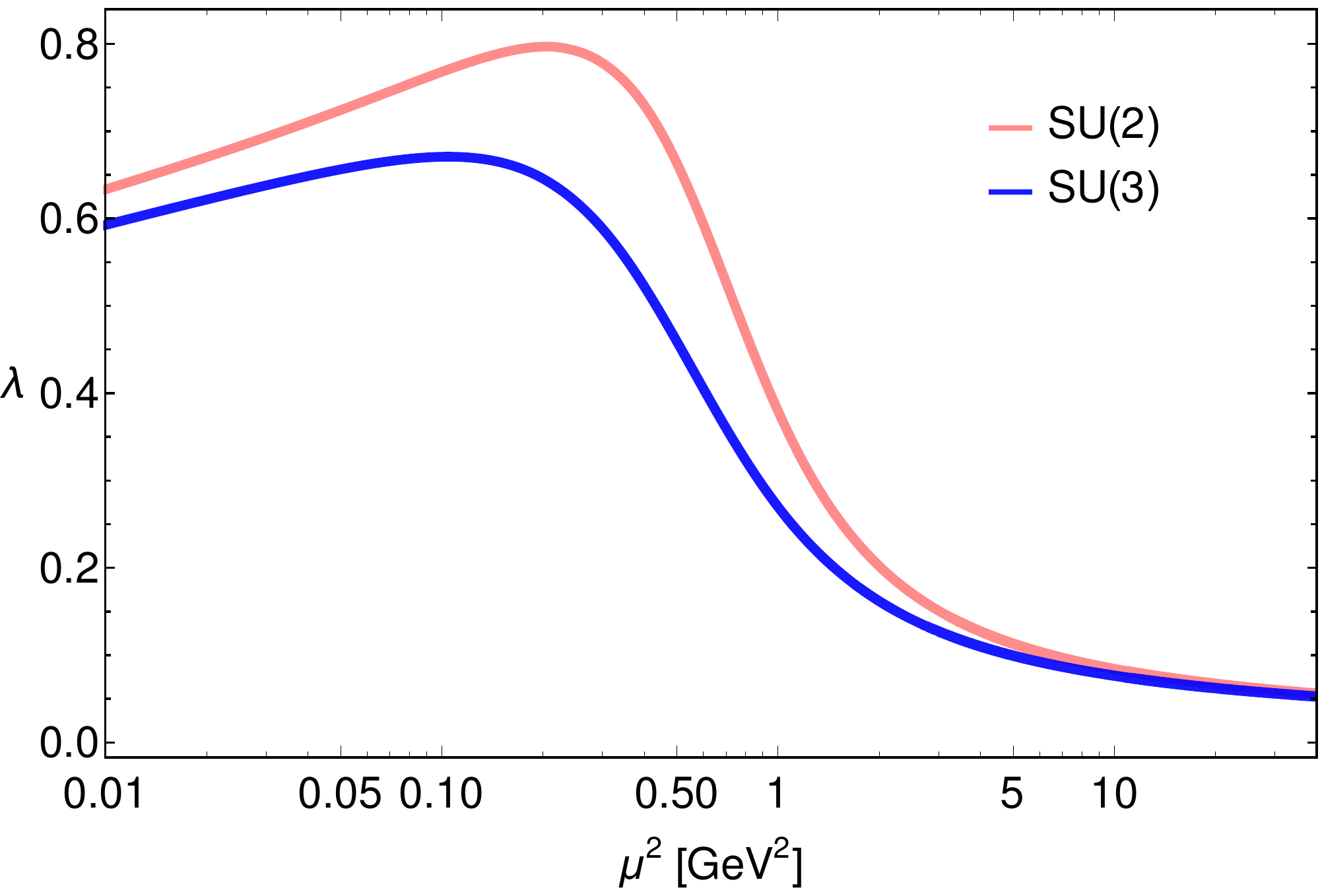}
\vglue6mm
\hglue-6mm\includegraphics[width=0.45\textwidth]{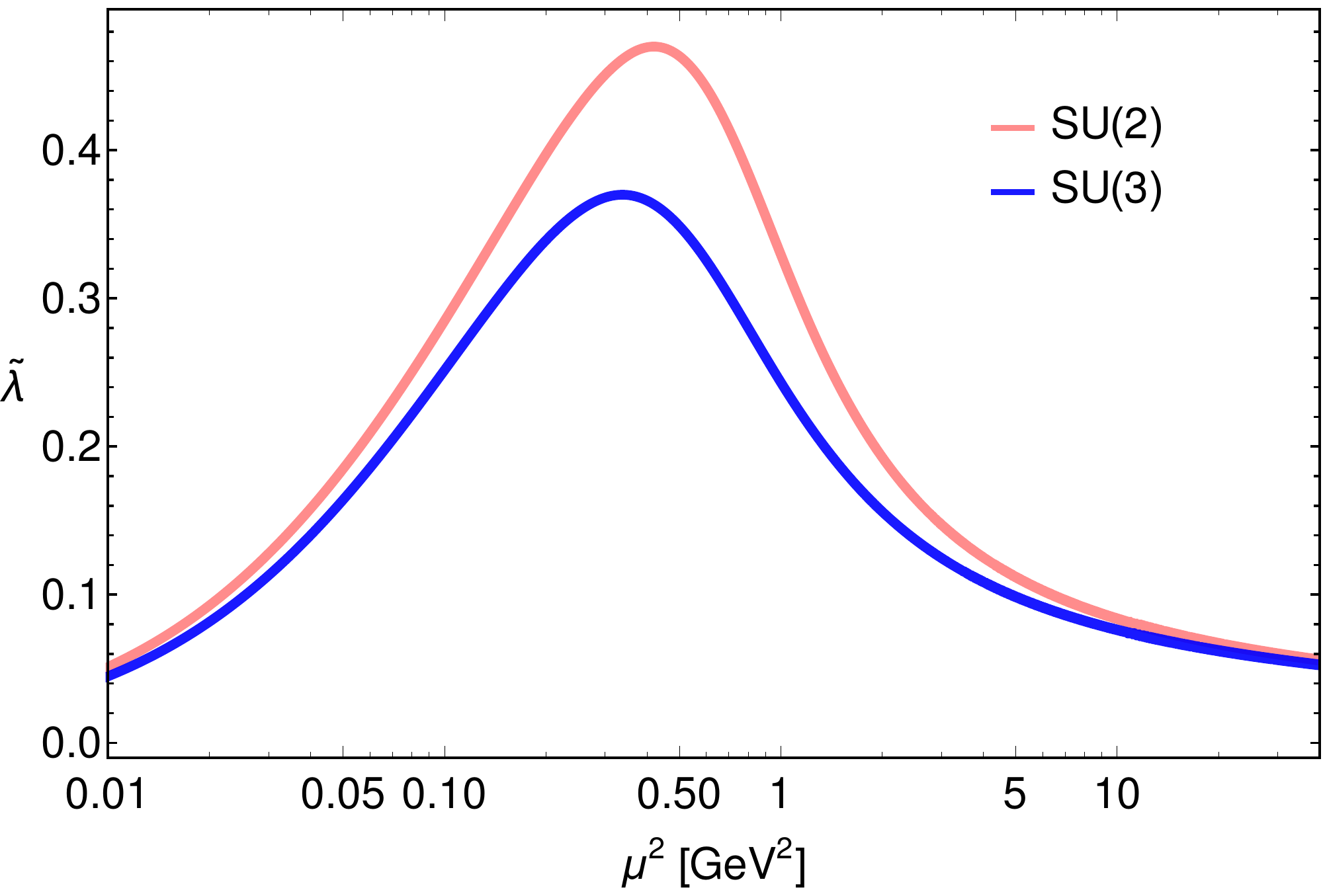}
\caption{Two-loop running of the expansion parameters $\lambda_{IS}(\mu^2)$ (top) and $\tilde \lambda_{IS}(\mu^2)$ (bottom) in
the IS scheme in the SU($2$) and SU($3$) cases.}
\label{Fig:lambda_IS}
\end{figure}

In particular, the two-point functions are reproduced to very good
accuracy using the CF model at one-loop order, and this agreement has improved to an impressive level in a recent two-loop
calculation \cite{Gracey:2019xom}. In general, the quality of the results is better for the SU($3$) gauge group than for SU($2$). An explanation could be that the expansion parameter $\lambda\equiv g^2N/(16\pi^2)$ in the IS scheme at two-loop order is bounded by $\simeq 0.6$ in the SU($3$) case but the corresponding parameter in the SU($2$) case overpasses $0.8$ in some region along the flow and thus approaches the limit of validity of the perturbative expansion, see Fig.~\ref{Fig:lambda_IS}. The apparent convergence of perturbation theory in the SU($3$) case has also been tested by comparing
the results in the IS scheme to results in other renormalization schemes, such as the family of vanishing momentum (VM) schemes, see
below. Although the scheme-dependences remain sizeable at one-loop order, they are considerably reduced at two-loop order in the SU($3$) case.

Let us point out, however, that the precise expansion parameter of the loop expansion in this model is not exactly known.
Indeed, most loop diagrams are rather controlled by an improved perturbative expansion
parameter $\tilde \lambda=\lambda \mu^2/(\mu^2+m^2)$ that takes into account that  most perturbative corrections including
internal gluon lines are suppressed in the infrared by at least one factor of order $\mu^2/m^2$ (with $\mu \ll m$). The parameter $\tilde \lambda$ is considerably
smaller than $\lambda$ in the infrared (as can be seen in Fig.~\ref{Fig:lambda_IS}) and its order of magnitude seems to
agree better with the observed errors of perturbative
calculations of most vertex functions in YM theory in the CF model. Which of the two expansion parameters $\lambda$ or 
$\tilde \lambda$ is the one that controls the perturbative expansion is not completely clear and possibly depends on the considered renormalization
scheme. Our calculations in the IS scheme to be presented below reveal that the theoretical error bars of the two-loops results typically are
governed by a parameter between $\tilde\lambda$ and $\lambda$.

As for the three-point vertex functions, both the three-gluon vertex and the ghost-antighost-gluon vertex were studied in \cite{Pelaez:2013cpa} in the SU($2$) case for arbitrary tensorial structures and for arbitrary configurations of momenta. The results were compared with the lattice data of \cite{Cucchieri:2008qm} with again a very good agreement, although not as good as in the case of the two-point functions.
 
It must be
stressed that the calculation of the three-point functions in \cite{Pelaez:2013cpa} is a {\it pure prediction} of the model since all parameters were fixed
by fitting the two-point functions, with no free parameter left to adjust the three-point functions.\footnote{The only exception is the overall normalization of the three-gluon vertex. This parameter corresponds just to
the renormalization factor for the bare lattice vertex.} Therefore, a direct comparison of the respective qualities of the two- and three-point functions
is a little bit biased because the parameters of the model were adjusted to best reproduce the lattice data for the two-point functions, and any inaccuracy at the level of the two-point functions impacts the determination of the parameters and, therefore, the prediction of the vertices. This point will be relevant below when we investigate the ghost-antighost-gluon vertex at two-loop order. Another relevant observation when matching one-loop vertices with lattice data is that the quality of the agreement with the lattice results is not uniform over all configurations of momenta. The agreement is far better for configurations where all external momenta are typically of the
same order, as compared to configurations where one of the gluon momenta vanishes.

As announced in the Introduction, we here initiate a systematic analysis of the two-loop corrections to the three-point functions in the CF model, similar to what has been done for the two-point functions in \cite{Gracey:2019xom}. We will address the case of the ghost-antighost-gluon vertex, leaving the more involved three-gluon vertex for a future analysis.  Moreover, since the analysis for an arbitrary configuration of momenta being too demanding at two-loop order,\footnote{Even with the standard FP Lagrangian,
the calculation remains quite technical and has been carried out only in the same configuration that we consider here \cite{Davydychev:1997vh}.} we focus on the particular configuration where
the momentum of the gluon vanishes.\footnote{Configurations where the ghost or antighost momentum is taken to zero are also
technically simpler. In the Landau gauge, they are even trivial due to
Taylor non-renormalization theorem (as recalled above) \cite{Taylor71}.} The calculations in this configuration are of the same order of complexity than those for the two-point functions. We stress, however, that this is precisely the configuration which lead to the least accurate results at one-loop order.
Therefore, the quality to be expected sets most probably a lower bound on the quality to be expected for two-loop calculations in the CF model (at least for this particular vertex).

\section{Ghost-antighost-gluon vertex}\label{sec:vertex}
In what follows, the ghost-antighost-gluon vertex will be written as
\beq
-V^{abc}_\mu(k,\ell)\equiv\begin{gathered}
  \includegraphics[width=0.40\linewidth]{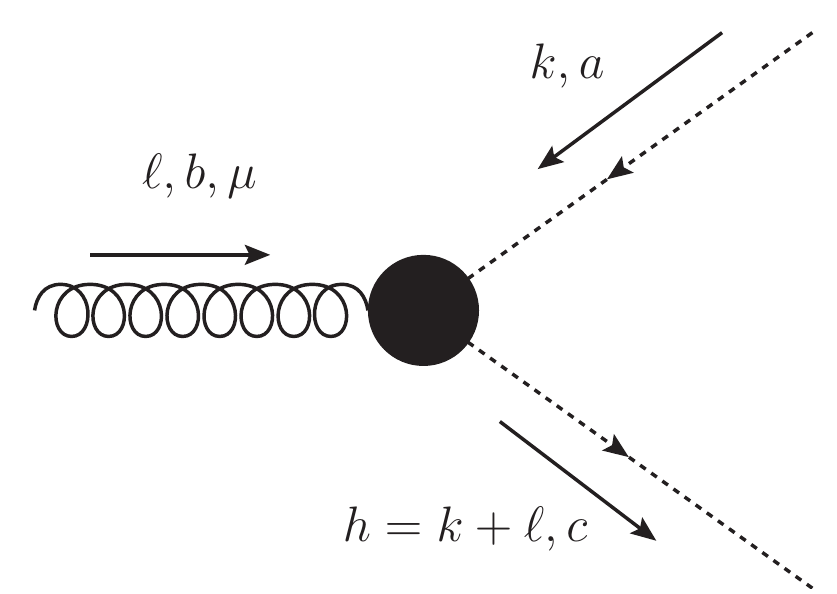}
 \end{gathered}\,,
 \eeq
 with $k$, $\ell$ and $\smash{h=k+\ell}$, the (incoming) ghost, (incoming) gluon and (outgoing) antighost momenta, respectively.
 From Lorentz symmetry, the vertex has {\it a priori} two tensor components:
 \beq
 V^{abc}_\mu(k,\ell)\!=\!k_\mu V^{abc}(k^2,k\cdot\ell,\ell^2)\!+\!\ell_\mu W^{abc}(k^2,k\cdot \ell,\ell^2).\nonumber\\
 \eeq
 However, in the limit of zero sources, the equation of motion for the Nakanishi-Lautrup field $ih^a$ in (\ref{eq_lagrang}) reads
 $\partial_\mu A_\mu^a=0$, which means that, once this constraint is imposed, the effective action,
 and then the vertex functions, should be restricted to transverse gauge field configurations.
 In particular, the only component of $V^{abc}_\mu(k,\ell)$ that contributes to connected correlation functions is
 \beq
 V^{abc}_{\perp,\mu}(k,\ell) & \!\!\equiv\!\! & P^\perp_{\mu\nu}(\ell)\,V^{abc}_\mu(k,\ell)\nonumber\\
 & \!\!=\!\! & P^\perp_{\mu\nu}(\ell)\,k_\nu V^{abc}(k^2,k\cdot \ell,\ell^2)\,,
 \eeq
 that is essentially $V^{abc}(k^2,k\cdot \ell,\ell^2)$. Furthermore, owing to the symmetry 
 \beq
 c^a\to\bar c^a, \quad \bar c^a\to -c^a, \quad ih^a\to ih^a-f^{abc}\bar c^bc^c\,,
 \eeq
 which applies when $ih^a$ is on-shell (that is in the absence of an associated source), it is easily deduced
 that $V^{abc}_{\perp,\mu}(k,\ell)=V^{cba}_{\perp,\mu}(-h,-\ell)$, from which it follows that
 \beq
 V^{abc}(k^2,k\cdot \ell,\ell^2)=-V^{cba}(h^2,h\cdot \ell,\ell^2)\,.
 \eeq
 In this work, we are interested in the limit of vanishing gluon momentum, in which case the previous identity means
 that $\smash{V^{abc}(k^2)\equiv V^{abc}(k^2,0,0)}$ is antisymmetric under $a\leftrightarrow c$ and can thus be parametrized as
 \beq
 V^{abc}(k^2)=ig_B\mu^\epsilon f^{abc}v(k^2)\,,
 \eeq
 since the other possible color tensor $d^{abc}$ is symmetric under $a\leftrightarrow c$.\footnote{More generally,
 the tensor $d^{abc}$ can be discarded using charge conjugation invariance \cite{Smolyakov:1980wq}.}
 
It is easily seen that the scalar function $v(k^2)$ renormalizes as $\smash{v(k^2)\to \sqrt{Z_A}Z_c Z_g v(k^2)}$. It is then finite,
owing to the non-renormalization theorem alluded to above, and even invariant under the RG-flow in the renormalization scheme considered here.
Moreover, since the vanishing of the loop corrections to the ghost-antighost-gluon vertex in the limit $k\to 0$ (see the
previous section) originates both from the vertex attached to the ghost leg and from the vertex attached to the antighost leg in the case
where the gluon momentum vanishes,\footnote{This is because the external antighost momentum is also equal to $k$ in this case and
multiplies naturally any diagram.} we find that the loop corrections to $V^{abc}_\mu(k,0)$ vanish at least like $k^2$ when $k\to 0$ and
thus that
\beq
V^{abc}(k^2)=\frac{k_\mu}{k^2}V^{abc}_\mu(k,0)\,
\eeq
approaches its tree-level value as $k^2\to 0$. In other words, $v(k^2\to 0)=1$. We mention that, in the scheme considered here,
this property is valid both for the bare and the renormalized $v(k^2)$, which are in fact equal to each other. In any scheme where
the finite part of $\sqrt{Z_A}Z_cZ_g$ is not fixed to $1$, the renormalized $v(k^2)$ obeys instead $v(k^2\to 0)=\sqrt{Z_A}Z_cZ_g$.
We also note that the above argument is only valid in the absence of infrared divergences in the limit $k\to 0$, which is made
possible here by the presence of a mass in the gluon propagator. This is an important difference with respect to standard perturbative
calculations in the FP model, where $v(k^2)$ diverges as $k\to 0$, in obvious disagreement with lattice results, as we recall below.

The function $v(k^2)$ has been computed at one-loop order in the CF model in \cite{Pelaez:2013cpa} and compared to lattice
simulations \cite{Cucchieri:2008qm}. Here, we would like to evaluate the two-loop corrections to this quantity to further constrain
the validity of the CF
model as an effective description of YM theory in the infrared. 

\subsection{Diagrams}
For later convenience, we write the two-loop expression for $v(k^2)$ at bare level as
\beq\label{eq:v_bare}
v(k^2)=1+\lambda_B\,v_1(k^2,m_B^2)+\lambda_B^2\,v_2(k^2,m_B^2)\,,
\eeq
where $v_n(k^2,m_B^2)$ with $n=1$ or $2$ represent the sum of one-loop and two-loop Feynman diagrams respectively. By writing $\lambda_B^n$ in front of $v_n(k^2,m_B^2)$,
we have naturally factored out the corresponding power of $g_B$ (which is nothing but $g_B^{2n}$) as well as the color
factor (which is nothing but $N^n$). Moreover, as it is customary, see for instance \cite{jag10}, we have absorbed a
factor $(16\pi^2)^n$ in $v_n(k^2,m_B^2)$, together with the factor $\mu^{2n\epsilon}$ that comes along with $g_B^{2n}$, see the remark at the end
of Sec.~\ref{ssec:IS}. In practice, this means that, in computing Feynman diagrams, the $d$-dimensional momentum integrals are replaced by
\beq
\int \frac{d^dp}{(2\pi)^d}\to\int_p\equiv 16\pi^2\mu^{2\epsilon}\int \frac{d^dp}{(2\pi)^d}\,.
\eeq 
We emphasize that, despite the presence of the factors $\mu^{2\epsilon}$, they all recombine into the $\mu$-independent dimensionful bare coupling $g_B\mu^\epsilon$, as it should be the case since $v(k^2)$ is a bare quantity and is, therefore, $\mu$-independent.

\begin{figure}
\includegraphics[width=0.5\linewidth]{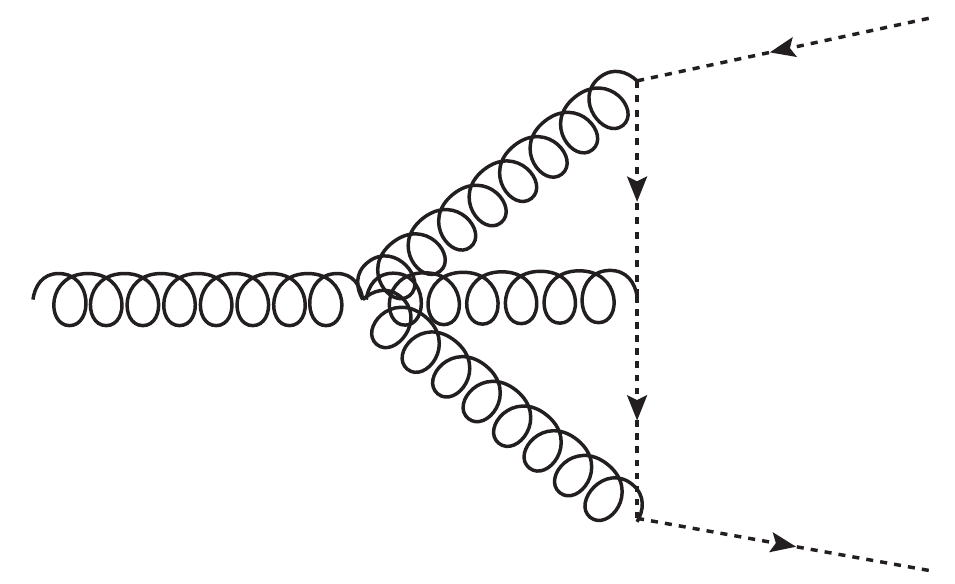}
\caption{Planar diagram that can not be seen neither  as a self-energy correction nor as a vertex correction to
the one-loop ghost-antighost-gluon vertex.}\label{fig:planar}
\end{figure}

\begin{figure}[h]
\includegraphics[width=0.42\linewidth]{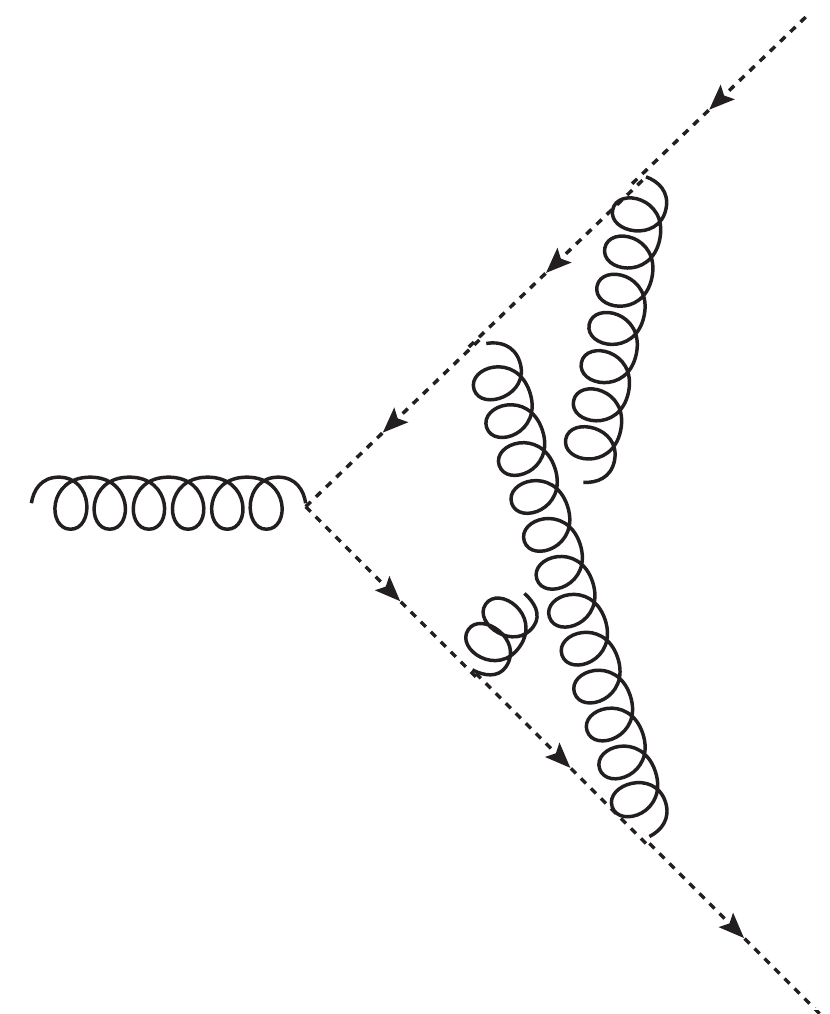}\\
\vspace{0.5cm}

\includegraphics[width=0.47\linewidth]{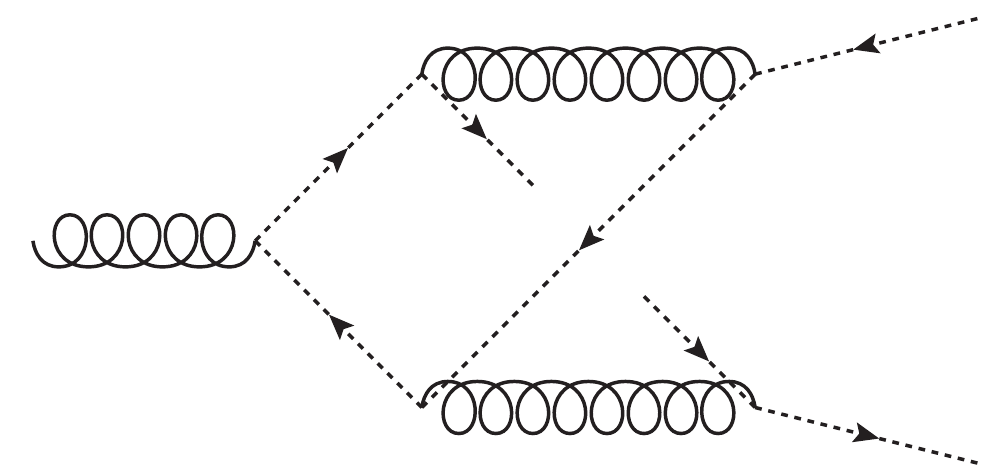}\\
\vspace{0.5cm}

\includegraphics[width=0.47\linewidth]{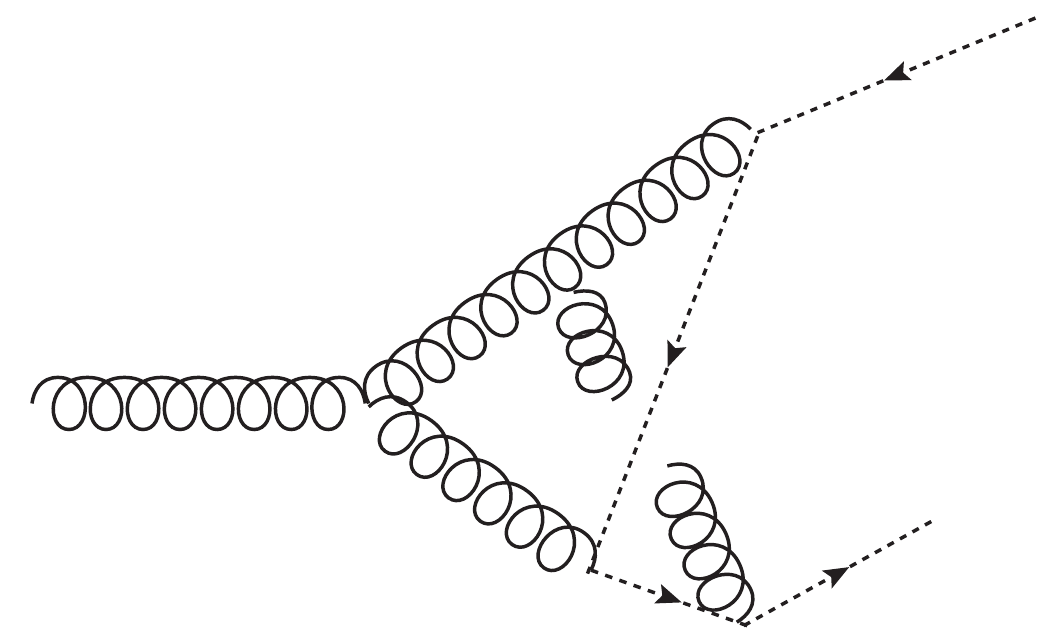}\\
\vspace{0.5cm}

\includegraphics[width=0.47\linewidth]{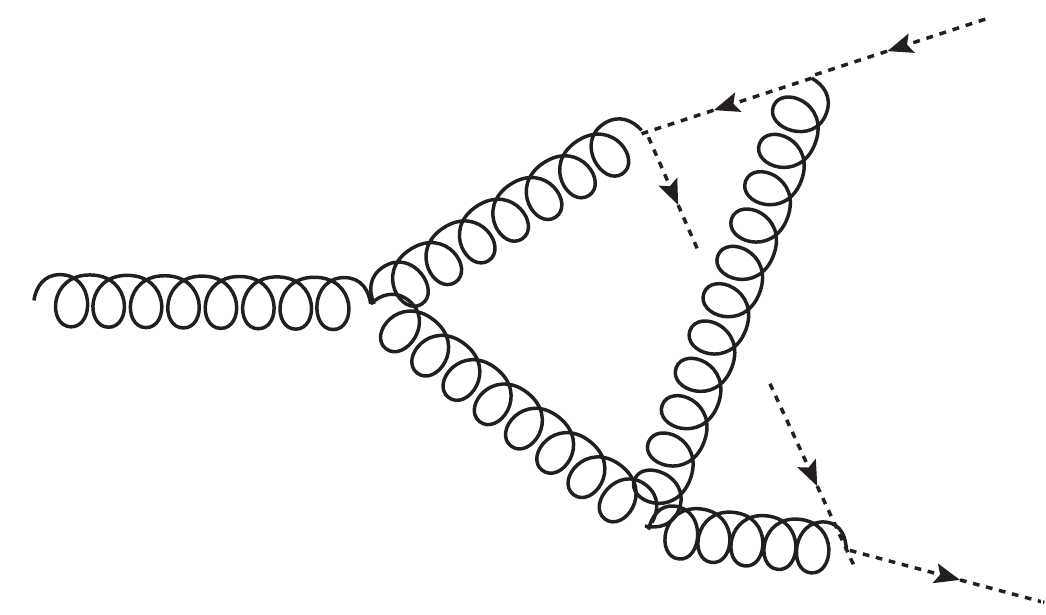}
\caption{Non-planar diagrams at two-loop order. All of them vanish as explained in the text.}\label{fig:non_planar}
\end{figure}

The Feynman diagrams contributing to $v_1(k^2,m_B^2)$ have been computed in \cite{Pelaez:2013cpa}. In order to ease the computation of the
Feynman
diagrams contributing to $v_2(k^2,m_B^2)$ while using the earlier one-loop results, it is convenient to organize the various diagrams
contributing to the ghost-antighost-gluon vertex in three categories:  i) those corresponding to self-energy corrections,
ii) those corresponding
to vertex corrections, and iii) the rest.  The diagrams corresponding to the categories (i) and (ii) are gathered in Appendix~\ref{twoloopdiags}.
Among those of category (iii), we need only to evaluate the planar diagram of Fig.~\ref{fig:planar} since the other (non-planar) diagrams,
see Fig.~\ref{fig:non_planar}, all vanish \cite{Davydychev:1997vh}. Indeed their color factor is
\beq
& & f^{eaf}f^{fhg}f^{gic}f^{bhd}f^{die}\nonumber\\
& & \hspace{0.5cm}=\,-f^{ade}f^{eif}f^{fhg}f^{gic}f^{bhd}-f^{aie}f^{efd}f^{fhg}f^{gic}f^{bhd}\,,\nonumber\\
\eeq
where we have used Jacobi identity. In the first term, we can identify the color loop $f^{eif}f^{fhg}f^{gic}=f^{ief}f^{fhg}f^{gci}=-(N/2)f^{ehc}$, while in the second term, we have the loop $f^{efd}f^{fhg}f^{bhd}=f^{def}f^{fgh}f^{hbd}=-(N/2)f^{egb}$. It follows that the color factor of the non-planar diagrams reads
\beq
& & \frac{N}{2}\big[f^{ade}f^{bhd}f^{ehc}+f^{aie}f^{gic}f^{egb}\big]\nonumber\\
& & \hspace{0.5cm}=\,\frac{N^2}{4}\big[f^{abc}-f^{abc}\big]=0\,,
\eeq
as announced.

\pagebreak

\subsection{Reduction to master integrals}
After each diagram contributing to $v(k^2)$ has been written in terms of the corresponding Feynman integral, we proceed to reducing the
latter into one-loop master integrals,
\beq
A_m & \!\!\equiv\!\! & \int_p G_m(p)\,,\label{eq:A}\\
 B_{m_1m_2}(k^2) & \!\!\equiv\!\! & \int_pG_{m_1}(p)G_{m_2}(p+k)\,,
\eeq
with
\beq
G_m(p)\equiv \frac{1}{p^2+m^2}\,,
\eeq 
and two-loop master integrals,
\beq
& & S_{m_1m_2m_3}(k^2)\nonumber\\
& & \hspace{0.6cm}\equiv\,\int_pG_{m_1}(p)B_{m_2m_3}((p+k)^2)\,,\\
& & U_{m_1m_2m_3m_4}(k^2)\nonumber\\
& & \hspace{0.6cm}\equiv\,\int_pG_{m_2}(p)G_{m_1}(p+k)B_{m_3m_4}(p^2)\,,\\
& & M_{m_1m_2m_3m_4,m_5}(k^2)\nonumber\\
& & \hspace{0.6cm}\equiv \int_pG_{m_1}(p)G_{m_3}(p+k)\nonumber\\
&  & \hspace{0.8cm}\times\,\int_qG_{m_2}(q)G_{m_4}(q+k)G_{m_5}(q-p)\,,\label{eq:M}
\eeq
that can then be efficiently computed numerically using the TSIL package \cite{jag10}. 

The reduction into master integrals was performed using the FIRE package \cite{Smirnov:2008iw}. The output of the procedure
is an expression for $v_1(k^2,m_B^2)$ as a sum of integrals of the type $A$ and $B$ multiplied by rational fractions
involving $k^2$, $m_B^2$ and $d$, and, similarly, an expression for $v_2(k^2,m_B^2)$ as a sum of integrals of the type $S$, $U$, $M$, or
products of the integrals $A$ and $B$, multiplied again by rational fractions.

It is worth noting that all the master integrals given above can be obtained formally from $M$ by switching off some of the propagators.
In our reduction of $v(k^2)$, we also found some integrals obtained from $M$ by elevating, in addition, one of the propagators to the
power $-1$. Fortunately, these integrals can be reduced to the master integrals listed above. We illustrate this reduction in
Appendix \ref{app:minus_one}.

\subsection{UV divergences}
As already mentioned above, $v(k^2)$ is UV finite. Of course, this is explicit only after one expresses $v(k^2)$ in terms of renormalized parameters. We thus plug the rescalings (\ref{eq:ren_param}) into Eq.~(\ref{eq:v_bare}) and expand to order $g^4$ using that $\delta Z_\lambda\equiv Z^2_g-1$ and $\delta Z_{m^2}\equiv Z_{m^2}-1$ are both of order $\lambda\propto g^2$. We find
\beq\label{eq:v_renorm}
v(k^2) & = & 1+\lambda\,v_1(k^2,m^2)+\lambda^2\,v_2(k^2,m^2)\nonumber\\
& + & \lambda\left(\delta Z_\lambda+\delta Z_{m^2}m^2\frac{\partial}{\partial m^2}\right)\!v_1(k^2,m^2)\,.
\eeq
The derivative $\partial v_1/\partial m^2$ generates integrals of the type $\partial A_m/\partial m^2$ and $\partial B_{m0}(k^2)/\partial m^2$.
The latter can be re-expressed in terms of the master integrals $A_m$ and $B_{m0}$ as
\beq
\frac{\partial A_m}{\partial m^2} & = & \left(\frac{d}{2}-1\right)\frac{A_m}{m^2}\,,\\
\frac{\partial B_{m0}(k^2)}{\partial m^2} & = & \frac{(d-3)B_{m0}(k^2)+\partial A_m/\partial m^2}{k^2+m^2}\,,
\eeq
\vglue1mm
\noindent{where the first identity is easily obtained using dimensional analysis and the second from integration by parts techniques, more precisely by writing the two identities}
\beq
0 & = & \int_p \frac{\partial}{\partial p_\mu}\frac{p_\mu}{(p^2+m^2)(p+k)^2}\,,\\
0 & = & \int_p \frac{\partial}{\partial p_\mu}\frac{k_\mu}{(p^2+m^2)(p+k)^2}\,,\label{eq:mark}
\eeq
as a linear system for $\partial B_{m0}(k^2)/\partial m^2$ and a second integral that is not needed here.

The first two terms in Eq.~(\ref{eq:v_renorm}) correspond to the one-loop result. They do not involve any counterterm which means that the integrals
entering $v_1(k^2,m^2)$ should combine into a UV finite contribution. This is easily verified. In fact, each one-loop diagram is easily seen to be
finite. This is because, at the level of the vertex attached to the ghost leg, one
has $P^\perp_{\rho\sigma}(q)(q+k)_\sigma=P^\perp_{\rho\sigma}(q)k_\sigma$, where $q$ denotes the momentum of the gluon propagator attached to the vertex.
Owing to the contraction with the transverse projector, one power of $q$ is lost in the power-counting, yielding a superficial degree of divergence
equal to $-1$.

The same reasoning applies to each diagram contributing to $v_2(k^2,m^2)$ which consequently also have a superficial degree of divergence
equal to $-1$. However, this does not mean that the diagrams are finite since, being
two-loop diagrams,
they can also contain subdivergences. The latter should be precisely killed by the second line of Eq.~(\ref{eq:v_renorm}) with $\delta Z_\lambda$ and $\delta Z_{m^2}$ taken at one-loop accuracy (i.e. order $g^2$ or $\lambda$).
This is a non-trivial check of our reduction of $v_2(k^2,m^2)$ using FIRE. Indeed the individual terms contributing to $v_2(k^2,m^2)$ after the reduction into master integrals contain simple, double and even triple poles in $1/\epsilon$. The simple and double poles come from the two-loop master integrals or from products of one-loop master integrals.
The triple poles originate from the fact that the reduction into master integrals generates certain terms with an extra prefactor $(4-d)^{-1}$. More precisely, those are
\begin{widetext}
\beq\label{eq:spurious}
& & \frac{(4-d)^{-1}}{96}\left[-\left(14+11\frac{k^2}{m^2}\right)\frac{A_m}{m^2}B_{00}-3\left(2+\frac{k^2}{m^2}\right)B_{m0}B_{00}-\left(2-11\frac{k^2}{m^2}\right)\left(1+\frac{m^2}{k^2}\right)\frac{S_{m00}}{m^2}\right.\nonumber\\
& & \left.\hspace{1.5cm}-\,\left(6+13\frac{k^2}{m^2}\right)\frac{S_{000}}{m^2}+\left(5+2\frac{m^2}{k^2}\right)\frac{I_{m00}}{m^2}-8\left(1+\frac{k^2}{m^2}\right)U_{00m0}+3\left(2+\frac{k^2}{m^2}\right)U_{0m00}\right],
\eeq
\end{widetext}
where $I_{m_1m_2m_3}$ stands for $\smash{S_{m_1m_2m_3}(k^2=0)}$. We have verified that the triple poles cancel among the various terms in this formula, as it should be the case since there is no other source of triple poles. The double poles cancel against the other contributions to $v_2(k^2,m^2)$, as it should happen if one wants to have a chance of cancelling the subdivergences with the second line of Eq.~(\ref{eq:v_renorm}) which contains only simple poles.\footnote{We mention that the absence of multiple poles applies in fact to each diagrammatic contribution to $v_2(p^2,m^2)$ taken separately, since their superficial degree of divergence is $\delta=-1$, meaning that there are at most (one-loop) subdivergences, and thus at most simple poles in $1/\epsilon$.} Finally, we have checked that the remaining simple poles in $v_2(k^2,m^2)$ are exactly opposite to those in the second line of Eq.~(\ref{eq:v_renorm}). We recall that $\delta Z_\lambda$ and $\delta Z_{m^2}$ are already known from the renormalization of
the two-point functions and the two non-renormalization theorems. At one-loop order we have
\beq
\delta Z_\lambda & = & \lambda\left(\frac{z_{\lambda11}}{\epsilon}+z_{\lambda10}\right),\label{eq:zlambda}\\
\delta Z_{m^2} & = & \lambda\left(\frac{z_{m^211}}{\epsilon}+z_{m^210}\right),\label{eq:zm2}
\eeq
with $z_{\lambda11}=-11/3$, $z_{m^211}=-35/12$, and where $z_{\lambda10}$ and $z_{m^210}$ depend on the considered scheme.

Other similar checks that test non-trivial cancellations between the many terms generated by the FIRE reduction will be presented in Sec.~\ref{sec:checks}.

\subsection{Finite parts}
Beyond the UV divergences, we are of course interested in the UV finite contributions to $v(k^2)$. Since the counterterms, the master integrals and even some prefactors multiplying these integrals contain poles in $1/\epsilon$, it is a question to which order in $\epsilon$ one should expand both the counterterms and the master integrals in order not to miss any contribution of order $\epsilon^0$. 

Consider first the counterterms $\delta Z_\lambda$ and $\delta Z_{m^2}$ that multiply $v_1(k^2,m^2)$ in Eq.~(\ref{eq:v_renorm}).
Because the latter is UV finite, its $\epsilon$-expansion starts at order $\epsilon^0$. This means that it is enough to consider the counterterms at
order $\epsilon^0$ as well.\footnote{This is in contrast to what happened in the case of the two-point functions
where certain counterterms needed to be expanded to order $\epsilon^1$ because they multiplied UV divergent integrals \cite{Gracey:2019xom}.} On the contrary, $v_1(p^2,m^2)$
itself, and therefore the integrals $A$ and $B$ that appear linearly within it,\footnote{We mention, for completeness, that the prefactors of these integrals do not contain any extra factor $(4-d)^{-1}$ in this case.} need to be expanded to order $\epsilon^1$. In the case of $v_2(p^2,m^2)$, in any term that does not contain the extra prefactor $(4-d)^{-1}$, the master integrals $A$ and $B$ need to be expanded to order $\epsilon^1$ while the others need to be expanded only to order $\epsilon^0$. In contrast, for those terms in Eq.~(\ref{eq:spurious}), $A_m$, $B_{m0}$, $B_{00}$ need to be expanded to order $\epsilon^2$, whereas $S_{000}$, $S_{m00}$, $I_{m00}$, $U_{0m00}$ and $U_{00m0}$ need to be expanded to order $\epsilon^1$. These expansions are given in Appendix \ref{app:epsilon}.

We mention finally that, because the renormalized expression (\ref{eq:v_renorm}) is just an expansion to order $g^4$ of the $\mu$-independent
expression (\ref{eq:v_bare}), it should be $\mu$-independent up to contributions of order $g^6$. The $\mu$-independence is
crucial since it allows us to choose $\smash{\mu=k}$ in practice, and, therefore, to obtain a controlled perturbative estimate of the IR and
UV tails, where an evaluation at a fixed $\mu$ would generate large logarithms $\ln(k^2/\mu^2)$ spoiling the validity of the
perturbative expansion.\footnote{We stress that this procedure can only be applied directly along renormalization group trajectories without a Landau pole,
such as those in the IS scheme below the separatrix. We refer to Sec.~\ref{schemedep} for the appropriate modification
of the procedure in those schemes that suffer from a Landau pole.} We shall implement this choice of scale when presenting our results in the IS scheme in comparison to the lattice results. We note, nonetheless, that most of the tests that we perform in the next section are valid for a fixed value of the renormalization scale $\mu$.

\section{Crosschecks}\label{sec:checks}
As we have mentioned above, the reduction of $v(k^2)$ into master integrals generates an expression with many terms, which it is wise
to test in as many ways as possible. The tests we consider are always of the same type: we use properties of $v(k^2)$ that are not obeyed
by the individual many terms making the reduced expression for $v(k^2)$ but which emerge as the result of cancellations between these many
terms.\footnote{Most of the properties apply, however, to the individual diagrams entering $v(k^2)$, so we could in principle perform
a finer test by checking them for each individual diagram.} We have already seen one example of such cancellations: the cancellation
of triple and double poles in $1/\epsilon$, and the cancellation of simple poles against the counterterm contributions in
Eq.~(\ref{eq:v_renorm}). We next discuss various other properties that rely on similar cancellations: the asymptotic UV and IR behaviors, the
regularity of $v(k^2)$ for $\smash{k^2=m^2}$, and the regularity and correctness of the $\smash{m^2\to 0}$ limit.

\subsection{UV behavior}
We have seen above that each diagram contributing to $v(k^2)$ has a superficial degree of divergence $\delta=-1$. From Weinberg theorem
we would naively expect that $v(k^2)$ behaves like $1/k$ (up to logarithms) as $k^2\to\infty$. However, one should not forget that the
reduction of the superficial degree of divergence leaves a factorized extra factor of $k$, see the discussion
below Eq.~(\ref{eq:mark}), meaning
that $v(k^2)$ should behave logarithmically at large $k^2$. 

On the other hand, the individual terms that make $v(k^2)$ after the FIRE reduction, can grow much faster. In order to check that
these unwanted contributions cancel, we used UV expansions for the various master integrals given above, obtained using our own
implementation of the algorithm
described in Ref.~\cite{Davydychev:1993pg} and which exploits Weinberg theorem. As compared to our earlier implementation \cite{Gracey:2019xom}, where we needed only to determine the UV expansions of the $\epsilon^0$ contributions to these integrals, here we needed to extend the routine to obtain the UV expansion of the corresponding $\epsilon^1$ contributions,
when necessary. In particular, this meant computing $I_{m00}$ at order $\epsilon^1$ which is easily done using Eq.~(\ref{eq:Im}).

At leading order, we find
\beq\label{eq:pprev}
v(k^2\to\infty) & \!\!=\!\! & 1+\frac{3\lambda}{4}+\lambda^2\left(\frac{11+3z_{\lambda11}}{4\epsilon}+\frac{317}{32}+z_{\lambda11}\right.\nonumber\\
& \!\!+\!\! & \left.\frac{3z_{\lambda10}}{4}+\frac{22+3z_{\lambda11}}{4}\ln\frac{\bar\mu^2}{k^2}\right)\!+\!{\cal O}\left(\frac{m^2}{k^2}\right)\!,\nonumber\\
\eeq
where $z_{\lambda11}$ and $z_{\lambda10}$ were defined in Eq.~(\ref{eq:zlambda}) and $\bar\mu^2\equiv 4\pi\mu^2e^{-\gamma}$, with $\gamma$ the Euler constant.
Upon using the value of $z_{\lambda11}$, this becomes
\beq\label{eq:previous}
v(k^2\to\infty) & \!\!=\!\! & 1+\frac{3\lambda}{4}+\lambda^2\left(\frac{599}{96}+\frac{3z_{\lambda10}}{4}-\frac{11}{4}\ln\frac{k^2}{\bar\mu^2}\right)\nonumber\\
& \!\!+\!\! & {\cal O}\left(\frac{m^2}{k^2}\right)\!,
\eeq
The absence of logarithms in the contribution of order $\lambda$ is reminiscent of the fact that $v_1(k^2,m^2)$ is finite, whereas the presence of a simple logarithm in the contribution of order $\lambda^2$ comes from the fact that the diagrams in $v_2(p^2,m^2)$ have
only subdivergences but no global divergences. We also note that the running of $\lambda$ obeys
\beq
0 & = & \mu\frac{\partial}{\partial\mu}\ln(\lambda_B\mu^{2\epsilon})=\mu\frac{\partial\ln Z_\lambda}{\partial\mu}
+\mu\frac{\partial\ln\lambda}{\partial\mu}+2\epsilon\nonumber\\
& = & \frac{\lambda}{Z_\lambda}\mu\frac{\partial z_{\lambda10}}{\partial\mu}
+\left(1+\frac{\delta Z_\lambda}{Z_\lambda}\right)\mu\frac{\partial\ln\lambda}{\partial\mu}+2\epsilon\,,
\eeq
that is
\beq
\mu\frac{\partial\lambda}{\partial\mu}=\left(2z_{\lambda11}-\mu\frac{\partial z_{\lambda10}}{\partial\mu}\right)\lambda^2+{\cal O}(\lambda^3)\,.\label{eq:lambda_running}
\eeq
From this, it is easily checked that the $\mu$-dependence in Eq.~(\ref{eq:previous}) appears formally only at order $\lambda^3\propto g^6$, as already anticipated above. The corrections of order $m^2/k^2$ also contain logarithms and involve the finite part $z_{m^210}$ of $\delta Z_{m^2}$.

We mention finally that the choice $\smash{\mu=k}$ that we shall eventually make in the IS scheme does not jeopardize the ordering
in powers of $m^2/k^2$ in Eq.~(\ref{eq:previous}) because $m(k)$ runs to $0$ in the UV \cite{Tissier:2011ey,Reinosa:2017qtf}. Moreover, each
term in the expansion is dominated by the one with less powers of $\lambda(k)$. We conclude that, once the running is included, $v(k^2)$ approaches $1$ logarithmically in the UV.

\subsection{IR behavior}
Similar remarks apply in the opposite $k^2\to 0$ limit. We have seen that $v(k^2\to 0)\to 1$. However, this property is not necessarily
true for the individual terms contributing to $v(k^2)-1$. In order to check that the appropriate cancellations occur, we used IR expansions
for the various master integrals listed above,
obtained by implementing the algorithm in Ref.~\cite{Davydychev:1992mt}. In certain cases, the algorithm cannot be applied
and one needs to resort to a more sophisticated version described in Ref.~\cite{Berends:1994sa}. In the present case, for most
of the problematic integrals, we could circumvent the difficulty using the fact that these integrals are known analytically. For a few
of them which do not have a known analytic expressions, in particular for the order $\epsilon^1$ contributions to $U_{0m00}$ and $U_{00m0}$, we
implemented our own strategy which we detail in Appendix \ref{appsec:lowk}.

\begin{widetext}
At first non-trivial order, we find
\beq
v(k^2\to 0) & \!\!=\!\! & 1+\left\{\left(\frac{17}{48}-\frac{1}{8}\ln\frac{k^2}{m^2}\right)\lambda+\left(\frac{2323}{1152}-\frac{29}{1152}\pi^2-\frac{999}{128}S_2+\frac{17}{48}z_{\lambda10}-\frac{11}{48}z_{m^210}+\frac{3}{32}\ln\frac{m^2}{\bar\mu^2}\ln\frac{k^2}{m^2}\right.\right.\nonumber\\
& & \hspace{1.4cm}\left.\left.+\left[-\frac{5}{64}-\frac{z_{\lambda10}}{8}+\frac{z_{m^210}}{8}\right]\ln\frac{k^2}{\bar\mu^2}+\left[-\frac{53}{96}+\frac{z_{\lambda10}}{8}-\frac{z_{m^210}}{8}\right]\ln\frac{m^2}{\bar\mu^2}\right)\lambda^2\right\}\frac{k^2}{m^2}+{\cal O}\left(\frac{k^4}{m^4}\right)\!,\label{eq:IR}
\eeq
\end{widetext}
where $z_{m^210}$ was defined in Eq.~(\ref{eq:zm2}) and 
\begin{equation}
S_2=\frac{4}{9\sqrt{3}} \mathrm{Im}\big(\mathrm{Li_2}(e^{i\pi/3})\big).
\end{equation}
In order to test the $\mu$-independence of this expression, we need the running of the mass, which we derive by writing
\beq
0=\mu\frac{\partial}{\partial\mu}\ln(Z_{m^2}m^2)
\eeq
which leads to
\beq
\mu\frac{\partial m^2}{\partial\mu} & = & -m^2\mu\frac{\partial}{\partial\mu}\ln Z_{m^2}\nonumber\\
& = & -m^2\mu\frac{\partial}{\partial\mu}\left\{\lambda\left(\frac{z_{m^211}}{\epsilon}+z_{m^210}\right)\right\}\nonumber\\
& = & -m^2\mu\frac{\partial}{\partial\mu}\left\{\lambda_B\mu^{2\epsilon}\mu^{-2\epsilon}\left(\frac{z_{m^211}}{\epsilon}+z_{m^210}\right)\right\}\nonumber\\
& = & \lambda m^2\left(2z_{m^211}-\mu\frac{\partial z_{m^210}}{\partial \mu}\right).
\eeq
Together with Eq.~(\ref{eq:lambda_running}), this allows one to check that the asymptotic behavior (\ref{eq:IR}) is $\mu$-independent up to higher order contributions ($\sim \lambda^3\sim g^6$).

We mention finally that, as it is obvious from Eq.~(\ref{eq:IR}), $v(k^2\to 0)\to 1$, in line with the argumentation given at the beginning
of Sec.~\ref{sec:vertex} in the case where $m\neq 0$. This property is not affected by the running in the IS scheme  since both the mass and
the coupling run logarithmically to zero in the infrared.

\subsection{Regularity at $\smash{k^2=m^2}$}
The function $v(k^2)$ is not only regular for $\smash{k^2=0}$ but in fact for any other Euclidean momentum. However, the various contributions that
enter in the reduction into master integrals might be singular at some values of $k^2$ and it is thus necessary to check that the corresponding residue
vanishes. Aside from the cancelling singular contributions at $k^2=0$ that we treated in the previous section, we only found intermediate singular
contributions at $\smash{k^2=m^2}$. When adding all the contributions, the corresponding residue writes
\beq
 \frac{\lambda^2}{64}\hspace{-.4cm}&&\Big((d-2)\big(A_m B_{00}(m^2)+I_{m00}\big)\!+\!(d-3)m^2\big(B_{00}(m^2)\big)^2\nonumber\\
& & \hspace{0.4cm}+\,(8-3d)S_{m00}(m^2)\!+\!\frac{d-4}{2}m^4M_{0000m}(m^2)\,\Big).
\eeq
Fortunately, all these integrals are known exactly and it is easily checked that the residue indeed vanishes, as
expected. Similar singularities (although at a different value of $k^2$) appeared in the intermediate steps leading to the evaluation of the gluon and ghost two-point functions at two-loop order  \cite{Gracey:2019xom}. 

We stress that we are here implicitly assuming that $m\neq 0$. The case $m=0$ yields a true singularity at $k^2=m^2=0$, as we recall in the next section.

\subsection{Zero mass limit}
A final check involves the limit $m\to 0$. This limit is regular for any $k^2>0$ and the expression for $v_{m^2=0}(k^2>0)$ has
been determined
in \cite{Davydychev:1997vh}. That the limit is not regular for $k^2=0$ can be simply seen from the fact that $v_{m^2\neq 0}(k^2\to0)\to 1$,
whereas $v_{m^2=0}(k^2\to0)\to \infty$. As already mentioned above, this is an important difference with regard to the comparison with the lattice data.
Putting these considerations aside, investigating the $m\to 0$ limit of our result represents a double check of the reduction into master integrals
since 1) individual terms in the reduction are not necessarily regular in the limit $m\to 0$ and cancellations should occur in order to ensure
the regularity of the
limit, and, 2) the limit should coincide with the result of \cite{Davydychev:1997vh}.

We can envisage taking the limit $m\to 0$ using various strategies. One possibility is to exploit dimensional analysis to write any of the master
integrals given above as
\beq\label{eq:dim_analysis}
(\mu^{2\epsilon})^L F(p^2,m^2)=(\mu^{2\epsilon})^L (m^2)^{D/2}F(p^2/m^2,1)\,,
\eeq 
where $L$ is the number of loops and $D$ the mass dimension of the integral (letting aside the powers of $\mu$ that multiply it).
It is clear from this relation that the low mass expansion of any master integral can be obtained using the large momentum
expansion, as discussed above. Consequently, the zero mass limit of $v_{m^2}(k^2)$ is nothing but
the leading term in the expansion (\ref{eq:pprev}), which we checked coincides with the result of \cite{Davydychev:1997vh} in the
Landau gauge (up to the fact that we consider general renormalization factors).

Another possible strategy, which we used as a further crosscheck, is to Taylor expand the master integrals in powers of $m^2$. Although simpler a priori, the reason
why this approach works is a little bit subtle as we discuss at the end of this section and in Appendix~\ref{app:low_m}. We can proceed in two
ways depending on which of the $\epsilon$ and $m$ expansions is considered first. In both cases, we have to deal with the fact that certain
contributions to $v_{m^2}(k^2)$ are not regular and the correct limit $m\to 0$ is reached only after the corresponding singularities
have been cancelled. It turns out that these cancellations are more easy to handle if we first expand in $m$ for an arbitrary
dimension $d$, and only then expand in $\epsilon$. In fact, the regularity of the $m\to 0$ limit must take place for all
dimensions $d>2$ \cite{Tissier:2011ey}.

We find potentially singular terms proportional to $m^{-4}$ and $m^{-2}$. The contribution diverging as $m^{-4}$ is proportional to
\beq
(8-3d)S_{000}(k^2)+(d-4)\big[k^2U_{0000}(k^2)-I_{000}\big]\,,
\eeq
but this quantity vanishes fortunately ($I_{000}$ vanish trivially by itself). Similarly, the contribution diverging
as $m^{-2}$ is proportional to
\beq
& & (d-4)\big[2(d-3)k^2B_{00}^2(k^2)+(d-4)k^4 M_{00000}(k^2)\big]\nonumber\\
& & \hspace{0.5cm}-\,2(3d-8)(3d-10)S_{000}(k^2)\,,
\eeq
which turns out to be zero as well. To finally compare with the result of Ref.~\cite{Davydychev:1997vh}, we extract the $m^0$ term, which is proportional to
\beq
& & (d-4)(d-6)(d-8)\Big[(d-1)(d-4)k^4 M_{00000}(k^2)\nonumber\\
& & +\,2(2d^4-28d^3+134d^2-252d+147) k^2 B_{00}(k^2)\Big]\nonumber\\
& &  +\,2 (88832 - 224384 d + 223348 d^2 - 113336 d^3\nonumber\\
& &  \hspace{0.5cm}+\,31705 d^4 - 4895 d^5 + 386 d^6 - 12 d^7)S_{000}(k^2).\nonumber\\
\eeq
After expansion in powers of $\epsilon$, this leads again to the result of Ref.~\cite{Davydychev:1997vh} in the Landau gauge.

At first sight, it may seem suspicious that we were able to obtain the correct $m\to 0$ limit of $v_{m^2}(k^2)$ from
a na\"ive Taylor expansion of the master integrals in powers of the mass. Indeed, from Eq.~(\ref{eq:dim_analysis}) and Weinberg theorem, we
expect the low mass expansion of a given master integral to involve more terms than those that arise from a simple Taylor expansion. On the other
hand, because $v_{m^2}(k^2)$ is regular in the limit $m\to 0$, it turns out that the terms that are missed by using the na\"\i ve Taylor
expansion cancel each other and one ends up with the correct result. We illustrate these various features in Appendix~\ref{app:low_m}.\\

\begin{table}[htp]
\begin{tabular}{|c||l|c|c||c|c|c|c|}
 \hline
Group & \multicolumn{3}{| c ||}{$N=2$} & \multicolumn{3}{| c |}{$N=3$}\\
\hline
 \hline
Params.  & $\,\,\lambda_0$ & $m_0$ (GeV) & \hspace{1mm}$ \chi $ \hspace{1mm}& $\lambda_0$ & $m_0$ (GeV) & \hspace{1mm}$ \chi $\hspace{1mm} \\
\hline
\hline
1-loop & 0.34 & 0.45 & 10\% & 0.24 & 0.35 & 7\% \\
\hline
 2-loop & 0.38 & 0.39 & 6\% & 0.27 & 0.33 & 4\% \\
 \hline
\end{tabular}
\caption{\label{Table:parameters} Parameters in the IS scheme, as obtained from fitting the lattice results for the two-point functions, together with the corresponding error.}
\end{table}

\section{Results}
\label{sec:results}
In what follows we discuss our results for $v(k^2)$ in comparison to available
lattice simulations \cite{Cucchieri:2008qm,Ilgenfritz:2006he,Sternbeck:2006rd,Maas:2019ggf}. Except when explicitly stated, we work in the IS scheme and we employ two different strategies.

 First, we fix the parameters by fitting the
lattice data for the gluon and ghost propagators with the corresponding expressions in the CF model. In that case,
the values of the parameters $g$ and $m$ at a reference scale $\bar\mu_0={\rm 1}$\,GeV have been determined 
independently of the vertex and the vertex becomes then a pure prediction of the model. We stress that we get different parameters depending
on the considered accuracy of the propagators. In Table \ref{Table:parameters}, we summarize the obtained values in the
IS scheme and quote the corresponding errors.\footnote{We mention here that due to an unfortunate coding typo
in our determination of the error for the one-loop SU(2) results, the one-loop error of $7\%$ quoted in \cite{Gracey:2019xom} is in
fact an error of $10\%$. Interestingly, because the two-loop error was correctly estimated, the observed improvement from one-loop order to two-loop order is higher than what was originally claimed in that reference.}

As shown below, this procedure turns out to give excellent results in the SU($3$) case but gives much poorer results
for SU($2$). For this reason, we consider a second strategy where we first perform an independent fit of the various functions
and we then look for optimal parameters for which all functions are reproduced to a reasonable accuracy.

\subsection{CF prediction for the function $v(k^2)$ and comparison to the lattice data}
\begin{figure}[t]
\hglue-6mm\includegraphics[width=0.51\textwidth]{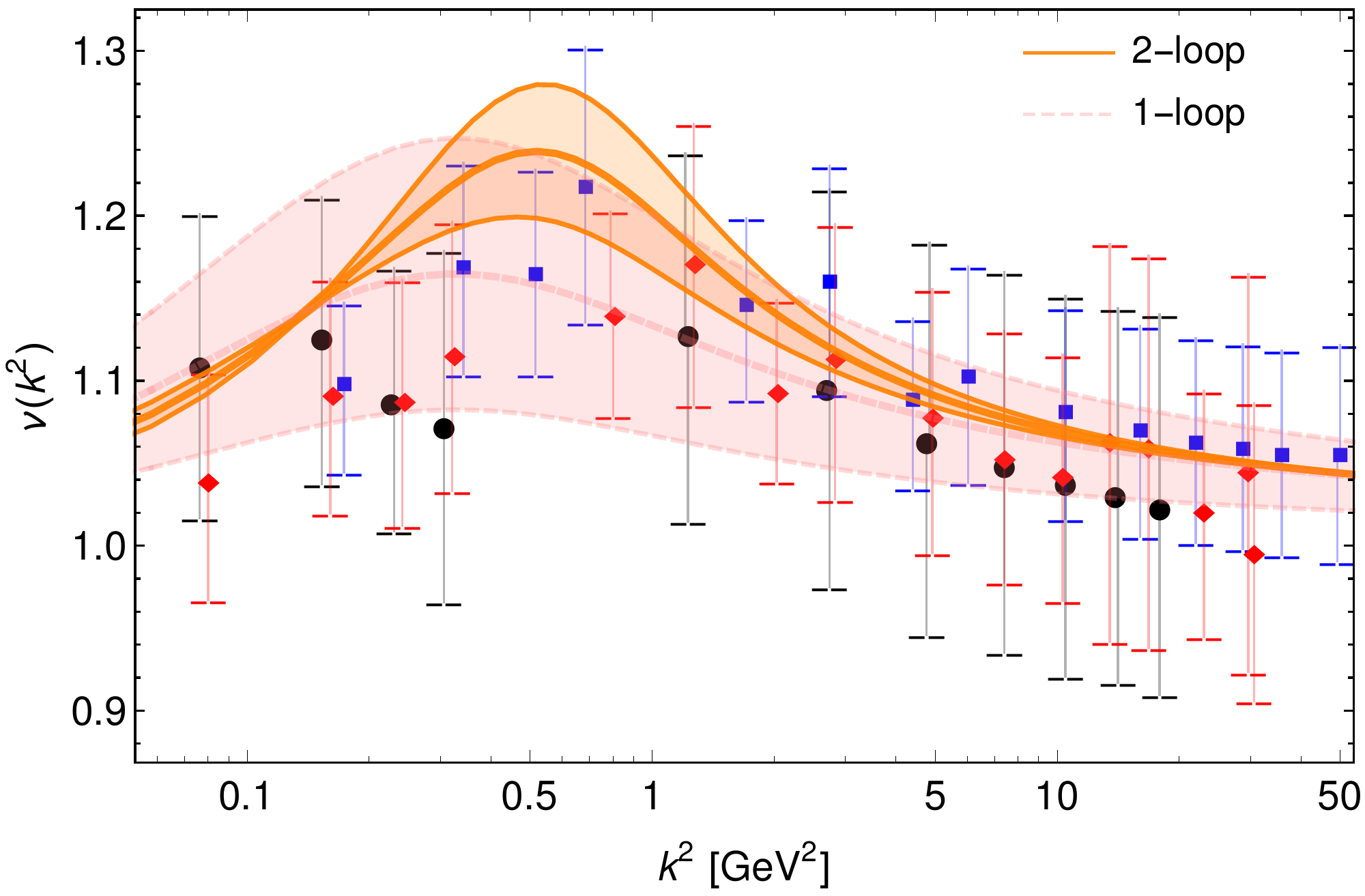}
\caption{CF prediction for the the function $v(k^2)$ in the SU(3) case and in the IS scheme, compared to the lattice data in the
Taylor scheme \cite{Ilgenfritz:2006he,Sternbeck:2006rd}. The parameters $m$ and $g$ at the initial scale $\bar\mu_0$ are those previously
determined from the fits of the gluon and ghost propagators. The lattice data were extracted manually from the plots in \cite{Ilgenfritz:2006he,Sternbeck:2006rd} using WebPlotDigitizer \cite{wpd}. We estimated the error related to the extraction procedure to be at most $0.8\%$.}
\label{Fig:v_SU(3)}
\end{figure}

\begin{figure}
\hglue-6mm\includegraphics[width=0.5\textwidth]{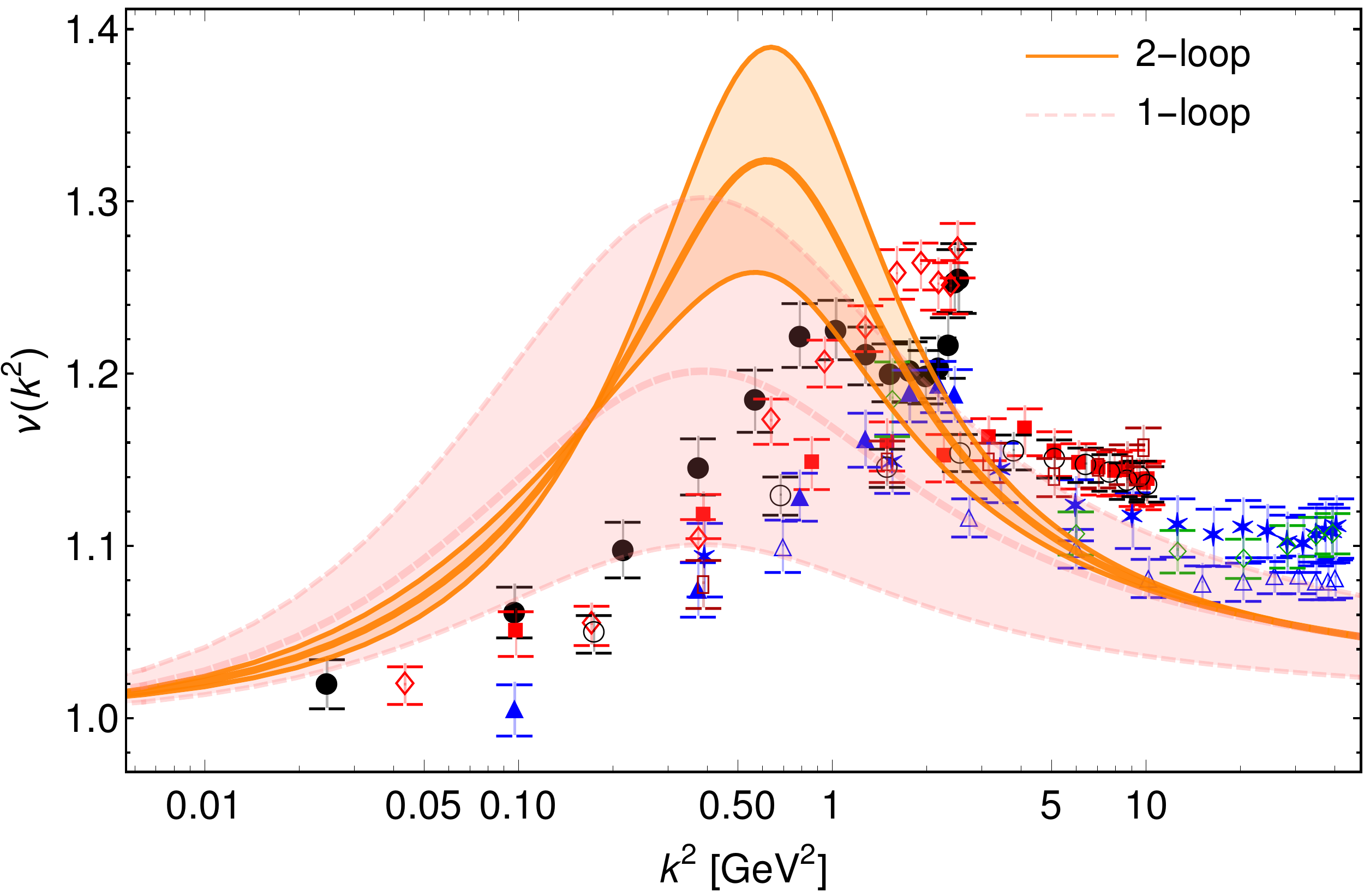}
\caption{CF prediction for the the function $v(k^2)$ in the SU(2) case and in the IS scheme, compared to the lattice data in the Taylor
scheme \cite{Maas:2019ggf}. The parameters $m$ and $g$ at the initial scale $\bar\mu_0$ are those previously determined from the fits
of the gluon and ghost propagators.}
\label{Fig:v_SU(2)}
\end{figure}

In the first strategy, since the parameters are already fixed, we can evaluate $v(k^2)$ with no further adjustment and
compare it directly with the lattice data. Our results
are shown in Fig.~\ref{Fig:v_SU(3)} for the SU($3$) case and in Fig.~\ref{Fig:v_SU(2)} for the SU($2$) case. The colored
bands display a simple estimate of our theoretical error defined by the absolute difference between central values at a given order and the previous one.

In the SU($3$) case, we observe that, except for a tiny region in the IR where the two-loop corrections accidentally vanish (preventing us from estimating the error), our two-loop results are compatible with the lattice data.
Moreover, the theoretical error diminishes when going from one-loop to two-loop order, indicating that perturbation theory shows a good apparent convergence for those parameters.

The situation is drastically different in the SU($2$) case where, even though the theoretical error still diminishes from one-loop to
two-loop order, our results are far from the lattice data. In particular the scale at which $v(k^2)$ reaches a maximum is underestimated by a
factor of $2$. Given the large error bars and the dispersion of the results with the various lattice parameters,
one can not exclude the possibility that this discrepancy originates in lattice artefacts, at least partially.
As already mentioned, another explanation could be the size of the expansion parameter in the SU($2$) case that lies in the limit of validity of perturbation theory.
Finally, a third source of discrepancy (possibly complementary to the previous ones) is that the parameters have been adjusted to best reproduce the two-point
functions. Therefore, any inaccuracy in the determination of the two-point functions (be it numerical or originating from the fact that perturbation theory is maybe not as justified as in the SU($3$) case), necessarily impacts the determination of the parameters and in turn the prediction of the vertex. 

For this last reason, it is interesting to consider independent fits of the various vertex functions in view of finding
the optimal choice of parameters that reproduce each function at best. We proceed to this analysis in the next section. Given
that the lattice SU($3$) error bars for the vertex are quite large, this analysis only makes sense for the SU($2$) case.

\begin{figure}
\hglue-6mm\includegraphics[width=0.45\textwidth]{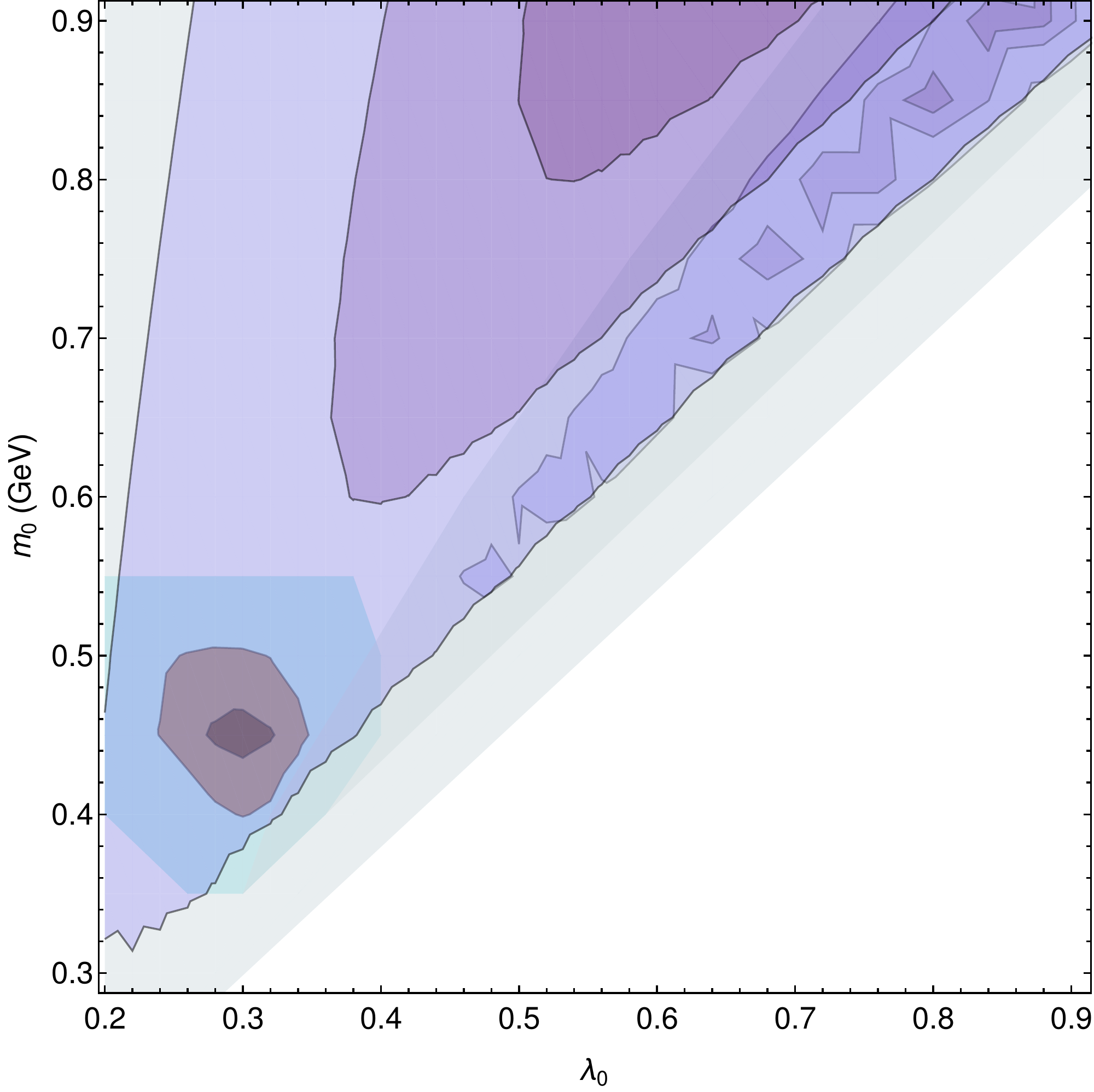}
\vglue2mm
\hglue-6mm\includegraphics[width=0.45\textwidth]{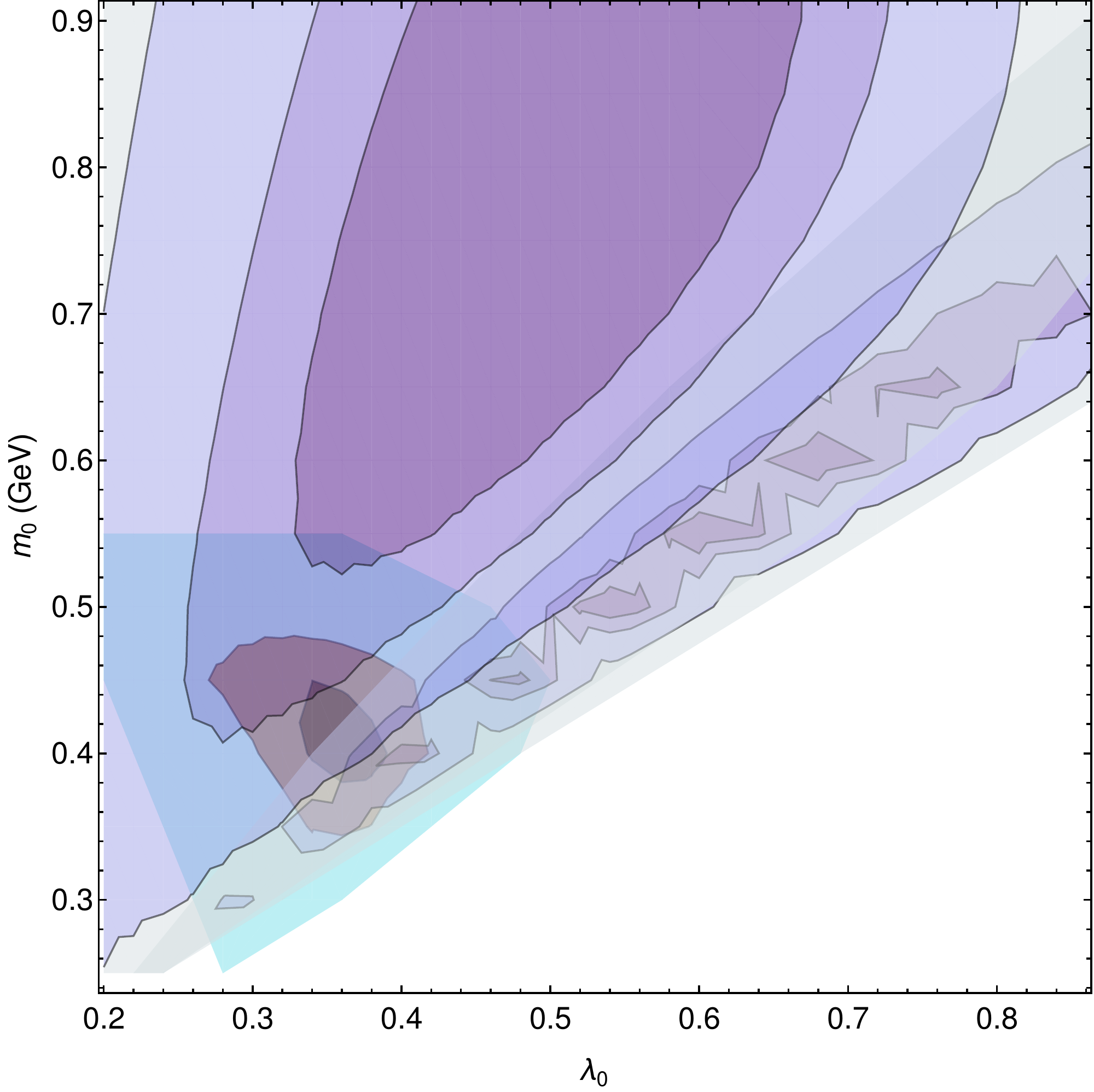}
\caption{Error regions with, respectively, 10\%, 7\%, 5\% and 4\% accuracy obtained when fitting the ghost dressing and the
vertex function $v(k^2)$ in the SU(2) case and in the IS scheme to the lattice data in the Taylor
scheme \cite{Maas:2019ggf}. The wide region corresponds to the vertex and the cigarre-like region to the ghost dressing function. As for the gluon propagator, its error region corresponds to the small round region at the bottom left and is subdivided in subregions representing 20\%, 10\% and 7\% accuracy. The parameters $m$ and $g$ are fixed at the initial scale $\mu_0$. Top: one-loop case. Bottom: two-loop case.}
\label{Fig:errorzones_SU(2)}
\end{figure}

\subsection{Independent fit of the various vertex functions}
As we have just mentioned, the parameters that optimize the gluon and ghost propagators in the SU($2$) case give
poor results for the vertex. We analyze here the error bars for these functions independently.

 In Fig.~\ref{Fig:errorzones_SU(2)}, the error regions associated to the estimation of parameters are shown for various confidence intervals. The successive regions correspond,
for the ghost propagator and vertex, to fits to lattice data with, respectively,
10\%, 7\%, 5\% and 4\% accuracy. The gluon propagator is much more demanding and the regions correspond to fits to
lattice data with, respectively, 20\%, 10\% and 7\% accuracy. We show the error regions both at one-loop order and at two-loop order.
It is seen that the optimal fitting parameters do not coincide for the various functions but the tension is considerably reduced when going from one-loop to two-loop order. This may explain the disappointing results obtained in the previous subsection for the SU($2$) case. 

\begin{figure}
\hglue-6mm\includegraphics[width=0.51\textwidth]{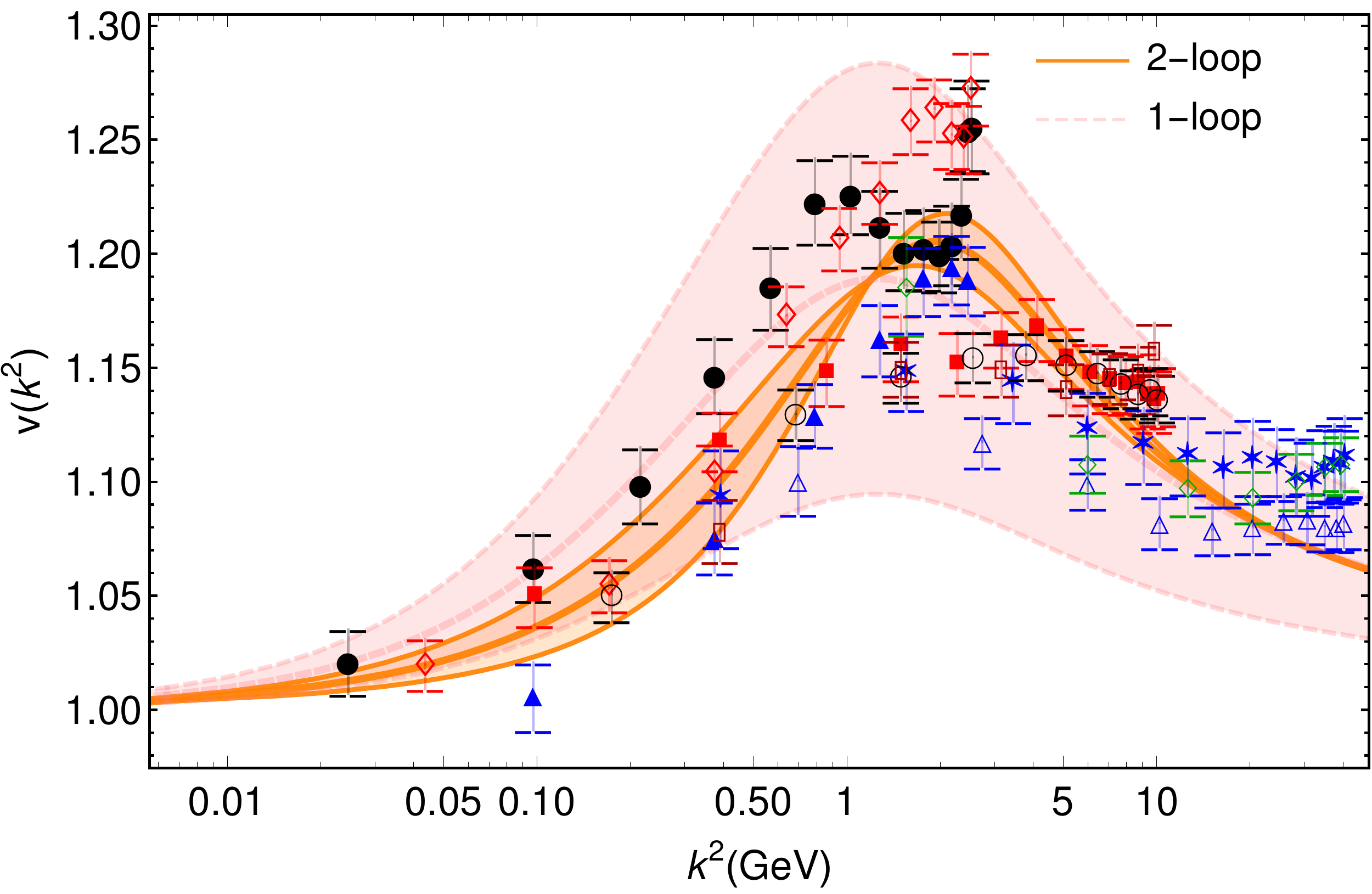}
\caption{Best fit for the vertex function $v(k^2)$ in the SU(2) case and in the IS scheme,
when compared to the lattice data in the Taylor
scheme \cite{Maas:2019ggf}.}
\label{Fig:bestfit_SU(2)}
\end{figure}

On the other hand, if one fits only the vertex function (as it has been done previously in other approaches) without simultaneously optimizing the two-point functions, one can obtain an excellent fit, as we illustrate in Fig.~\ref{Fig:bestfit_SU(2)}. Therefore, by only fitting the vertex, one can have the incorrect impression of finding excellent
agreement with the data. But one must recognize that the lattice data for two-point functions have a much better
precision since both statistical and systematic errors are manifestly more under control. As a consequence,
an excellent agreement when fitting the vertex to the data without a similar fit of the two-point functions must be taken with serious skepticism.

\begin{figure}
\hglue-6mm\includegraphics[width=0.5\textwidth]{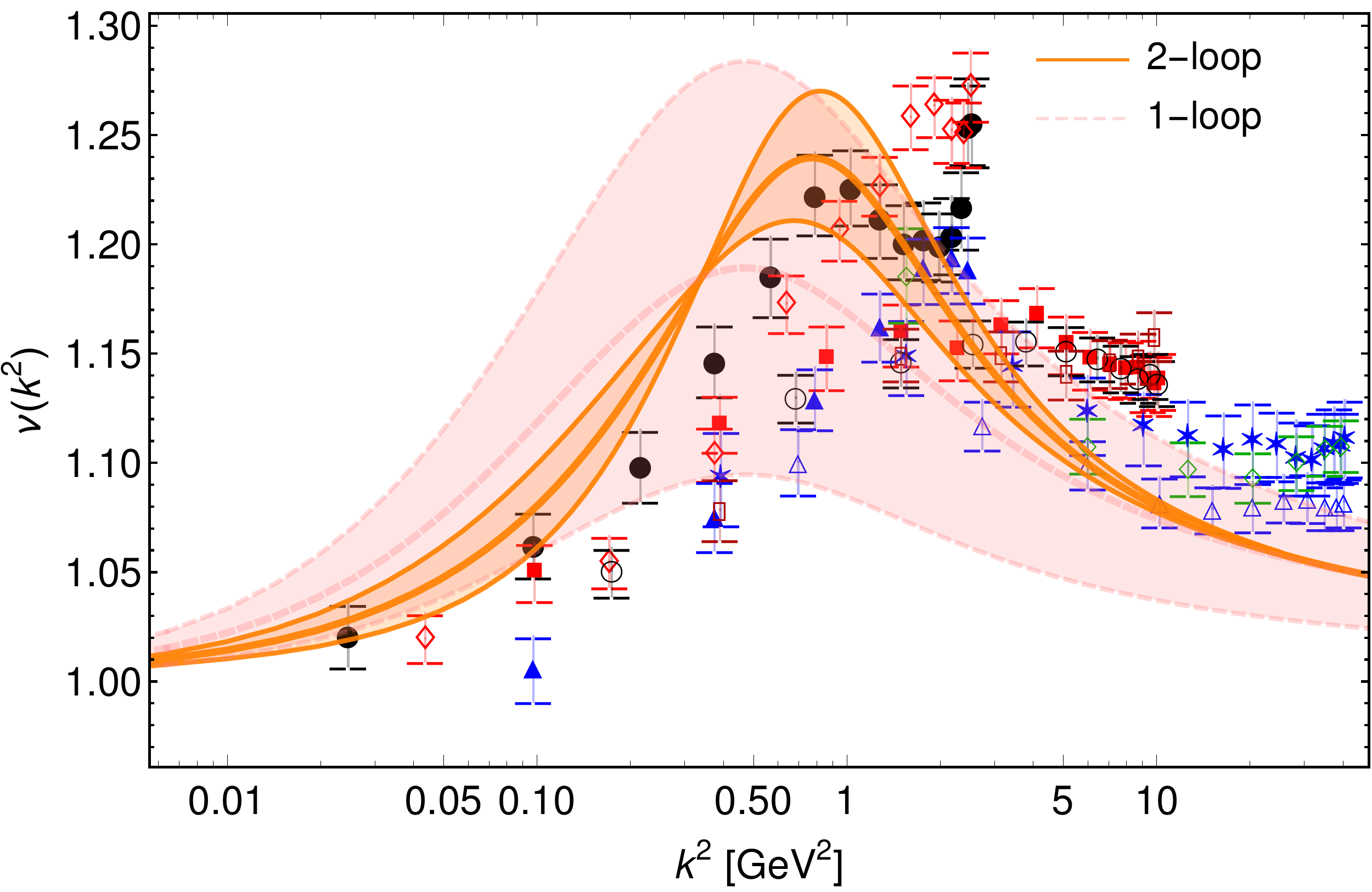}
\vglue6mm
\hglue-5mm\includegraphics[width=0.47\textwidth]{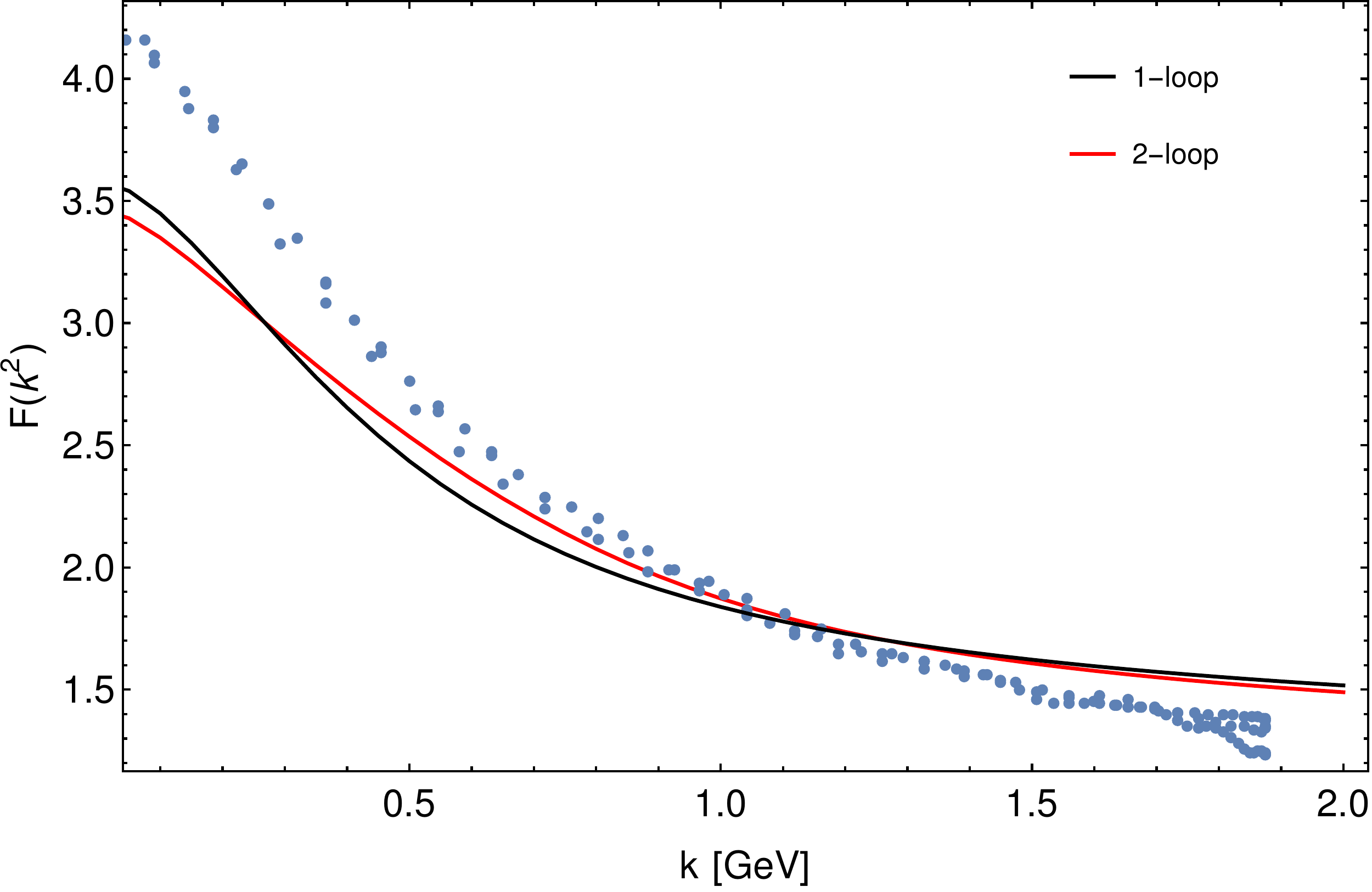}
\vglue6mm
\hglue-5mm\includegraphics[width=0.49\textwidth]{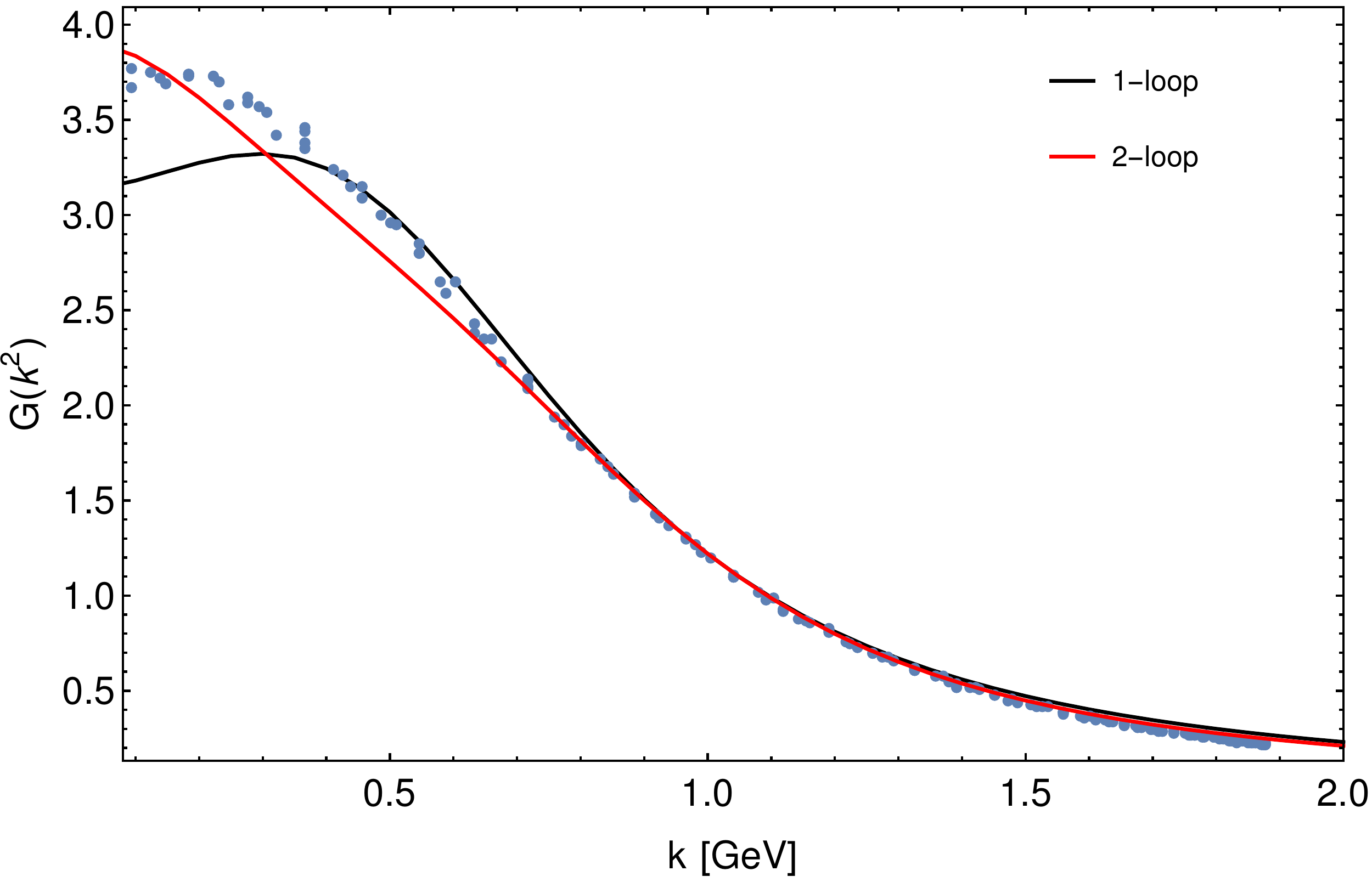}
\caption{Ghost-antighost-gluon vertex, ghost dressing function $F(k^2)\equiv k^2 D(k^2)$ and gluon propagator $G(k^2)$ for a choice of
parameters that reproduces the three functions to a reasonable accuracy. Lattice data from \cite{Cucchieri:2008qm}.}
\label{Fig:c}
\end{figure}

We mention finally that one can try to locate parameters for which each all functions are reproduced to a reasonable accuracy
by minimizing a joint error function, see Fig.~\ref{Fig:c}.

\subsection{Scheme dependence}
\label{schemedep}
Another possible way to test the validity of the perturbative approach is to study the dependence with respect to a change in the
renormalization scheme. Indeed, it is usually expected that the more convergent a perturbative expansion is,
the less dependent it should be to such changes. To test this in the present context, we compared the IS scheme to the so-called vanishing momentum (VM) scheme obtained by replacing the constraint (\ref{eq:nonrenorm}) fixing $Z_{m^2}$ with the condition
\beq
G^{-1}(k=0)=m^2(\mu)\,.
\eeq
Unfortunately, this scheme suffers from the presence of an IR 
Landau singularity \cite{Tissier:2010ts,Tissier:2011ey}. We can cure the problem by stopping the flow at scale $\mu=\sqrt{k^2+\alpha m^2}$ with $\alpha=1$ or $2$, with the price however of introducing a systematic error in the deep IR.

Our results are displayed in Table \ref{tab:scheme}, where we show our estimate for the relative error between the IS scheme evaluation of $v(k^2)$ and the corresponding evaluations in the VM scheme with $\alpha=1$ or $\alpha=2$.\\

\begin{table}[htp]
\begin{tabular}{|c||l|c||c|c|c|}
 \hline
Group & \multicolumn{2}{| c ||}{$N=2$} & \multicolumn{2}{| c |}{$N=3$}\\
\hline
 \hline
 VM & $\alpha=1$ & $\alpha=2$ \hspace{1mm}& $\alpha=1$ & $\alpha=2$\hspace{1mm} \\
\hline
\hline
1-loop & \,\,0.9\% & 1.1\% & 0.9\% & 1.1\% \\
\hline
 2-loop & \,\,1.2\% & 2.0\% & 0.8\% & 0.7\% \\
 \hline
\end{tabular}
\caption{\label{tab:scheme} Relative difference between the IS scheme and the VM schemes for $\alpha=1$ and $\alpha=2$.}
\end{table}

We note first that the relative variation when changing scheme is systematically smaller than the relative error from the determination of the parameters, which indicates that the comparison displayed in Table \ref{tab:scheme} is meaningful. We then observe that the scheme dependence diminishes in the SU($3$) case as one increases the number of loops, specially for $\alpha=2$. Although the effect is not as strong for $\alpha=1$, the result remains compatible with the scenario that perturbation theory within the CF model provides a good grasp on YM correlation functions in the SU($3$) case. On the contrary, the scheme dependence increases in the SU($2$) case, in line once more with the earlier observation that, in this case, we are closer to the limit of validity of the perturbative expansion.

\section{Conclusions} \label{sec:conclusion}
In the present article, we computed the two-loop ghost-antighost-gluon vertex in Landau-gauge Yang-Mills
theory using the
CF model and we compared the results with available lattice simulations
\cite{Ilgenfritz:2006he,Sternbeck:2006rd,Cucchieri:2008qm,Maas:2019ggf}. In order to keep the calculations manageable,
we restricted our analysis to the case where the gluon momentum vanishes.

As discussed in the Introduction, the perturbative expansion of the CF, as proposed originally in
\cite{Tissier:2010ts,Tissier:2011ey}, has been shown to reproduce
accurately many correlation functions (both in the vacuum and at finite temperature and density) in Yang-Mills theory.
It incorporates in the perturbative analysis two main ingredients observed in lattice simulations in Landau gauge:
the gluon propagator displays a massive-like behavior in the
infrared \cite{Bonnet:2000kw,Cucchieri_08b,Bogolubsky09,Bornyakov09,Iritani:2009mp,Maas:2011se,Oliveira:2012eh}
and the Yang-Mills coupling constant takes moderate
values for all momenta \cite{Bogolubsky09,Boucaud:2011ug} allowing for a perturbative analysis even in the infrared.

The present work generalizes previous studies in many ways. First, it pursues the analysis of the two-point correlation functions
at two loop order in
the CF model \cite{Gracey:2019xom} and applies it to one of the Yang-Mills three-point correlation functions.
Second, it extends the well-known two-loops result for the same vertex in the case of a 
massless gluon in the same momentum configuration \cite{Davydychev:1997vh}. Third,
it refines the previous one-loop result obtained in the CF model \cite{Pelaez:2013cpa}.\footnote{It must be stressed,
however, that in that reference the three-point vertices were calculated for arbitrary momentum configurations and not only
for a zero momentum gluon.}

When compared with lattice data, our two-loop results improve the previous one-loop results, both for SU($2$) and
SU($3$). However, the improvement is much more spectacular in the SU($3$) case where an excellent agreement is achieved.
We stress that this result is, in a sense, a pure prediction of the model for it was obtained without adjusting any parameter.
Indeed, the parameters of the model had already been fixed independently by fitting the two-point functions \cite{Gracey:2019xom}.

In the SU($2$) case the same procedure does not give such an excellent agreement. In particular, the position of the maximum
of the vertex is shifted by a factor of two approximatively. Even though the results improve when going from one-loop order to
two loop order, the improvement is not as impressive as that for two-point functions or that for the SU($3$) vertex.
Let us note, however, that a very good fit can be achieved by fitting the vertex alone.
What seems to be in tension are the parameters obtained from the two-point functions versus the parameters needed
to reproduce the vertex.
This fact is important to be taken into account when comparing to other studies where the vertex has been fitted directly without
imposing that the associated parameters must also fit
the two-point functions with at least the same accuracy.

When discussing the quality of the present results, various pieces of information must be taken into account. First, the analyzed
momentum configuration was the most challenging, at least for the one-loop analysis \cite{Pelaez:2013cpa}. This is
to be expected. If one of the momenta is small, the results become much more sensitive to the sector of
the theory with higher coupling. Accordingly, the present analysis should probably be seen as the ``worst case scenario'',
at least as far as the considered ghost-antighost-gluon vertex is concerned. In the same vein, the expansion parameter
is larger in the SU($2$) case which necessarilty impacts the quality of our perturbative estimate.

Second, the lattice data for the three-point functions are much less accurate than those for the two-point functions. This is again expected since it is of course harder to simulate three-point functions than two-point functions. This simple remark has however important consequences for the present analysis. In contrast to the case of the two-point functions where we considered that the main source of error in the fit to lattice data was the internal precision
of the perturbative calculation in the CF model, it is not clear in the present case whether the errors coming
from the lattice simulation can be neglected. Indeed, the lattice systematics are clearly visible when comparing lattice
data with different parameters. As such, it could happen that part of our discrepancies with lattice data have their
origin in the simulations themselves. In order to discard this possible source of error, more precise lattice simulations
for the three-point vertices would be extremely valuable.

A third point must be stressed regarding our results:
the inclusion of the gluon mass turns out to be crucial in order to obtain a good agreement with lattice data, even at a qualitative level.
The massless case \cite{Davydychev:1997vh} features a divergence in the present momentum configuration when the ghost
momentum goes to zero. This is at odds with lattice data and the CF result
(both in the SU($2$) and in the SU($3$) case) which, instead, saturate to their bare value in the far infrared. This again
strongly supports the use of a modified perturbation theory
in the presence of a gluon mass.

The present study can be extended in many ways. First, we are currently including quarks at two-loop order in order to look
at unquenching effect in two-point functions (not only in the gluon and ghost propagators but also in the quark propagator). As
mentioned in the Introduction, the use of perturbation theory in the quark sector seems to be much more problematic
\cite{Pelaez:2014mxa,Pelaez:2015tba} particularly in the chiral limit \cite{Pelaez:2017bhh}. Let us note, however, that
some aspects of the quark self-energy seem to be dominated by two-loop perturbative effects \cite{Pelaez:2014mxa} and
we plan to test if the inclusion of those contributions improves our understanding of these questions. Second, the present study
concerning the ghost-antighost-gluon can be extended to the (more intricate) three-gluon vertex in the Yang-Mills case which
we plan to evaluate in the near future (also with one vanishing gluon momentum).

\acknowledgements{We would like to thank J.~A. Gracey and M. Tissier for useful discussions related to the present project, as well as A. Maas for sharing the results of his Monte-Carlo simulations and for providing us with valuable insight on the interpretation of the data.
We also acknowledge the  financial support from PEDECIBA program and from the ANII-FCE-1-126412  project.  Part of this work also benefited from the
support of a CNRS-PICS project ``irQCD''. Finally, we thank the Laboratoire International Associ\'e of the CNRS, Institut Franco-Uruguayen de Physique.}

\appendix

\section{Reducing integrals with inverted propagators}\label{app:minus_one}
When implementing the FIRE reduction package, it may happen that not all the resulting integrals belong to the list of master integrals (\ref{eq:A})-(\ref{eq:M}). In some instances, it can happen that one of the propagators is elevated to the power $-1$. We now discuss a generic example as an illustration of how these integrals are dealt with in practice. It will be convenient for the following discussion to introduce the notation
\beq
& & I_{12345}(n_1,n_2,n_3,n_4,n_5)\nonumber\\
& & \hspace{0.05cm}\equiv\!\!\int_p\!\int_q\!G^{n_1}_1(p)G^{n_2}_2(q)G^{n_3}_3(k-p)G^{n_4}_4(k-q)G^{n_5}_5(p-q)\,,\nonumber\\
\eeq 
with $G_i(\ell)\equiv 1/(\ell^2+m^2_i)$. Some of the masses can be zero, in which case we replace the corresponding index by $0$ and $m_0^2=0$.

Let us consider the integral $I_{10045}(1,-1,0,1,1)$ which has one propagator elevated to the power $-1$. Using
\beq
q^2=(k-q)^2+m^2_4-k^2-m^2_4+2(k\cdot q)\,,
\eeq
the integral rewrites
\beq
& & I_{10045}(1,-1,0,1,1)\nonumber\\
&  & \hspace{0.4cm}=\,I_{15000}(1,1,0,0,0)\nonumber\\
& & \hspace{0.6cm} -\,(k^2+m^2_4)I_{10045}(1,0,0,1,1)\nonumber\\
& & \hspace{0.6cm}+\int_p\!\int_q 2(k\cdot q)G_1(p)G_4(k-q)G_5(p-q)\,.\label{eq:A3}
\eeq
We next perform the change of variables $p\to k-p$ and $q\to k-q$, followed by $p\leftrightarrow q$, which basically replaces $k\cdot q$ by $k\cdot(k-p)$, while exchanging the role of the indices $1$ and $4$. We then arrive at
\beq
& & I_{10045}(1,-1,0,1,1)\nonumber\\
&  & \hspace{0.2cm}=\,I_{15000}(1,1,0,0,0)\nonumber\\
& & \hspace{0.6cm} +\,2k^2 I_{40015}(1,0,0,1,1)\nonumber\\
& & \hspace{0.6cm} -\,(k^2+m^2_4)I_{10045}(1,0,0,1,1)\nonumber\\
& & \hspace{0.6cm}-\,\frac{1}{2}\int_q 2k\,G_1(k-q)\!\cdot\!\!\!\int_p 2p\,G_4(p)G_5(p-q).
\eeq
The benefit of this form with respect to (\ref{eq:A3}) is that now the inner integral is a vector depending only on $q$. It follows that
\beq
& & I_{10045}(1,-1,0,1,1)\nonumber\\
&  & \hspace{0.2cm}=\,I_{15000}(1,1,0,0,0)\nonumber\\
& & \hspace{0.6cm} +\,2k^2 I_{40015}(1,0,0,1,1)\nonumber\\
& & \hspace{0.6cm} -\,(k^2+m^2_4)I_{10045}(1,0,0,1,1)\nonumber\\
& & \hspace{0.6cm}-\,\frac{1}{2}\int_q 2(k\cdot q)G_0(q)G_1(k-q)\nonumber\\
& & \hspace{1.4cm}\times\,\int_p 2(p\cdot q)G_4(p)G_5(p-q).
\eeq
Then, inserting the identities
\beq
 2(k \cdot q) & = & k^2+m^2_1+q^2-(k-q)^2-m^2_1\,,\label{eq:id1}\\
 2(p\cdot q) & = & q^2+m^2_5-m^2_4+p^2+m^2_4-(p-q)^2-m^2_5\,,\label{eq:id2}\nonumber\\
\eeq
and identifying master integrals, we find
\beq
& & 2I_{10045}(1,-1,0,1,1)+I_{40015}(1,-1,0,1,1)\nonumber\\
&  & \hspace{0.2cm}=\,A_1A_4+A_4A_5+A_5A_1\nonumber\\
& & \hspace{0.6cm}-\,(k^2+m^2_1)(A_5-A_4)B_{01}\nonumber\\
& & \hspace{0.6cm}+\,(m^2_5-m^2_4)I_{045}\nonumber\\
& & \hspace{0.6cm} +\,(k^2-m^2_1-m^2_4-m_5^2) S_{145}\nonumber\\
& & \hspace{0.6cm}-\,(k^2+m^2_1)(m^2_5-m^2_4)U_{1045}\,.\label{eqapp}
\eeq
In the case where $m_1=m_4$, we have obtained an explicit expression of $I_{10045}(1,-1,0,1,1)$ in terms of the master integrals. In the case where $m_1\neq m_4$, we can consider the same equation with $1\leftrightarrow 4$:
\beq
& & I_{10045}(1,-1,0,1,1)+2I_{40015}(1,-1,0,1,1)\nonumber\\
&  & \hspace{0.2cm}=\,A_1A_4+A_4A_5+A_5A_1\nonumber\\
& & \hspace{0.6cm}-\,(k^2+m^2_4)(A_5-A_1)B_{04}\nonumber\\
& & \hspace{0.6cm}+\,(m^2_5-m^2_1)I_{015}\nonumber\\
& & \hspace{0.6cm} +\,(k^2-m^2_1-m^2_4-m_5^2) S_{145}\nonumber\\
& & \hspace{0.6cm}-\,(k^2+m^2_4)(m^2_5-m^2_1)U_{4015}\,.
\eeq
Together with Eq.~(\ref{eqapp}), this provides an invertible linear system that can be solved in order to obtain $I_{10045}(1,-1,0,1,1)$ in terms of the master integrals.\\

\section{Feynman integrals to order $\epsilon$}\label{app:epsilon}
Let us start gathering some formulas that will be useful in this section and the next one. First we recall the well known integrals
\beq\label{eq:Ja}
& & J_\alpha(m^2)\equiv\int_q\frac{1}{(q^2+m^2)^\alpha}\nonumber\\
& & \hspace{0.2cm}=\,\frac{(m^2)^{2-\alpha-\epsilon}}{(4\pi\mu^2)^{-\epsilon}}\frac{\Gamma(\alpha-2+\epsilon)}{\Gamma(\alpha)}
\eeq 
and
\beq\label{eq:III}
& & I_{\alpha,\beta}(p^2)\equiv\int_q\frac{1}{(q^2)^\alpha((q+p)^2)^\beta}=I_{\beta,\alpha}(p^2)\nonumber\\
&  & \hspace{0.2cm}=\,\frac{(p^2)^{2-\alpha-\beta-\epsilon}}{(4\pi\mu^2)^{-\epsilon}}\nonumber\\
& & \hspace{0.2cm}\times\,\frac{\Gamma(2-\alpha-\epsilon)\Gamma(2-\beta-\epsilon)\Gamma\big(\alpha+\beta-2+\epsilon\big)}{\Gamma(\alpha)\Gamma(\beta)\Gamma(4-\alpha-\beta-2\epsilon)}\,,\nonumber\\
\eeq
obtained by a simple application of the Feynman trick. One also has
\beq\label{eq:Jab}
& & J_{\alpha,\beta}(m^2)\equiv\int_q\frac{1}{(q^2+m^2)^\alpha(q^2)^{\beta}}\nonumber\\
& & \hspace{0.2cm}=\,\frac{(m^2)^{2-\alpha-\beta-\epsilon}}{(4\pi\mu^2)^{-\epsilon}}\frac{\Gamma(2-\beta-\epsilon)\Gamma(\alpha+\beta-2+\epsilon))}{\Gamma(2-\epsilon)\Gamma(\alpha)}\,,\nonumber\\
\eeq
which can be obtained by interpreting $J_{\alpha,\beta}(m^2)$ as the integral $J_\alpha(m^2)$ in $d-2\beta$ dimensions, up to some appropriate normalization factor. Combining this result together with Eq.~(\ref{eq:III}), one also finds
\begin{widetext}
\beq\label{eq:Im}
& & I_{\alpha,\beta,\gamma}(m^2)\equiv\int_p\int_q\frac{1}{(p^2+m^2)^\alpha(q^2)^\beta((q+p)^2)^\gamma}=I_{\alpha,\gamma,\beta}(m^2)\nonumber\\
& & \hspace{0.2cm}=\,\frac{(m^2)^{4-\alpha-\beta-\gamma-2\epsilon}}{(4\pi\mu^2)^{-2\epsilon}}\frac{\Gamma(2-\beta-\epsilon)\Gamma(2-\gamma-\epsilon)\Gamma(\beta+\gamma-2+\epsilon)\Gamma(\alpha+\beta+\gamma-4+2\epsilon)}{\Gamma(\alpha)\Gamma(\beta)\Gamma(\gamma)\Gamma(2-\epsilon)}\,.
\eeq
Finally, we quote the following result by Berends et al. \cite{Berends:1994sa}:
\beq\label{eq:Imm}
& & I_{\alpha,\beta,\gamma}(m^2,m^2)\equiv\int_p\int_q \frac{1}{(p^2+m^2)^\alpha(q^2+m^2)^\beta((q+p)^2)^\gamma}=I_{\beta,\alpha,\gamma}(m^2,m^2)\nonumber\\
& & \hspace{0.2cm}=\,\frac{(m^2)^{4-\alpha-\beta-\gamma-2\epsilon}}{(4\pi\mu^2)^{-2\epsilon}}\frac{\Gamma(2-\gamma-\epsilon)\Gamma(\alpha+\gamma-2+\epsilon)\Gamma(\beta+\gamma-2+\epsilon)\Gamma(\alpha+\beta+\gamma-4+2\epsilon)}{\Gamma(\alpha)\Gamma(\beta)\Gamma(2-\epsilon)\Gamma(\alpha+\beta+2\gamma-4+2\epsilon)}\,.
\eeq
\end{widetext}

As described in the main text, we need to expand the master integrals $A_m$, $B_{m0}$ and $B_{00}$ to
order $\epsilon^2$, as well as $S_{000}$, $S_{m00}$, $I_{m00}$, $U_{0m00}$ and $U_{00m0}$ to order $\epsilon^1$.
The integrals $A_m$ and $B_{00}$ are easily handled from (\ref{eq:Ja}) and (\ref{eq:III}).
Similarly, $S_{000}$ can be handled by using (\ref{eq:III}) twice. We can also easily deal with $I_{m00}$ using (\ref{eq:Im}). 

To deal with $B_{m0}(k^2)$, we use the Feynman trick to write it as
\beq
B_{m0}(k^2)=\frac{\Gamma(\epsilon)}{(4\pi\mu^2)^{-\epsilon}}\int_0^1dx(xm^2+x(1-x)k^2)^{-\epsilon}\,.\nonumber\\
\eeq
Since the prefactor of the integral diverges as $1/\epsilon$, we need to expand the later up to and including order $\epsilon^2$. The integrals appearing as coefficients of $\epsilon^0$, $\epsilon^1$ and $\epsilon^2$ can all be performed analytically in terms of logarithms and di-logarithms.

To deal with $S_{m00}(k^2)$, we write it as
\beq
& & S_{m00}(k^2)=\int_p\frac{I_{1,1}((p+k)^2)}{p^2+m^2}\\
& & \hspace{0.1cm}=\,\frac{\Gamma(1-\epsilon)^2\Gamma(\epsilon)}{(4\pi\mu^2)^{-\epsilon}\Gamma(2-2\epsilon)}\int\frac{d^dp}{(2\pi)^d}\frac{((p+k)^2)^{-\epsilon}}{p^2+m^2}\nonumber\\
& & \hspace{0.1cm}=\,\frac{\Gamma\big(1-\epsilon\big)^2\Gamma(-1+2\epsilon)}{(4\pi\mu^2)^{-2\epsilon}\Gamma(2-2\epsilon)}\nonumber\\
& & \hspace{0.2cm}\times\int_0^1\!dx\,(1-x)^{-1+\epsilon}(xm^2)^{1-2\epsilon}\left(1+(1-x)\frac{k^2}{m^2}\right)^{1-2\epsilon}\!\!\!\!\!,\nonumber
\eeq
where we have once more made use of the Feynman trick in the last step. One should refrain from expanding the integrand in $\epsilon$ at this stage because this would generate a singularity $\int^1 dx\,(1-x)^{-1}$. In fact, $\epsilon$ plays the role of a regulator for the integral which diverges as $\epsilon\to 0$. Instead, we first write
\beq
& & \left(1+(1-x)\frac{k^2}{m^2}\right)^{1-2\epsilon}=\nonumber\\
& & \hspace{0.5cm}1\,+\left[\left(1+(1-x)\frac{k^2}{m^2}\right)^{1-2\epsilon}-1\right].
\eeq
The first term leads to an analytically computable integral which contains the divergence of the integral as $\epsilon\to 0$. In contrast, the second term leads to an integral that is regular in the limit $\epsilon$ and whose integrand can be safely expanded. We arrive at
\beq
S_{m00}(k^2) & = & \frac{(m^2)^{1-2\epsilon}}{(4\pi\mu^2)^{-2\epsilon}}\frac{\Gamma(1-\epsilon)^2\Gamma(-1+2\epsilon)\Gamma(\epsilon)}{\Gamma(2-\epsilon)}\nonumber\\
& + & \frac{(m^2)^{1-2\epsilon}}{(4\pi\mu^2)^{-2\epsilon}}\frac{\Gamma(1-\epsilon)^2\Gamma(-1+2\epsilon)}{\Gamma(2-2\epsilon)}\nonumber\\
& & \times\,\int_0^1dx\,x^{1-2\epsilon}(1-x)^{-1+\epsilon}\nonumber\\
& & \hspace{0.3cm}\times\,\left[\left(1+(1-x)\frac{k^2}{m^2}\right)^{1-2\epsilon}-1\right].
\eeq
Since the prefactor of the integral diverges as $1/\epsilon$ as $\epsilon\to 0$, we need to expand the integrand to order $\epsilon^2$. Again the integrals appearing as coefficients of $\epsilon^0$, $\epsilon^1$ and $\epsilon^2$ can all be performed analytically in terms of logarithms and di-logarithms.

Next, we consider $U_{0m00}(k^2)$ which we write
\beq
& & U_{0m00}(k^2)=\int_p \frac{I_{1,1}(p^2)}{p^2+m^2}\frac{1}{(p-k)^2}\nonumber\\
& & \hspace{0.1cm}=\,\frac{\Gamma(1-\epsilon)^2\Gamma(\epsilon)}{(4\pi\mu^2)^{-\epsilon}\Gamma(2-2\epsilon)}\int_p\frac{(p^2)^{-\epsilon}}{p^2+m^2}\frac{1}{(p-k)^2}\nonumber\\
& & \hspace{0.1cm}=\,\frac{\Gamma(1-\epsilon)^2\Gamma(2\epsilon)}{(4\pi\mu^2)^{-2\epsilon}\Gamma(2-2\epsilon)}(m^2)^{-2\epsilon}\int_0^1\!dx\,x^{-1+\epsilon} \nonumber\\
& & \hspace{0.5cm}\times\int_0^{1-x}\!dy\,\left(1-x-y+y(1-y)\frac{k^2}{m^2}\right)^{-2\epsilon}\!\!\!\!\!\!\!.
\eeq
As before, expanding the integrand in $\epsilon$ is incorrect because of the appearance of a divergence at $x=0$. Instead, we add and subtract to the integral over $y$, its value at $x=0$. The added term can be computed analytically, while the subtracted term is regular in the limit $\epsilon\to 0$ and the corresponding integrand can be expanded. We find (we use that $\epsilon>0$)
\beq
& & U_{0m00}(k^2)\nonumber\\
& & \hspace{0.1cm}=\,\frac{\Gamma(1-\epsilon)^2\Gamma(2\epsilon)}{(4\pi\mu^2)^{-2\epsilon}\Gamma(2-2\epsilon)}\frac{(m^2)^{-2\epsilon}}{\epsilon} \nonumber\\
& & \hspace{0.5cm}\times\int_0^1\!dy\,\left(1-y+y(1-y)\frac{k^2}{m^2}\right)^{-2\epsilon}\nonumber\\
& & \hspace{0.1cm}+\,\frac{\Gamma(1-\epsilon)^2\Gamma(2\epsilon)}{(4\pi\mu^2)^{-2\epsilon}\Gamma(2-2\epsilon)}(m^2)^{-2\epsilon}\int_0^1\!dx\,x^{-1+\epsilon} \nonumber\\
& & \hspace{0.5cm}\times\left[\int_0^{1-x}\!dy\,\left(1-x-y+y(1-y)\frac{k^2}{m^2}\right)^{-2\epsilon}\right.\nonumber\\
& & \hspace{0.9cm}\left.-\int_0^1\!dy\,\left(1-y+y(1-y)\frac{k^2}{m^2}\right)^{-2\epsilon}\right].
\eeq
Due the presence of a double pole that multiplies the first integral, the latter needs to be expanded to order $\epsilon^3$. Since the integral multiplying the $\epsilon^3$ term is not computable analytically, we resorted to a numerical evaluation. As for the subtracted integral, it needs to be expanded to order $\epsilon^2$ for it is multiplied by a simple pole. Again we evaluated the corresponding expansion coefficients numerically.

Finally, we consider $U_{00m0}(k^2)$ which we write
\beq
& & U_{00m0}(k^2)=\int_q\frac{1}{q^2+m^2}\int_p\frac{1}{p^2(p-q)^2(p-k)^2}\nonumber\\
& & \hspace{0.1cm}=\,\frac{\Gamma(1+\epsilon)}{(4\pi\mu^2)^{-\epsilon}}\int_0^1dx\int_0^{1-x}\!\!\!\!dy\,\nonumber\\
& & \hspace{0.1cm}\times\int_q\frac{(x(1-x)q^2+y(1-y)k^2-2xyq\cdot k)^{-1-\epsilon}}{q^2+m^2}\nonumber\\
\eeq
Pulling out a factor $(x(1-x))^{-1-\epsilon}$ in the numerator of the second integral, we can interpret the latter as a propagator to the power $1+\epsilon$. Then, applying once more the Feynman trick, we find
\begin{widetext}
\beq
U_{00m0}(k^2) & \!\!\!=\!\!\! &\frac{\Gamma(2+\epsilon)}{(4\pi\mu^2)^{-\epsilon}}\int_0^1\!\!dx\,x^{-1-\epsilon}(1-x)^{-1-\epsilon}\int_0^{1-x}\!\!\!\!\!\!dy\int_0^1\!\!dz\,z^\epsilon\int_q\frac{1}{\left(q^2+(1-z)m^2+\frac{yz(1-x-y+xy(1-z))}{x(1-x)^2}k^2\right)^{2+\epsilon}}\nonumber\\
& \!\!\!=\!\!\! & \frac{\Gamma(2\epsilon)}{(4\pi\mu^2)^{-2\epsilon}}\int_0^1\!\!dx\,x^{-1+\epsilon}(1-x)^{-1+3\epsilon}\int_0^{1-x}\!\!\!\!\!\!dy\int_0^1\!\!dz\,z^\epsilon\left(x(1-x)^2(1-z)m^2+yz(1-x-y+xy(1-z))k^2\right)^{-2\epsilon}\!\!\!\!\!.\nonumber\\
\eeq
\end{widetext}
When setting $\epsilon\to 0$ in the integrand, we find a divergence at $x=0$. We proceed as above by adding and subtracting from the $yz$-integral, its value at $x=0$. The added integral can be evaluated analytically, whereas the subtracted integral is regular in the limit $\epsilon\to 0$ and its integrand can be expanded. We find
\begin{widetext}
\beq
U_{00m0}(k^2) & = & \frac{(k^2)^{-2\epsilon}}{(4\pi\mu^2)^{-2\epsilon}}\frac{\Gamma(2\epsilon)\Gamma(\epsilon)\Gamma(3\epsilon)\Gamma(1-2\epsilon)^2}{\Gamma(4\epsilon)\Gamma(2-4\epsilon)(1+\epsilon)}+\frac{\Gamma(2\epsilon)}{(4\pi\mu^2)^{-2\epsilon}}\int_0^1dx\,x^{-1+\epsilon}(1-x)^{-1+3\epsilon}\nonumber\\
&  & \hspace{0.5cm}\times\left[\int_0^{1-x}\!\!\!\!dy\int_0^1dz\,z^\epsilon\left(x(1-x)^2(1-z)m^2+yz(1-x-y+xy(1-z))k^2\right)^{-2\epsilon}\right.\nonumber\\
& & \hspace{7.0cm}\left.-\,\int_0^1dy\int_0^1dz\,z^\epsilon\left(yz(1-y)k^2\right)^{-2\epsilon}\right].
\eeq
Due to the presence of a simple pole in the last term, we need to expand the corresponding triple integral to order $\epsilon^2$.\\

\end{widetext}

\section{Low momentum expansion}\label{appsec:lowk}
Here we discuss the low momentum expansion of some of the integrals for which we did not have an analytic expression. We make use of the integrals defined in the previous section. For convenience, we write them as $J_\alpha(m^2)=(m^2)^{2-\alpha-\epsilon} J_\alpha(1)$, $J_{\alpha,\beta}(m^2)=(m^2)^{2-\alpha-\beta-\epsilon}J_{\alpha,\beta}(1)$ and  $I_{\alpha,\beta}(p^2)=(p^2)^{2-\alpha-\beta-\epsilon}I_{\alpha,\beta}(1)$, where the `1' in the argument of each function means that we replace $m^2$ and $p^2$ formally by $1$ in the corresponding analytical expression. Similarly, $I_{\alpha,\beta,\gamma}(m^2)=(m^2)^{4-\alpha-\beta-\gamma-2\epsilon}I_{\alpha,\beta,\gamma}(1)$, $I_{\alpha,\beta,\gamma}(m^2,m^2)=(m^2)^{4-\alpha-\beta-\gamma-2\epsilon}I_{\alpha,\beta,\gamma}(1,1)$, and we mention that $I_{\alpha,\beta,\gamma}(1)=B_{\beta,\gamma}(1)J_{\alpha,\beta+\gamma-2+\epsilon}(1)$. 

A word of caution is in order before we start. The integrals $J_{\alpha,\beta}(m^2)$,
$I_{\alpha,\beta}(p^2)$, $I_{\alpha,\beta,\gamma}(m^2)$ and $I_{\alpha,\beta,\gamma}(m^2,m^2)$ are IR divergent when some of their
indices are large enough. Although these divergences are regularized in dimensional regularization (that is, the integrals admit
a well defined expression as long as $\epsilon\neq 0$),\footnote{The only exception is when one of those indices equals
exactly $d/2$, in which case dimensional regularization does not regularize the IR divergence, or when the sum of all
indices is equal to $d/2$, in which case dimensional regularization does not regularize the UV divergence.
We shall never encounter these undefined integrals.} they mix with the UV divergences and the integrals need to be manipulated
with care. In fact, the subtlety with these integrals is that they are not continuously connected to the related integrals in which
the IR divergence has been regularized by means of some momentum or mass scale. In many instances, however, the IR divergent
integrals arise precisely from expanding in powers of this IR regulating scales. Only if the original quantity is infrared
safe, does the expansion makes sense and one can use these IR divergent integrals. We shall give various examples below and also
in Appendix~\ref{app:low_m}.

\subsection{Small $k^2$ expansion of $U_{0m00}(k^2)$}
We have
\beq
U_{0m00}(k^2)=I_{1,1}(1) \int_p\frac{(p^2)^{-\epsilon}}{p^2+m^2}\frac{1}{(p+k)^2}\,.\label{eq:U1_short}
\eeq
We now would like to expand $U_{0m00}(k^2)/I_{1,1}(1)$ for small $k^2$. The first term in the expansion is the limit as $k^2\to 0$ and is easily computed to be
\beq
\frac{U_{0m00}(0)}{I_{1,1}(1)}=J_{1,1+\epsilon}(m^2)=(m^2)^{-2\epsilon}J_{1,1+\epsilon}(1)\,,
\eeq
and so
\beq
\frac{U_{0m00}(k^2)}{I_{1,1}(1)}=(m^2)^{-2\epsilon}J_{1,1+\epsilon}(1)+{\cal O}(k^2)\,,
\eeq
Na\"\i vely, in order to obtain the next term in the expansion, we would write
\beq
\frac{1}{(p+k)^2} & = & \frac{1}{p^2}\frac{1}{1+\frac{2(p\cdot k)+k^2}{p^2}}\nonumber\\
& = & \frac{1}{p^2}\left[1-\frac{2(p\cdot k)}{p^2}-\frac{k^2}{p^2}+\frac{4(p\cdot k)^2}{p^4}+\dots\right]\!.\label{eq:k_exp_0}\nonumber\\
\eeq
The first term in the bracket leads to the just computed leading order contribution, the second term vanishes upon angular integration and the third and fourth terms can be expressed in terms of $J_{1,2+\epsilon}$, yielding
\beq
\frac{U_{0m00}(k^2)}{I_{1,1}(1)} & = & (m^2)^{-2\epsilon}\left[J_{1,1+\epsilon}(1)+\frac{2\epsilon}{d}\frac{k^2}{m^2} J_{1,2+\epsilon}(1)\right]\nonumber\\
& + & {\cal O}(k^4)\,.\label{eq:wrong}
\eeq
The analytical expressions for $J_{1,1+\epsilon}(1)$ and $J_{1,2+\epsilon}(1)$ are well defined, see Eq.~(\ref{eq:Jab}), so it
seems that we obtain a meaningful expansion. However, this expansion is wrong because $dU_{0m00}(k^2)/dk^2$ has a logarithmic divergence
as $k\to 0$.
In fact, $J_{1,2+\epsilon}(1)$ is one of the IR divergent integrals discussed above and as already mentioned, it can only be used
when it arises from the expansion of an IR safe quantity. The quantity $U_{0m00}(k^2)/I_{1,1}(1)$ is certainly IR safe.
However, $U_{0m00}(k^2)/I_{1,1}(1)-(m^2)^{-2\epsilon}J_{1,1+\epsilon}(1)$, which we need to obtain the next term in the expansion,
is not.

To cope with this, the idea is to split the integrand in (\ref{eq:U1_short}) into a piece that is analytically tractable, and a piece where the expansion in $k$ can be pushed to one order higher without encountering any infrared divergence. To this purpose, we write
\beq\label{eq:trick}
\frac{1}{p^2+m^2} & = & \frac{1}{m^2}+\left[\frac{1}{p^2+m^2}-\frac{1}{m^2}\right]\nonumber\\
& = & \frac{1}{m^2}\left[1-\frac{p^2}{p^2+m^2}\right]
\eeq
and then
\beq
\frac{U_{0m00}(k^2)}{I_{1,1}(1)} & = & \frac{1}{m^2}\Bigg[(k^2)^{1-2\epsilon}I_{1,\epsilon}(1)\nonumber\\
& & \hspace{0.7cm}-\,\int_p\frac{(p^2)^{1-\epsilon}}{p^2+m^2}\frac{1}{(p+k)^2}\Bigg].\label{eq:U1_short_2}
\eeq
The first term in the RHS is known exactly, see Eq.~(\ref{eq:III}), whereas in the second term the small $k^2$ expansion can be pushed to one order further than above, before we meet an infrared divergence (since we have gained an extra power of $p^2$ in the integrand). Plugging (\ref{eq:k_exp_0}) in (\ref{eq:U1_short_2}), we arrive at
\beq
\frac{U_{0m00}(k^2)}{I_{1,1}(1)} & = & \frac{k^2}{m^2}(k^2)^{-2\epsilon}B_{1,\epsilon}(1)\nonumber\\
& - & (m^2)^{-2\epsilon}\left[J_{1,\epsilon}(1)+\frac{2\epsilon}{d}\frac{k^2}{m^2}J_{1,1+\epsilon}(1)\right]\nonumber\\
& + & {\cal O}(k^4)\,.\label{eq:U1_exp_2}
\eeq
 The first term is of dimension $2-4\epsilon$. It contains poles in $1/\epsilon$ associated with UV divergences. This leads to the appearance of terms of the form $k^2\ln k^2/\mu^2$, as expected. The remaining terms are regular in $k^2$. One could be surprised that the leading term now appears as $-(m^2)^{-2\epsilon}J_{1,\epsilon}(1)$ instead of $(m^2)^{-2\epsilon}J_{1,1+\epsilon}(1)$.
But this is no surprise since $J_{1,\beta}(1)=-J_{1,\beta-1}(1)$, as follows from Eq.~(\ref{eq:Jab}) insofar $J_{1,\beta}(1)$
and $J_{1,\beta-1}(1)$ are well defined. Finally, we mention that (\ref{eq:U1_exp_2}) can be expanded to any order in $\epsilon$.\\

The same strategy can be applied at any order. To this purpose, we iterate (\ref{eq:trick}) 
 \beq
\frac{1}{p^2+m^2} & = & \frac{1}{m^2}\left[1-\frac{p^2}{p^2+m^2}\right]\nonumber\\
& = & \frac{1}{m^2}\left[1-\frac{p^2}{m^2}+\frac{p^2}{m^2}\frac{p^2}{p^2+m^2}\right]\nonumber\\
& = & \frac{1}{m^2}\left[\sum_{j=0}^n\left(-\frac{p^2}{m^2}\right)^j-\left(-\frac{p^2}{m^2}\right)^n\frac{p^2}{p^2+m^2}\right].\nonumber\\
\eeq
Then
\beq
\frac{U_{0m00}(k^2)}{I_{1,1}(1)} & = & \frac{1}{m^2}\left[\sum_{j=0}^n(-1)^j \frac{(k^2)^{1+j-2\epsilon}}{(m^2)^j}I_{1,\epsilon-j}(1)\right.\nonumber\\
& & \left.-\,\frac{(-1)^n}{(m^2)^n}\int_p\frac{(p^2)^{1+n-2\epsilon}}{p^2+m^2}\frac{1}{(p+k)^2}\right]\!.\label{eq:U1_exp_n0}\nonumber\\
\eeq
The terms in the sum are known exactly and contribute up to and including order $(k^2)^{n+1-2\epsilon}$, which generate logarithms of $k^2$ since the $I_{1,\epsilon-j}(1)$'s are all UV divergent and contain poles in $1/\epsilon$. The integral in (\ref{eq:U1_exp_n0}) can be expanded up to order $(k^2)^{n+1}$ without encountering any IR divergence and the corresponding, regular expansion can be expressed in terms of the $J_{\alpha,\beta}(1)$'s. More precisely, extending (\ref{eq:k_exp_0}), we write
\beq
& & \frac{1}{(p+k)^2}=\frac{1}{p^2}\sum_{j=0}^{2(n+1)} (-1)^j \left(\frac{2(p\cdot k)+k^2}{p^2}\right)^j+\dots\nonumber\\
& & \hspace{0.2cm}=\,\sum_{j=0}^{2(n+1)}\!\!\frac{(-1)^j}{(p^2)^{j+1}} \sum_{\ell=0}^j \frac{j!}{\ell!(j-\ell)!} (2\,p\cdot k)^\ell(k^2)^{j-\ell}+\dots\label{eq:k_exp}\nonumber\\
\eeq
We consider the sum up to $2(n+1)$ to be sure that we generate all powers of $k^2$ up to $(k^2)^{n+1}$, but it is understood that we should truncate any term beyond. Plugging (\ref{eq:k_exp}) in the last term of (\ref{eq:U1_exp_n0}) and using the formula (for $\ell$ even, otherwise the integral vanishes)
\beq
\int_p\,f(p^2)\,(2\,p\cdot k)^\ell=\frac{\ell!}{(\ell/2)!}\frac{(k^2)^{\ell/2}}{(2-\epsilon)_{\ell/2}}\int_p\,f(p^2)\,(p^2)^{\ell/2}\,,\label{eq:Davy}\nonumber\\
\eeq 
see Ref.~\cite{Davydychev:1993pg}, we arrive at
\beq
& & {\color{red}} \int_p\frac{(p^2)^{1+n-\epsilon}}{p^2+m^2}\frac{1}{(p+k)^2}\nonumber\\
& & \hspace{0.5cm}=\,(m^2)^{1+n-2\epsilon}\sum_{j=0}^{2(n+1)}\!\!\sum_{\ell(even)=0}^j \!\!(-1)^j\frac{j!}{(\ell/2)!(j-\ell)!} \nonumber\\
& & \hspace{2.5cm} \left.\times\,\left(\frac{k^2}{m^2}\right)^{j-\ell/2}\frac{J_{1,\epsilon-n+j-\ell/2}(1)}{(2-\epsilon)_{\ell/2}}\right|_{(k^2)^{n+1}}\nonumber\\
& & \hspace{0.5cm}+\cdots
\eeq
and eventually
\beq
 \frac{U_{0m00}(k^2)}{I_{1,1}(1)} & = & (k^2)^{-2\epsilon}\sum_{j=0}^n(-1)^j \left(\frac{k^2}{m^2}\right)^{j+1} I_{1,\epsilon-j}(1)\nonumber\\
& - & (m^2)^{-2\epsilon}\sum_{j=0}^{2(n+1)}\sum_{\ell=0}^{[j/2]}  (-1)^{n+j} \frac{j!}{\ell!(j-2\ell)!}\nonumber\\
& & \hspace{1.0cm} \left.\times\,\left(\frac{k^2}{m^2}\right)^{j-\ell}\frac{J_{1,\epsilon-n+j-\ell}(1)}{(2-\epsilon)_\ell}\right|_{(k^2)^{n+1}}\nonumber\\
& + & {\cal O}((k^2)^{n+2})\,.
\eeq
We have checked that this formula leads to the known low-$k^2$ expansion for the $\epsilon^0$ contributions. We can then use it to evaluate the corresponding expansion for the contributions of order $\epsilon^1$. We have checked that the latter matches with a numerical evaluation of the corresponding $\epsilon^1$ contributions to $U_{0m00}(k^2)$ at large $k^2$.

\subsection{Small $k^2$ expansion of $U_{00m0}(k^2)$}
We next consider the integral
\beq
U_{00m0}(k^2)\!=\!\int_p\frac{1}{p^2}\frac{1}{(p+k)^2}\int_q\frac{1}{q^2+m^2}\frac{1}{(q+p)^2}.
\eeq
In this case, the leading term of the low-$k^2$ expansion is already delicate. We use
\beq\label{eq:trick2}
\frac{1}{(q+p)^2} & = & \frac{1}{q^2}+\left[\frac{1}{(q+p)^2}-\frac{1}{q^2}\right]\nonumber\\
& = & \frac{1}{q^2}\left[1-\frac{2(p\cdot q)+p^2}{(q+p)^2}\right]
\eeq
to arrive at
\beq\label{eq:jnsp}
U_{00m0}(k^2) & = & -\frac{1}{m^2}J_1(m^2)\,I_{1,1}(k^2)\nonumber\\
& - & \int_q\frac{1}{q^2(q^2+m^2)}\int_p\frac{2(q\cdot p)+p^2}{p^2(p+k)^2(q+p)^2}\,.\nonumber\\
\eeq
The first term is known exactly while the second is regular in the limit $k^2\to 0$. We can evaluate it by writing
\beq
\int_p\frac{2(q\cdot p)+p^2}{p^4(q+p)^2} & = & \int_p\frac{(q+p)^2-q^2}{p^4(q+p)^2}\nonumber\\
& = & \int_p\frac{1}{p^4}-q^2\int_p\frac{1}{p^4(q+p)^2}\,.\label{eq:trick3}
\eeq
We mention that (\ref{eq:trick3}) provides yet another example of the use of (dimensionally regularized) IR divergent integrals.
Since the LHS is infrared safe, the decomposition in terms of IR divergent integrals makes perfect sense. To evaluate the
second line and if one is not so sure about the value to give to $\int d^dp/p^4$, one can add a mass regulator to both quartic
propagators (since the LHS is infrared safe) as $1/p^4\to 1/(p^2+m^2)^2$ or even $1/p^2\to 1/(p^2(p^2+m^2))$ and complete the
calculation. We have checked that one obtains the same result by applying the Feynman trick directly to the LHS. We have also
checked that the same result is obtained by using the well known result $\int d^dp/p^4=0$ \cite{ZinnJustin:2002ru}, so eventually the
final result can be written as
\beq
\int_p\frac{2(q\cdot p)+p^2}{p^4(q+p)^2} & = & -q^2\int_p\frac{1}{p^4(q+p)^2}\nonumber\\
& - & \frac{(q^2)^{-\epsilon}}{(4\pi\mu^2)^{-\epsilon}}\Gamma(1+\epsilon)\frac{\Gamma(1-\epsilon)\Gamma(-\epsilon)}{\Gamma(1-2\epsilon)}\,,\nonumber\\
\eeq
which, once plugged back into (\ref{eq:jnsp}) leads to the known integral $J_{1,1+\epsilon}$.\\

We can compute higher orders by iterating (\ref{eq:trick2})
\beq
\frac{1}{(q+p)^2} & = & \frac{1}{q^2}\left[1-\frac{2(p\cdot q)+p^2}{(q+p)^2}\right]\nonumber\\
& = & \frac{1}{q^2}\left[1-\frac{2(p\cdot q)+p^2}{q^2}\right.\nonumber\\
& & \hspace{1.0cm}\left.+\,\frac{2(p\cdot q)+p^2}{q^2}\frac{2(p\cdot q)+p^2}{(q+p)^2}\right]\nonumber\\
& = & \frac{1}{q^2}\left[\sum_{j=0}^n \left(-\frac{2(p\cdot q)+p^2}{q^2}\right)^j\right.\nonumber\\
& & \hspace{1.0cm}\left.-\,\left(-\frac{2(p\cdot q)+p^2}{q^2}\right)^n\frac{2(p\cdot q)+p^2}{(q+p)^2}\right].\nonumber\\
\eeq
Then
\beq\label{eq:jnsp2}
U_{00m0}(k^2) & = & \sum_{j=0}^n (-1)^j\int_q\frac{(q^2)^{-j-1}}{(q^2+m^2)}\int_p\frac{(2(p\cdot q)+p^2)^j}{p^2(p+k)^2}\nonumber\\
& - & (-1)^n\int_q\frac{(q^2)^{-n-1}}{(q^2+m^2)}\int_p\frac{(2(p\cdot q)+p^2)^{n+1}}{p^2(p+k)^2(q+p)^2}\,.\nonumber\\
\eeq
The first line can be computed analytically while in the second line we can expand to the relevant order in $k^2$ before encountering any IR divergence and the corresponding expansion coefficients can again be determined analytically. For the first line, it is convenient to perform the $q$-integral first
\beq
& & \int_q\frac{(2(p\cdot q)+p^2)^j}{(q^2)^{j+1}(q^2+m^2)}\nonumber\\
& & \hspace{0,1cm}=\,\sum_{\ell=0}^j \frac{j!}{\ell!(j-\ell)!}(p^2)^{j-\ell}\int_q\frac{(2(p\cdot q))^\ell}{(q^2)^{j+1}(q^2+m^2)}\nonumber\\
& & \hspace{0,1cm}=\,(m^2)^{d/2-2}\sum_{\ell(even)=0}^j \frac{j!}{(\ell/2)!(j-\ell)!}\nonumber\\
& & \hspace{2.5cm}\times\,\left(\frac{p^2}{m^2}\right)^{j-\ell/2}\frac{J_{1,1+j-\ell/2}(1)}{(2-\epsilon)_{\ell/2}}\,,\nonumber\\
\eeq
where we have once again used formula (\ref{eq:Davy}). To treat the last line, we write
\beq
& & \frac{(2(p\cdot q)+p^2)^{n+1}}{p^2(p+k)^2}\nonumber\\
& & \hspace{0.1cm}=\,\left.\sum_{j=0}^{2n}\frac{(-1)^j}{(p^2)^{j+2}}(2(p\cdot q)+p^2)^{n+1}(2(p\cdot k)+k^2)^j\right|_{(k^2)^n}\nonumber\\
& & \hspace{0.1cm}+\,\dots\nonumber\\
\eeq
We are then lead to consider
\beq
& & \int\frac{d^dp}{(2\pi)^d}\frac{(2(p\cdot q)+p^2)^{n+1}(2(p\cdot k)+k^2)^j}{(p^2)^{j+2}(q+p)^2}\nonumber\\
& & \hspace{0.5cm}=\,-q^2\int\frac{d^dp}{(2\pi)^d}\frac{(2(p\cdot q)+p^2)^{n}(2(p\cdot k)+k^2)^j}{(p^2)^{j+2}(q+p)^2}\nonumber\\
& & \hspace{0.5cm}=(-q^2)^k\int\frac{d^dp}{(2\pi)^d}\frac{(2(p\cdot q)+p^2)^{n+1-k}(2(p\cdot k)+k^2)^j}{(p^2)^{j+2}(q+p)^2}\nonumber\\
& & \hspace{0.5cm}=(-q^2)^{n+1}\int\frac{d^dp}{(2\pi)^d}\frac{(2(p\cdot k)+k^2)^j}{(p^2)^{j+2}(q+p)^2}\,,
\eeq
where we have used similar tricks as in (\ref{eq:trick3}). We next plug these formulas into (\ref{eq:jnsp2}) and invert the order of the integrals leading to
\begin{widetext}
\beq
U_{00m0}(k^2) & = & (k^2m^2)^{-\epsilon}\sum_{j=0}^n (-1)^j\sum_{\ell=0}^{[j/2]} \frac{j!}{\ell!(j-2\ell)!}\left(\frac{k^2}{m^2}\right)^{j-\ell}\frac{J_{1,1+j-\ell}(1)}{(2-\epsilon)_{\ell}}I_{1,1+\ell-j}(1)\nonumber\\
& + & \left.\sum_{j=0}^{2n}(-1)^j\int_p\frac{(2(p\cdot k)+k^2)^j}{(p^2)^{j+2}}\int_q\frac{1}{(q^2+m^2)(q+p)^2}\right|_{(k^2)^n}+{\cal O}((k^2)^{n+1})\nonumber\\
& = & (k^2m^2)^{-\epsilon}\sum_{j=0}^n (-1)^j\sum_{\ell=0}^{[j/2]} \frac{j!}{\ell!(j-2\ell)!}\left(\frac{k^2}{m^2}\right)^{j-\ell}\frac{J_{1,1+j-\ell}(1)}{(2-\epsilon)_{\ell}}I_{1,1+\ell-j}(1)\nonumber\\
& + & \left.\sum_{j=0}^{2n}(-1)^j\sum_{\ell=0}^j\frac{j!}{\ell!(j-\ell)!}(k^2)^{j-\ell}\int_p\frac{(2\,p\cdot k)^\ell}{(p^2)^{j+2}}\int_q\frac{1}{(q^2+m^2)(q+p)^2}\right|_{(k^2)^n}\nonumber\\
& + & {\cal O}((k^2)^{n+1})\,.
\eeq
We notice that the inner integral is a function of $p^2$. We can then use (\ref{eq:Davy}) to obtain
\beq
U_{00m0}(k^2) & = & (k^2m^2)^{-\epsilon}\sum_{j=0}^n (-1)^j\sum_{\ell=0}^{[j/2]} \frac{j!}{\ell!(j-2\ell)!}\left(\frac{k^2}{m^2}\right)^{j-\ell}\frac{J_{1,1+j-\ell}(1)}{(2-\epsilon)_{\ell}}I_{1,1+\ell-j}(1)\nonumber\\
& + & \left.\sum_{j=0}^{2n}(-1)^j\sum_{\ell=0}^{[j/2]}\frac{j!}{\ell!(j-2\ell)!}\frac{(k^2)^{j-\ell}}{(d/2)_{\ell}}\int_p\frac{1}{(p^2)^{j+2-\ell}}\int_q\frac{1}{(q^2+m^2)(q+p)^2}\right|_{(k^2)^n}\nonumber\\
& + & {\cal O}((k^2)^{n+1})\,.
\eeq
The double integral is nothing but $I_{1,1,j+2-\ell}(m^2)=(m^2)^{d-4-j+\ell}J_{1,1+\epsilon+j-\ell}(1)B_{1,j+2-\ell}(1)$. It follows that
\beq
U_{00m0}(k^2) & = & (k^2m^2)^{-\epsilon}\sum_{j=0}^n (-1)^j\sum_{\ell=0}^{[j/2]} \frac{j!}{\ell!(j-2\ell)!}\left(\frac{k^2}{m^2}\right)^{j-\ell}\frac{J_{1,1+j-\ell}(1)}{(2-\epsilon)_{\ell}}B_{1,1+\ell-j}(1)\nonumber\\
& + & (m^4)^{-\epsilon}\left.\sum_{j=0}^{2n}(-1)^j\sum_{\ell=0}^{[j/2]}\frac{j!}{\ell!(j-2\ell)!}\left(\frac{k^2}{m^2}\right)^{j-\ell}\frac{J_{1,1+\epsilon+j-\ell}(1)}{(2-\epsilon)_\ell}B_{1,j+2-\ell}(1)\right|_{(k^2)^n}\nonumber\\
& + & {\cal O}((k^2)^{n+1})\,.
\eeq

\subsection{Small $k^2$ expansion of $U_{0mm0}(k^2)$}
We next consider the integral
\beq
{\color{red}} U_{0mm0}(k^2)\equiv\int_p\frac{1}{p^2}\frac{1}{(p+k)^2+m^2}\int_q\frac{1}{q^2}\frac{1}{(q+p)^2+m^2}\,.
\eeq
We write
\beq
\frac{1}{(q+k)^2+m^2} & = & \frac{1}{q^2+m^2+2(q\cdot k)+k^2}\sum_{j=0}^\infty (-1)^j\frac{(2(q\cdot k)+k^2)^j}{(q^2+m^2)^{j+1}}\label{eq:k_exp_m}
\eeq
and arrive at
\beq
{\color{red}} U_{0mm0}(k^2) & = & \sum_{j=0}^\infty (-1)^j\int_p\frac{1}{p^2}\frac{(2(p\cdot k)+k^2)^j}{(p^2+m^2)^{j+1}}\int_q\frac{1}{q^2}\frac{1}{(q+p)^2+m^2}\nonumber\\
& = & \sum_{j=0}^\infty (-1)^j\sum_{\ell=0}^j\frac{j!}{\ell!(j-\ell)!}(k^2)^{j-\ell}\int_p\frac{1}{p^2}\frac{(2\,p\cdot k)^\ell}{(p^2+m^2)^{j+1}}\int_q\frac{1}{q^2}\frac{1}{(q+p)^2+m^2}\,.
\eeq
With the help of (\ref{eq:Davy}), this becomes
\beq
U_{0mm0}(k^2) & \!\!=\!\! & \sum_{j=0}^\infty (-1)^j\sum_{\ell=0}^{[j/2]}\frac{j!}{\ell!(j-2\ell)!}\frac{(k^2)^{j-\ell}}{(d/2)_{\ell}}\int_p\frac{(p^2)^\ell}{p^2(p^2+m^2)^{j+1}}\int_q\frac{1}{q^2}\frac{1}{(q+p)^2+m^2}\nonumber\\
& \!\!=\!\! & \sum_{j=0}^\infty (-1)^j\!\sum_{\ell=0}^{[j/2]}\frac{j!}{(j-2\ell)!}\frac{(k^2)^{j-\ell}}{(d/2)_{\ell}}\!\sum_{h=0}^{\ell}\frac{(-1)^{\ell-h}(m^2)^{\ell-h}}{h!(\ell-h)!}\!\int_p\frac{1}{p^2(p^2+m^2)^{j-h+1}}\!\int_q\frac{1}{q^2}\frac{1}{(q+p)^2+m^2}\,.
\eeq
We then use
\beq
{\color{red}} \frac{1}{p^2(p^2+m^2)^{j-h+1}} & = & \frac{1}{m^2p^2(p^2+m^2)^{j-h}}-\frac{1}{m^2(p^2+m^2)^{j-h+1}}\nonumber\\
& = & \frac{1}{m^4p^2(p^2+m^2)^{j-h-1}}-\frac{1}{m^4(p^2+m^2)^{j-h}}-\frac{1}{m^2(p^2+m^2)^{j-h+1}}\nonumber\\
& = & \frac{1}{(m^2)^{j-h+1}p^2}-\sum_{i=0}^{j-h}\frac{1}{(m^2)^{i+1}(p^2+m^2)^{j-h+1-i}}\,,
\eeq
which leads to
\beq
{\color{red}} U_{0mm0}(k^2) & = & \sum_{j=0}^\infty (-1)^j\sum_{\ell=0}^{[j/2]}\frac{j!}{(j-2\ell)!}\frac{(k^2)^{j-\ell}}{(d/2)_{\ell}}\sum_{h=0}^{\ell}\frac{(-1)^{\ell-h}(m^2)^{\ell-j-1}}{h!(\ell-h)!}I_{1,1,1}(m^2)\nonumber\\
& - & \sum_{j=0}^\infty (-1)^j\sum_{\ell=0}^{[j/2]}\frac{j!}{(j-2\ell)!}\frac{(k^2)^{j-\ell}}{(d/2)_{\ell}}\sum_{h=0}^{\ell}\frac{(-1)^{\ell-h}(m^2)^{\ell-h}}{h!(\ell-h)!}\sum_{i=0}^{j-h}\frac{I_{j-h+1-i,1,1}(m^2,m^2)}{(m^2)^{i+1}}\,,
\eeq
that is
\beq
{\color{red}} U_{0mm0}(k^2) & = & (m^2)^{-2\epsilon}\sum_{j=0}^\infty (-1)^j\sum_{\ell=0}^{[j/2]}\frac{j!}{(j-2\ell)!}\left(\frac{k^2}{m^2}\right)^{j-\ell}\sum_{h=0}^{\ell}\frac{(-1)^{\ell-h}}{h!(\ell-h)!}\frac{J_{1,2-d/2}(1)B_{1,1}(1)}{(2-\epsilon)_\ell}\nonumber\\
& - & (m^2)^{-2\epsilon}\sum_{j=0}^\infty (-1)^j\sum_{\ell=0}^{[j/2]}\frac{j!}{(j-2\ell)!}\left(\frac{k^2}{m^2}\right)^{j-\ell}\sum_{h=0}^{\ell}\frac{(-1)^{\ell-h}}{h!(\ell-h)!}\sum_{i=0}^{j-h}\frac{I_{j-h+1-i,1,1}(1,1)}{(2-\epsilon)_\ell}\,.
\eeq
The first line contributes only for $\ell=0$ and we end up with
\beq
U_{0mm0}(k^2) & = & (m^2)^{-2\epsilon}\sum_{j=0}^\infty (-1)^j\left(\frac{k^2}{m^2}\right)^j\frac{J_{1,2-d/2}(1)B_{1,1}(1)}{(2-\epsilon)_\ell}\nonumber\\
& - & (m^2)^{-2\epsilon}\sum_{j=0}^\infty (-1)^j\sum_{\ell=0}^{[j/2]}\frac{j!}{(j-2\ell)!}\left(\frac{k^2}{m^2}\right)^{j-\ell}\sum_{h=0}^{\ell}\frac{(-1)^{\ell-h}}{h!(\ell-h)!}\sum_{i=0}^{j-h}\frac{I_{j-h+1-i,1,1}(1,1)}{(2-\epsilon)_\ell}\,.
\eeq

\subsection{Small $k^2$ expansion of $S_{m00}(k^2)$}
We have
\beq
S_{m00}(k^2)=\int_p\frac{1}{p^2+m^2}\int_q\frac{1}{q^2(q+p+k)^2}
\eeq
and thus
\beq
\frac{S_{m00}(k^2)}{I_{1,1}(1)}=\int_p\frac{(p+k)^{-\epsilon}}{p^2+m^2}\,.
\eeq
If we leave the momentum $k$ in the massless propagator, we have to pay attention to the infrared divergences, but we can use the previous considerations. We start by writing
\beq
\frac{S_{m00}(k^2)}{I_{1,1}(1)} & = & \frac{1}{m^2}\left[\sum_{j=0}^n(-1)^j \frac{(k^2)^{2+j-2\epsilon}}{(m^2)^j}B_{\epsilon,-j}(1)-\frac{(-1)^n}{(m^2)^n}\int_p((p+k)^2)^{-\epsilon}\frac{(p^2)^{n+1}}{p^2+m^2}\right]\nonumber\\
& = & \frac{1}{m^2}\left[\sum_{j=0}^n(-1)^j \frac{(k^2)^{2+j-2\epsilon}}{(m^2)^j}B_{\epsilon,-j}(1)\right.+{\cal O}((k^2)^{n+3})\nonumber\\
& - & \left.\frac{(-1)^n}{(m^2)^n}\left.\sum_{j=0}^{2n+4}\frac{\Gamma(1-\epsilon)}{\Gamma(j+1)\Gamma(1-j-\epsilon)}\int_p(2(p\cdot k)+k^2)^j\frac{(p^2)^{1+n-j-\epsilon}}{p^2+m^2}\right|_{(k^2)^{n+2}}\right]\nonumber\\
& = & \frac{1}{m^2}\left[\sum_{j=0}^n(-1)^j \frac{(k^2)^{2+j-2\epsilon}}{(m^2)^j}B_{\epsilon,-j}(1)\right.+{\cal O}((k^2)^{n+3})\nonumber\\
& - & \left.\frac{(-1)^n}{(m^2)^n}\left.\sum_{j=0}^{2n+4}\frac{\Gamma(1-\epsilon)}{\Gamma(1-j-\epsilon)}\sum_{\ell=0}^{[j/2]}\frac{1}{\ell!(j-2\ell)!}\frac{(k^2)^{j-\ell}}{(d/2)_\ell}\int_p\frac{(p^2)^{1+n-j+\ell-\epsilon}}{p^2+m^2}\right|_{(k^2)^{n+2}}\right],
\eeq
where we have used once again Eq.~(\ref{eq:Davy}). Then
\beq
\frac{S_{m00}(k^2)}{I_{1,1}(1)} 
& = & \frac{1}{m^2}\left[\sum_{j=0}^n(-1)^j \frac{(k^2)^{2+j-2\epsilon}}{(m^2)^j}B_{\epsilon,-j}(1)+{\cal O}((k^2)^{n+3})\right.\nonumber\\
& - & \left.(-1)^n(m^2)^{2-2\epsilon}\left.\sum_{j=0}^{2n+4}\frac{\Gamma(1-\epsilon)}{\Gamma(1-j-\epsilon)}\sum_{\ell=0}^{[j/2]}\frac{1}{\ell!(j-2\ell)!}\left(\frac{k^2}{m^2}\right)^{j-\ell}\frac{J_{1,-1-n+j-\ell+\epsilon}(1)}{(d/2)_\ell}\right|_{(k^2)^{n+2}}\right]\nonumber\\
& = & (k^2)^{1-2\epsilon}\sum_{j=0}^n(-1)^j \left(\frac{k^2}{m^2}\right)^{j+1}B_{\epsilon,-j}(1)+{\cal O}((k^2)^{n+3})\nonumber\\
& - & (-1)^n(m^2)^{1-2\epsilon}\left.\sum_{j=0}^{2n+4}\frac{\Gamma(1-\epsilon)}{\Gamma(1-j-\epsilon)}\sum_{\ell=0}^{[j/2]}\frac{1}{\ell!(j-2\ell)!}\left(\frac{k^2}{m^2}\right)^{j-\ell}\frac{J_{1,\epsilon-1-n+j-\ell}(1)}{(2-\epsilon)_\ell}\right|_{(k^2)^{n+2}}\,,
\eeq
In fact $B_{\epsilon,-i}=0$, see Eq.~(\ref{eq:III}). This is in line with the fact that we expect a regular expansion in $k^2$ in the case of $S_1(k^2)$. Then
\beq
\frac{S_{m00}(k^2)}{I_{1,1}(1)} & = & (-1)^{n+1}(m^2)^{1-2\epsilon}\left.\sum_{j=0}^{2n+4}\frac{\Gamma(1-\epsilon)}{\Gamma(1-j-\epsilon)}\sum_{\ell=0}^{[j/2]}\frac{1}{\ell!(j-2\ell)!}\left(\frac{k^2}{m^2}\right)^{j-\ell}\frac{J_{1,\epsilon-1-n+j-\ell}(1)}{(2-\epsilon)_\ell}\right|_{(k^2)^{n+2}}\nonumber\\
& + & {\cal O}((k^2)^{n+3})\,.
\eeq
We can now use the formula $J_{1,\beta}(1)=-J_{1,\beta-1}(1)$, see above, to arrive at
\beq
\frac{S_{m00}(k^2)}{I_{1,1}(1)} & = & (m^2)^{1-2\epsilon}\left.\sum_{j=0}^{2n+4}\frac{\Gamma(1-\epsilon)}{\Gamma(1-j-\epsilon)}\sum_{\ell=0}^{[j/2]}\frac{1}{\ell!(j-2\ell)!}\left(\frac{k^2}{m^2}\right)^{j-\ell}\frac{J_{1,\epsilon+j-\ell}(1)}{(2-\epsilon)_\ell}\right|_{(k^2)^{n+2}}\nonumber\\
& + & {\cal O}((k^2)^{n+3})\,,
\eeq
which is nothing but the order $(k^2)^{n+2}$ of
\beq\label{eq:final}
\frac{S_{m00}(k^2)}{I_{1,1}(1)} & = & (m^2)^{1-2\epsilon}\sum_{j=0}^\infty\frac{\Gamma(1-\epsilon)}{\Gamma(1-j-\epsilon)}\sum_{\ell=0}^{[j/2]}\frac{1}{\ell!(j-2\ell)!}\left(\frac{k^2}{m^2}\right)^{j-\ell}\frac{J_{1,\epsilon+j-\ell}(1)}{(2-\epsilon)_\ell}\,.
\eeq

\subsection{Small $k^2$ expansion of $S_{mm0}(k^2)$}
We have
\beq
S_{mm0}(k^2)=\int_p\frac{1}{p^2}\frac{1}{(q+p)^2+m^2}\int_q\frac{1}{(q+k)^2+m^2}\,.
\eeq
Using (\ref{eq:k_exp_m}), we arrive at
\beq
S_{mm0}(k^2) & = & \sum_{j=0}^\infty (-1)^j\int_p\int_q\frac{1}{p^2}\frac{1}{(q+p)^2+m^2}\frac{(2(q\cdot k)+k^2)^j}{(q^2+m^2)^{j+1}}\nonumber\\
& = & \sum_{j=0}^\infty (-1)^j\sum_{\ell=0}^j \frac{j!}{\ell!(j-\ell)!}(k^2)^{j-\ell}\int_q\frac{(2\,q\cdot k)^\ell}{(q^2+m^2)^{j+1}}\int_p\frac{1}{p^2}\frac{1}{(q+p)^2+m^2}\nonumber\\
& = & \sum_{j=0}^\infty (-1)^j\sum_{\ell=0}^{[j/2]} \frac{j!}{\ell!(j-2\ell)!}\frac{(k^2)^{j-\ell}}{(d/2)_\ell}\int_q\frac{(q^2)^\ell}{(q^2+m^2)^{j+1}}\int_p\frac{1}{p^2}\frac{1}{(q+p)^2+m^2}\nonumber\\
& = & \sum_{j=0}^\infty (-1)^j\sum_{\ell=0}^{[j/2]} \frac{j!}{(j-2\ell)!}\frac{(k^2)^{j-\ell}}{(d/2)_\ell}\sum_{h=0}^\ell\frac{(-1)^{\ell-h}(m^2)^{\ell-h}}{h!(\ell-h)!}\nonumber\\
& & \hspace{3.0cm}\times\,\int_q\frac{1}{(q^2+m^2)^{j+1-h}}\int_p\frac{1}{p^2}\frac{1}{(q+p)^2+m^2}\nonumber\\
& = &  (m^2)^{1-2\epsilon}\sum_{j=0}^\infty (-1)^j\sum_{\ell=0}^{[j/2]} \frac{j!}{(j-2\ell)!}\left(\frac{k^2}{m^2}\right)^{j-\ell}\sum_{h=0}^\ell\frac{(-1)^{\ell-h}}{h!(\ell-h)!}\frac{I_{j+1-h,1,1}(1,1)}{(2-\epsilon)_\ell}\,.
\eeq
We could have proceeded similarly in the case of $S_{m00}(k^2)$ with the difference that in the step
\beq
S_{m00}(k^2) & = & \sum_{j=0}^\infty (-1)^j\sum_{\ell=0}^{[j/2]} \frac{j!}{\ell!(j-2\ell)!}\frac{(k^2)^{j-\ell}}{(d/2)_\ell}\int_q\frac{(q^2)^\ell}{(q^2+m^2)^{j+1}}\int_p\frac{1}{p^2}\frac{1}{(q+p)^2}
\eeq
we recognize immediately
\beq
\frac{S_{m00}(k^2)}{I_{1,1}(1)} & = & \sum_{j=0}^\infty (-1)^j\sum_{\ell=0}^{[j/2]} \frac{j!}{\ell!(j-2\ell)!}\frac{(k^2)^{j-\ell}}{(d/2)_\ell}\int\frac{d^dq}{(2\pi)^d}\frac{(q^2)^{-\epsilon+\ell}}{(q^2+m^2)^{j+1}}\nonumber\\
& = & (m^2)^{1-2\epsilon}\sum_{j=0}^\infty (-1)^j\sum_{\ell=0}^{[j/2]} \frac{j!}{\ell!(j-2\ell)!}\left(\frac{k^2}{m^2}\right)^{j-\ell}\frac{J_{j+1,\epsilon-\ell}(1)}{(2-\epsilon)_\ell}\,.
\eeq
Owing to
\beq
\frac{J_{j+a,b}(1)}{J_{a,j+b}(1)}=\frac{\Gamma(d/2-b)}{\Gamma(d/2-j-b)}\frac{\Gamma(a)}{\Gamma(j+a)}\,,
\eeq
we have
\beq
\frac{J_{j+1,\epsilon-\ell}(1)}{J_{1,j+\epsilon-\ell}(1)} & = & \frac{\Gamma(2-2\epsilon+\ell)}{\Gamma(2-2\epsilon+\ell-j)}\frac{1}{j!}\,,
\eeq
and thus
\beq
\frac{S_{m00}(k^2)}{I_{1,1}(1)} & = & \sum_{j=0}^\infty (-1)^j\sum_{\ell=0}^{[j/2]} \frac{j!}{\ell!(j-2\ell)!}\frac{(k^2)^{j-\ell}}{(d/2)_\ell}\int\frac{d^dq}{(2\pi)^d}\frac{(q^2)^{-\epsilon+\ell}}{(q^2+m^2)^{j+1}}\nonumber\\
& = & (m^2)^{1-2\epsilon}\sum_{j=0}^\infty (-1)^j\sum_{\ell=0}^{[j/2]} \frac{1}{\ell!(j-2\ell)!}\frac{\Gamma(2-2\epsilon+\ell)}{\Gamma(2-2\epsilon+\ell-j)}\left(\frac{k^2}{m^2}\right)^{j-\ell}\frac{J_{1,j+\epsilon-\ell}(1)}{(2-\epsilon)_\ell}\,.
\eeq
This looks like (\ref{eq:final}) but not exactly so. However, we have checked with Mathematica that the two formulas coincide for various values of $n$.\\
\end{widetext}

\section{Low mass expansion}\label{app:low_m}
We here discuss why a na\"\i ve Taylor expansion in powers of the mass does not lead to the correct low mass expansion of the mass integrals and why, despite this issue, such na\"\i ve Taylor expansions can be used for low mass expansion of certain quantities. Let us start with the following example of master integral:
\beq
B_{m0}(p^2)\equiv\int\frac{d^dq}{(2\pi)^d}\frac{1}{q^2+m^2}\frac{1}{(q+p)^2}\,.
\eeq
Its na\"ive Taylor expansion in powers of the mass leads to
\beq
B_{m0}(p^2) & \!\!\to\!\!\! & \int\!\frac{d^dq}{(2\pi)^d}\frac{1}{q^2}\frac{1}{(q+p)^2}-m^2\!\!\int\!\frac{d^dq}{(2\pi)^d}\frac{1}{q^4}\frac{1}{(q+p)^2}\nonumber\\
& & +\,{\cal O}\left(\frac{m^4}{p^4}\right).\label{eq:D1}
\eeq
This expansion is clearly suspicious because, even though the last integral is well defined in dimensional regularization, it introduces an extra pole in $1/\epsilon$ (corresponding to an infrared divergence), the one present in $B_{m0}(p^2)$ being already accounted for by the first integral in (\ref{eq:D1}).\footnote{We are of course assuming here that the $\epsilon$ and low-$m$ expansions commute. We will check below that this assumption is correct, at least for the example considered here.}
This is clearly a misuse of (dimensionally regularized) IR divergent integrals, similar to the one we discussed around
Eq.~(\ref{eq:wrong}).

In order to obtain better control on the low $m$ expansion, we proceed as follows. We first write
\beq
\frac{1}{(q+p)^2}=\frac{1}{p^2}+\left[\frac{1}{(q+p)^2}-\frac{1}{p^2}\right]
\eeq
which leads to
\beq
B_{m0}(p^2) & \!\!=\!\! & \frac{1}{p^2}\int\frac{d^dq}{(2\pi)^d}\frac{1}{q^2+m^2}\nonumber\\
& \!\!+\!\! & \int\frac{d^dq}{(2\pi)^d}\frac{1}{q^2+m^2}\left[\frac{1}{(q+p)^2}-\frac{1}{p^2}\right].
\eeq
The first term will be left as it is because it is proportional to $(m^2)^{d/2-1}=(m^2)^{1-\epsilon}$ and, therefore, does not admit any Taylor expansion. In the second term, and contrary to what happened above, the na\"\i ve Taylor expansion can be pushed up to order $m^2$ without generating infrared divergences. We find
\beq
B_{m0}(p^2) & \!\!=\!\! & \frac{1}{p^2}\int\frac{d^dq}{(2\pi)^d}\frac{1}{q^2+m^2}\nonumber\\
& \!\!+\!\! & \int\frac{d^dq}{(2\pi)^d}\frac{1}{q^2}\left[\frac{1}{(q+p)^2}-\frac{1}{p^2}\right]\nonumber\\
& \!\!-\!\! & m^2\int\frac{d^dq}{(2\pi)^d}\frac{1}{q^4}\left[\frac{1}{(q+p)^2}-\frac{1}{p^2}\right]\nonumber\\
& \!\!+\!\! & {\cal O}\left(\frac{m^4}{p^4}\right).
\eeq
After cancelling some dimensional regularization zeros, this rewrites
\beq
B_{m0}(p^2) & \!\!=\!\! & \int\!\frac{d^dq}{(2\pi)^d}\frac{1}{q^2}\frac{1}{(q+p)^2}-m^2\!\!\int\!\frac{d^dq}{(2\pi)^d}\frac{1}{q^4}\frac{1}{(q+p)^2}\nonumber\\
& \!\!+\!\! & \frac{1}{p^2}\int\frac{d^dq}{(2\pi)^d}\frac{1}{q^2+m^2}+{\cal O}\left(\frac{m^4}{p^4}\right),\label{eq:D2}
\eeq
which differs from (\ref{eq:D1}) by the presence of the last term. We mention that the integral we dubbed
problematic in (\ref{eq:D1}) is also present here. However, its pole in $1/\epsilon$ is exactly cancelled by the one in the last integral,
in such a way that the only pole in $1/\epsilon$ comes from the first integral, as it should be.

Another way to check that (\ref{eq:D2}) is the correct low mass expansion of $B_{m0}(p^2)$ is to obtain this expansion by an alternative method.
As already mentioned in the main text, from dimensional analysis, see Eq.~(\ref{eq:dim_analysis}), it is clear that the low mass expansion
can be obtained from the UV expansion by exploiting Weinberg theorem. The latter classifies the various contributions that make
the large momentum asymptotic expansion according to the possible ways the large momentum $p$ can flow inside the diagram.
For any such contribution, it is possible to expand in powers of any scale (momentum or mass, except $p$ of course) that appears in a propagator
whose total momentum is large. For the present example, these contributions are
\beq
B_{m0}(p^2) & \!\!\to\!\! & \left[\int\frac{d^dq}{(2\pi)^d}\frac{1}{q^2+m^2}\frac{1}{(q+p)^2}\right]_m\nonumber\\
& \!\!+\!\! & \int\frac{d^dq}{(2\pi)^d}\frac{1}{q^2+m^2}\left[\frac{1}{(q+p)^2}\right]_{q}\nonumber\\
& \!\!+\!\! & \int\frac{d^dq}{(2\pi)^d}\frac{1}{q^2}\left[\frac{1}{(q+p)^2+m^2}\right]_{q,m},\label{eq:D7}
\eeq
where $\left[\dots\right]_{\mu,\nu,\cdots}$ means that one should expand in powers of the scales $\mu$, $\nu$, \dots. It is easily
checked that expanding each term in Eq.~(\ref{eq:D7}) accordingly leads indeed to the expansion (\ref{eq:D2}) and not to (\ref{eq:D1}).

Yet another way to confirm (\ref{eq:D2}) is to compare its $\epsilon$ expansion with the low mass expansion of the analytic result
for the $\epsilon$-expansion of $B_{m0}(p^2)$
\beq
B_{m0}(p^2) & \!\!=\!\! & \frac{1}{16\pi^2}\left[\frac{1}{\epsilon}+2+\ln\frac{\bar\mu^2}{m^2}\right.\nonumber\\
& & \hspace{0.9cm}\left.-\!\left(1+\frac{m^2}{p^2}\right)\ln\!\left(1+\frac{p^2}{m^2}\right)\right]\!.
\eeq
We find again that (\ref{eq:D2}) is the correct starting point whereas (\ref{eq:D1}) misses one contribution, illustrating that the Taylor expansion does not lead to the correct low mass expansion.\\

To conclude this section, let us now show that despite the previous warnings concerning the validity of the Taylor expansion of the master integrals, it can be put to good use in some instances. Consider the following integral
\beq
{\cal I}_m(p^2)\equiv\int\frac{d^dq}{(2\pi)^d}\frac{1}{q^2+m^2}\frac{3q^2+2(p\cdot q)}{(q+p)^2(2q+p)^2}\,.
\eeq
Because of the presence of enough powers of $q$ in the numerator, it can be Taylor expanded up to order $m^2$ and one finds
\beq
{\cal I}_m(p^2) & = & \int\frac{d^dq}{(2\pi)^d}\frac{1}{q^2}\frac{3q^2+2(p\cdot q)}{(q+p)^2(2q+p)^2}\nonumber\\
& - & m^2\int\frac{d^dq}{(2\pi)^d}\frac{1}{q^4}\frac{3q^2+2(p\cdot q)}{(q+p)^2(2q+p)^2}\nonumber\\
& + & {\cal O}\left(\frac{m^4}{p^4}\right)\,.\label{eq:D10}
\eeq
Next, we notice that it can be decomposed in terms of master integrals as
\beq
{\cal I}_m(p^2) & = & \int\frac{d^dq}{(2\pi)^d}\frac{1}{q^2+m^2}\frac{1}{(q+p)^2}\nonumber\\
& - & \int\frac{d^dq}{(2\pi)^d}\frac{1}{q^2+m^2}\frac{1}{(2q+p)^2}\,.\label{eq:D11}
\eeq
As we have seen above, for each of these master integrals, the Taylor expansion cannot be pushed to order $m^2$. It is easily checked,
however, that this wrong Taylor expansions lead to the correct expansion (\ref{eq:D10}). The reason is that the same contribution for both
integrals, namely $(1/p^2)\int d^dq/(2\pi)^d 1/(q^2+m^2)$, which cancels in the difference (\ref{eq:D11}).
In general, we could imagine the following rule: suppose that a quantity $Q_m$ is regular in the limit $m\to 0$, together with its
first $n$ derivatives $\partial^k Q_m/\partial (m^2)^k$, and suppose that $Q_m$ is split into many pieces $Q_m=\sum_i Q_{m}^{i}$, with
the $Q_m^{(i)}$ (which are basically master integrals times some prefactors) not as regular as $Q_m$. Then the mass expansion
of $Q_m$ to order $n$ is nothing but its Taylor expansion to order $n$, and it can be obtained by Taylor expanding formally
the $Q_m^{(i)}$ to the same order, even though for the latter this does not correspond to their mass expansion.
We believe that this is the reason why we could obtain the correct limit $\lim_{m\to 0} v_{m^2}(k^2)$ while using na\"\i ve Taylor expansions.

\section{Two-loops diagrams}\label{twoloopdiags}
We classified two-loops diagrams in three categories: i) Those corresponding to self-energy corrections in one-loop diagrams,
ii) Those corresponding to vertex corrections in one-loop diagrams. iii) The rest. Diagrams in category (iii) were already depicted in
Figs.~\ref{fig:planar} and \ref{fig:non_planar}. Here, we list the diagrams in the other two categories.

\begin{figure}[h]
\includegraphics[width=0.32\linewidth]{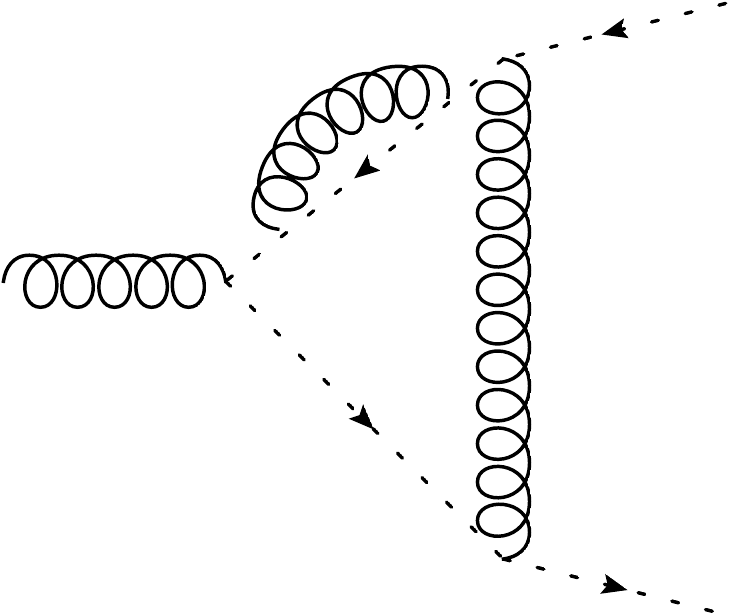}
\includegraphics[width=0.32\linewidth]{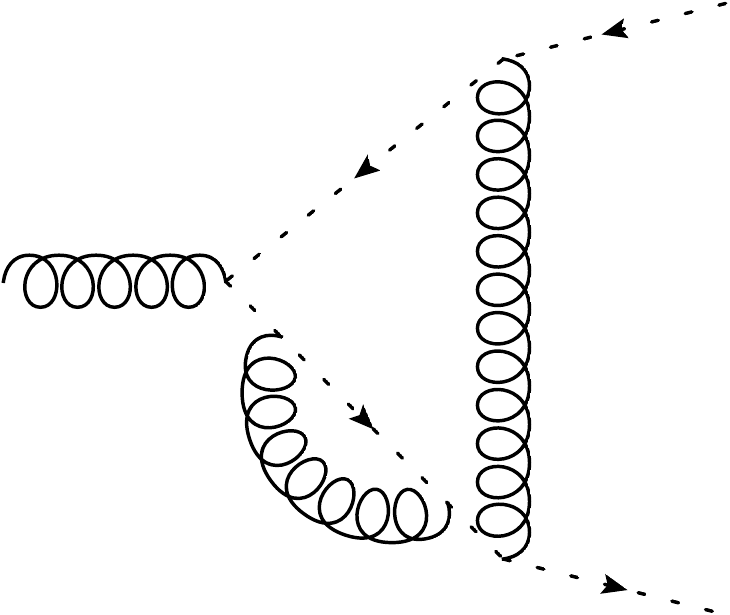}\\
\vspace{0.5cm}

\includegraphics[width=0.37\linewidth]{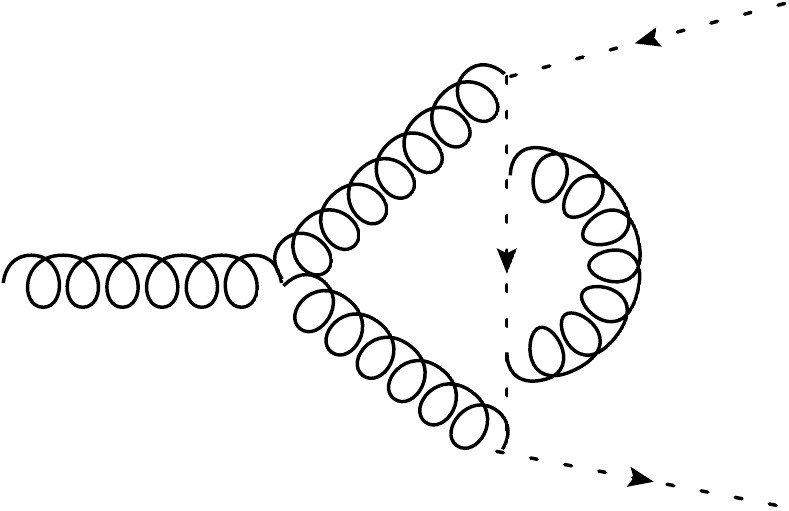}

\caption{Two-loop diagrams corresponding to ghost self-energy corrections inserted in one-loop diagrams.}\label{fig:ghost_self_corr}
\end{figure}

\begin{figure}[h]
\includegraphics[width=0.32\linewidth]{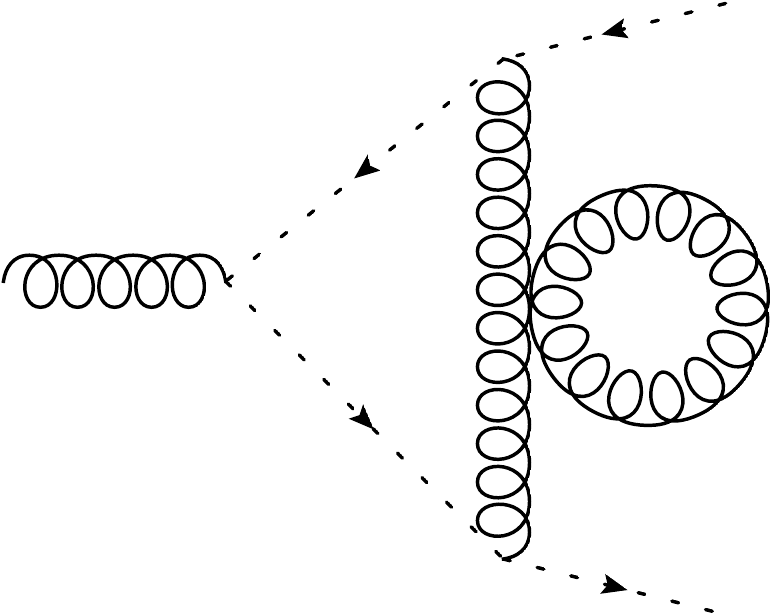}
\includegraphics[width=0.32\linewidth]{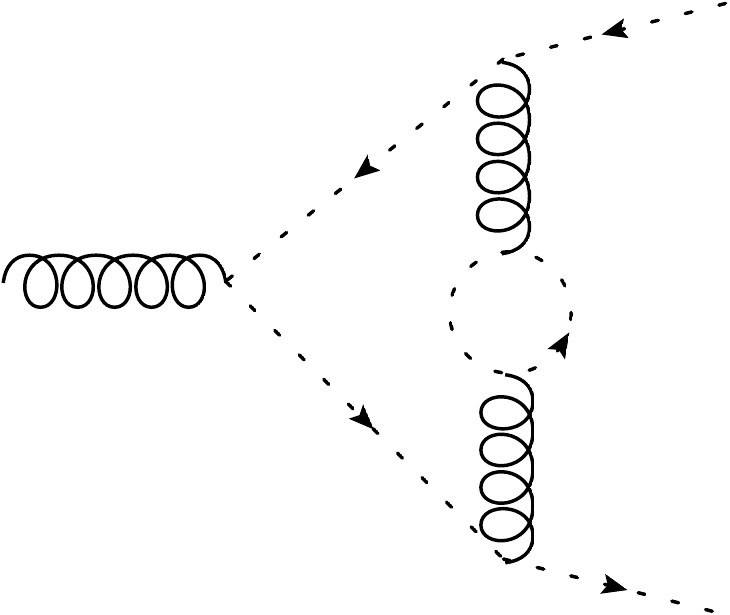}\\
\vspace{0.5cm}
\includegraphics[width=0.32\linewidth]{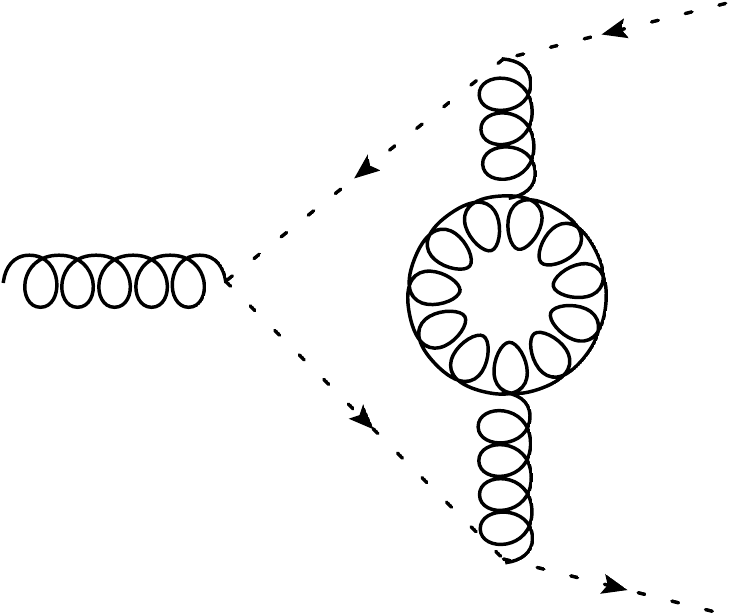}
\includegraphics[width=0.32\linewidth]{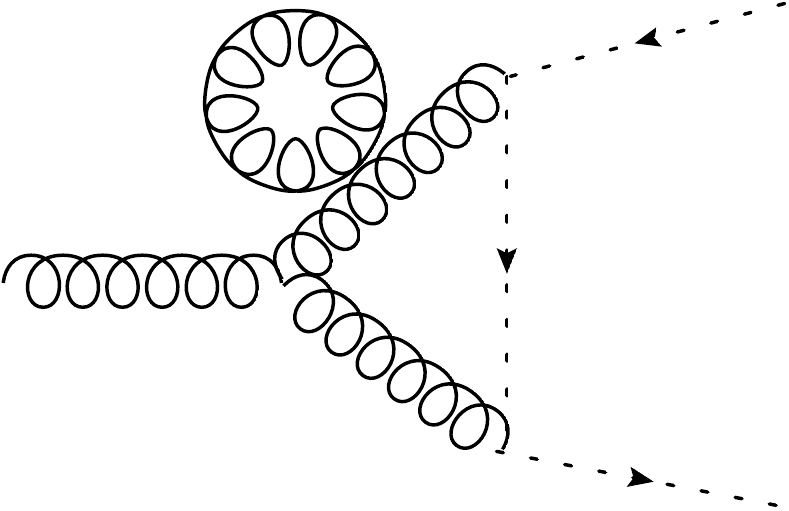}\\
\vspace{0.5cm}

\includegraphics[width=0.32\linewidth]{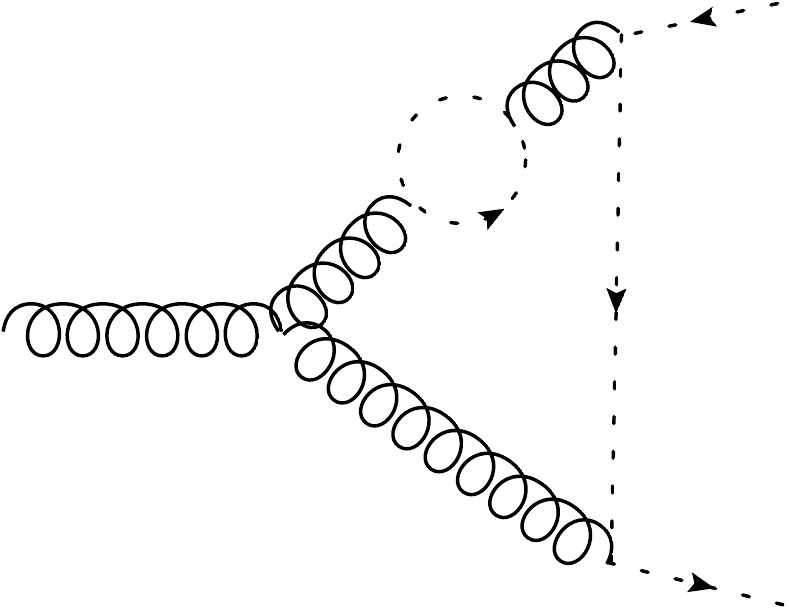}
\includegraphics[width=0.32\linewidth]{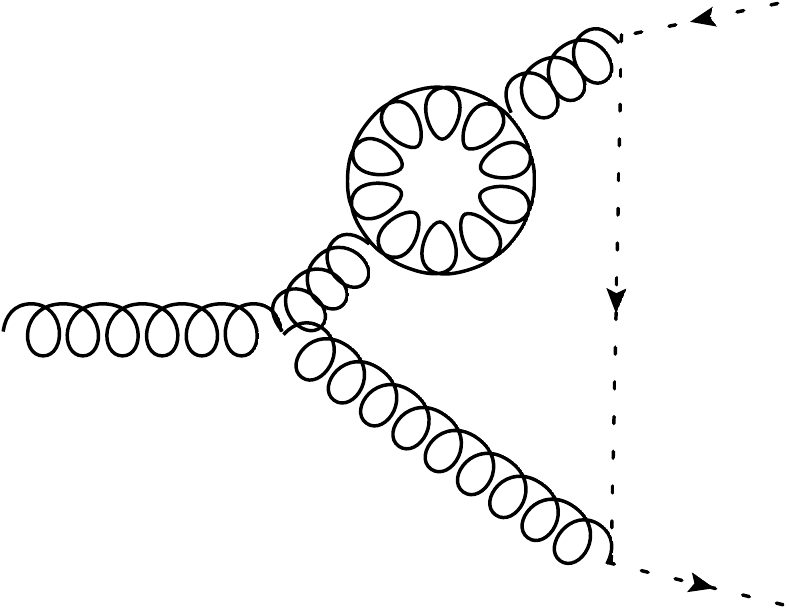}\\
\vspace{0.5cm}

\includegraphics[width=0.32\linewidth]{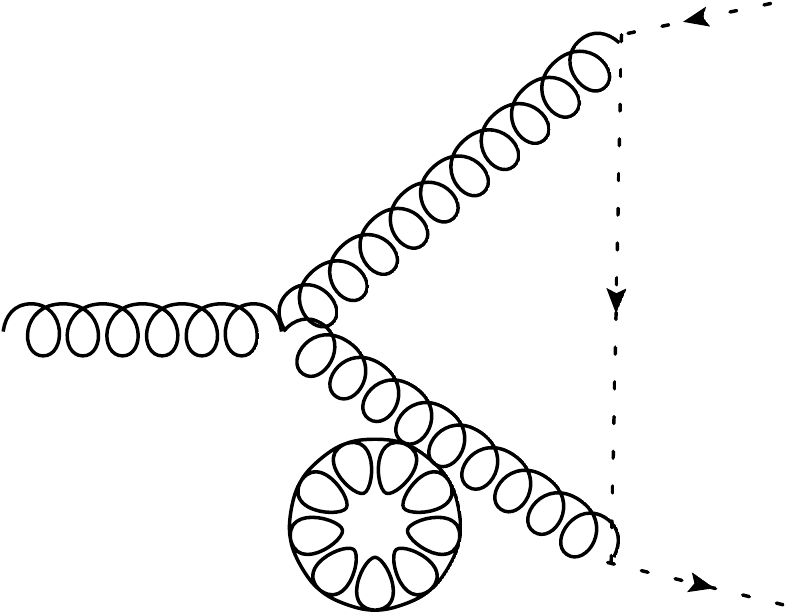}
\includegraphics[width=0.32\linewidth]{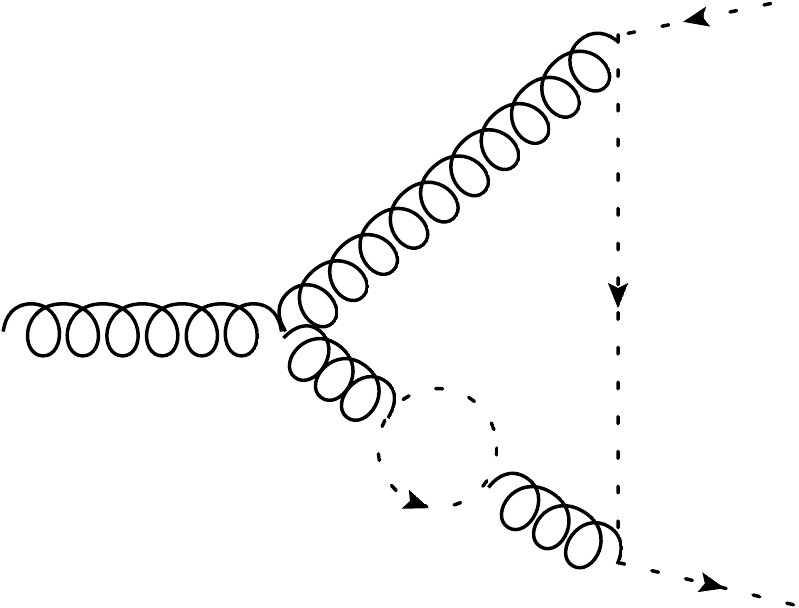}\\
\vspace{0.5cm}

\includegraphics[width=0.32\linewidth]{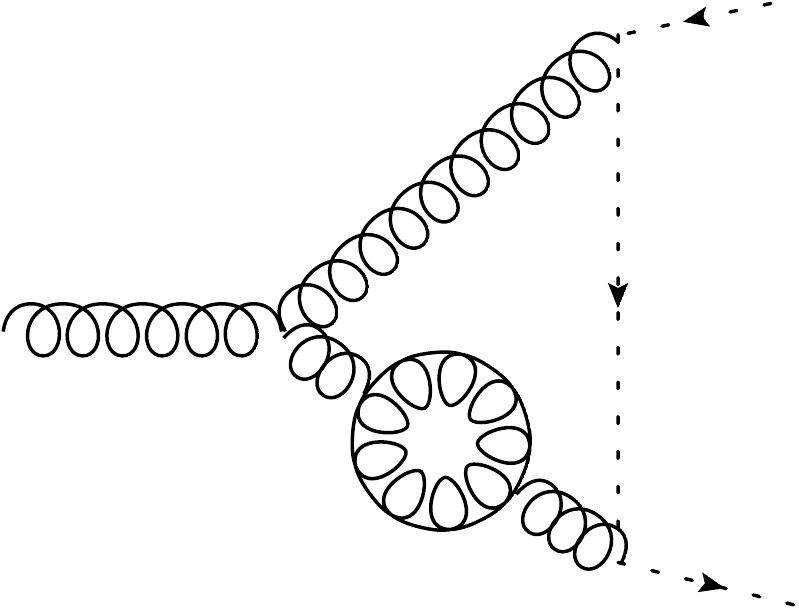}

\caption{Two-loop diagrams corresponding to gluon self-energy corrections inserted in one-loop diagrams.}\label{fig:gluon_self_corr}
\end{figure}

\begin{figure}[h]
\includegraphics[width=0.32\linewidth]{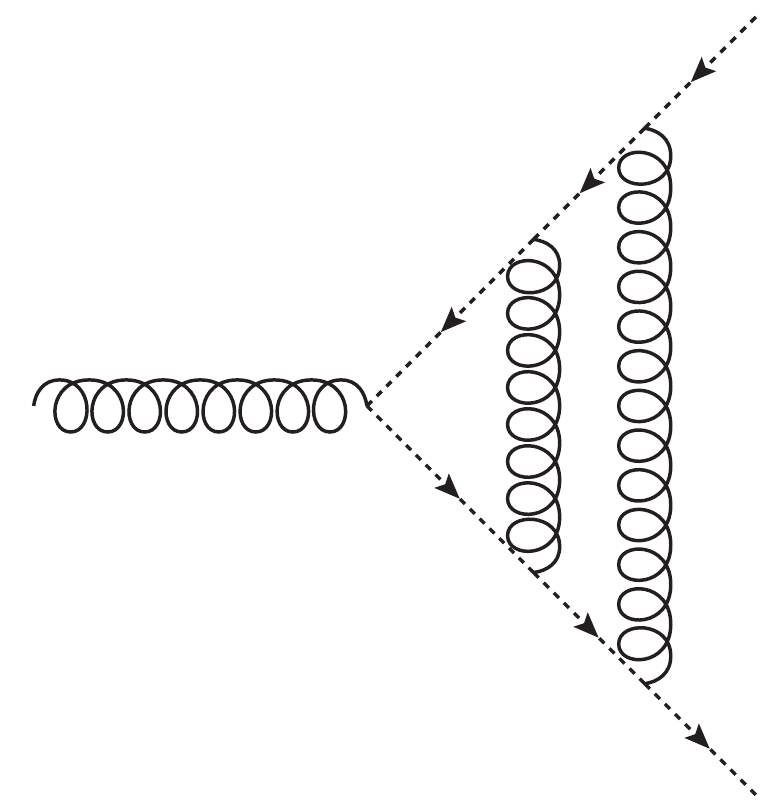}
\includegraphics[width=0.32\linewidth]{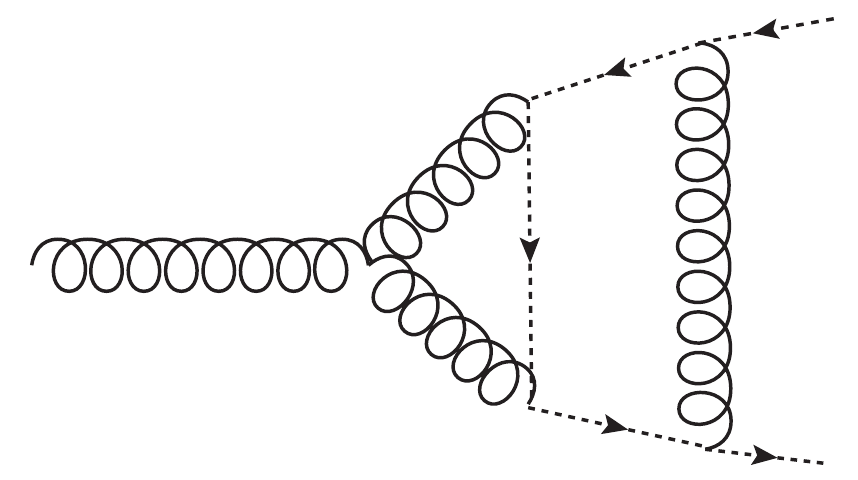}\\
\vspace{0.5cm}

\includegraphics[width=0.37\linewidth]{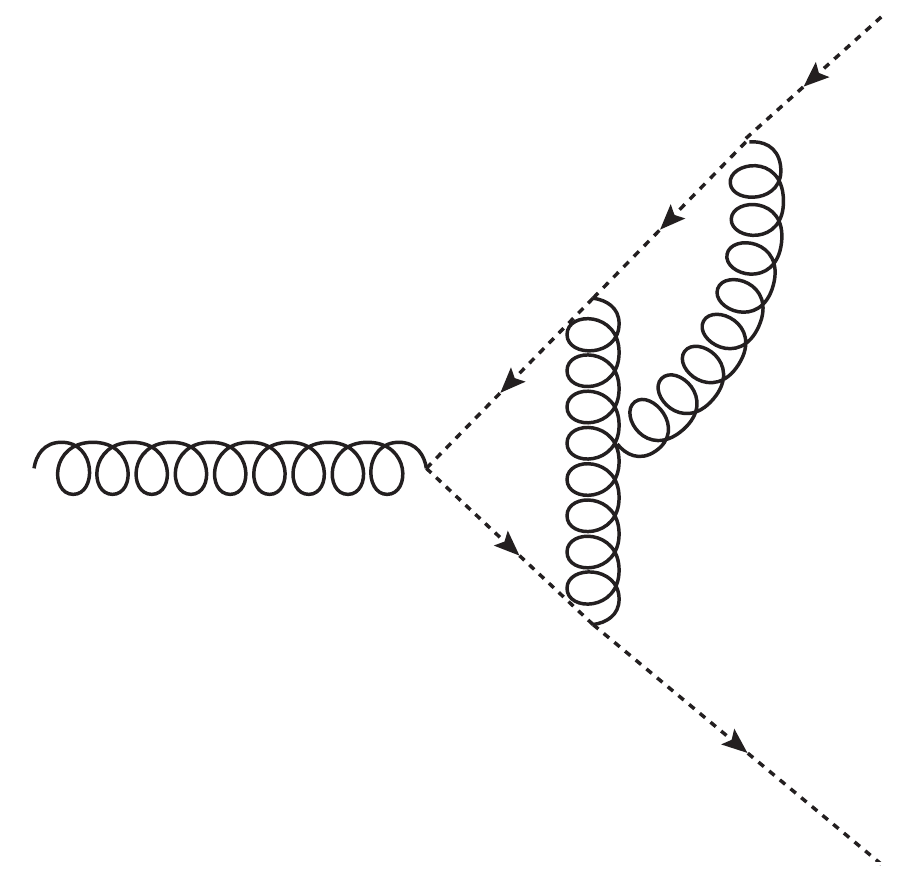}
\includegraphics[width=0.32\linewidth]{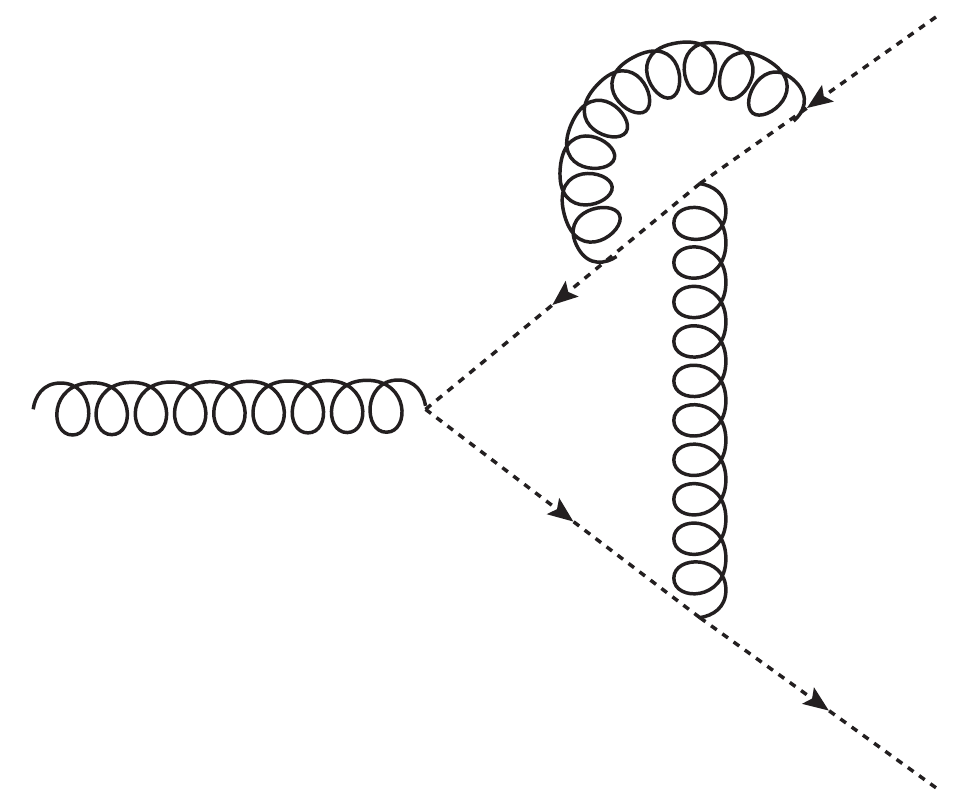}\\
\vspace{0.5cm}

\includegraphics[width=0.37\linewidth]{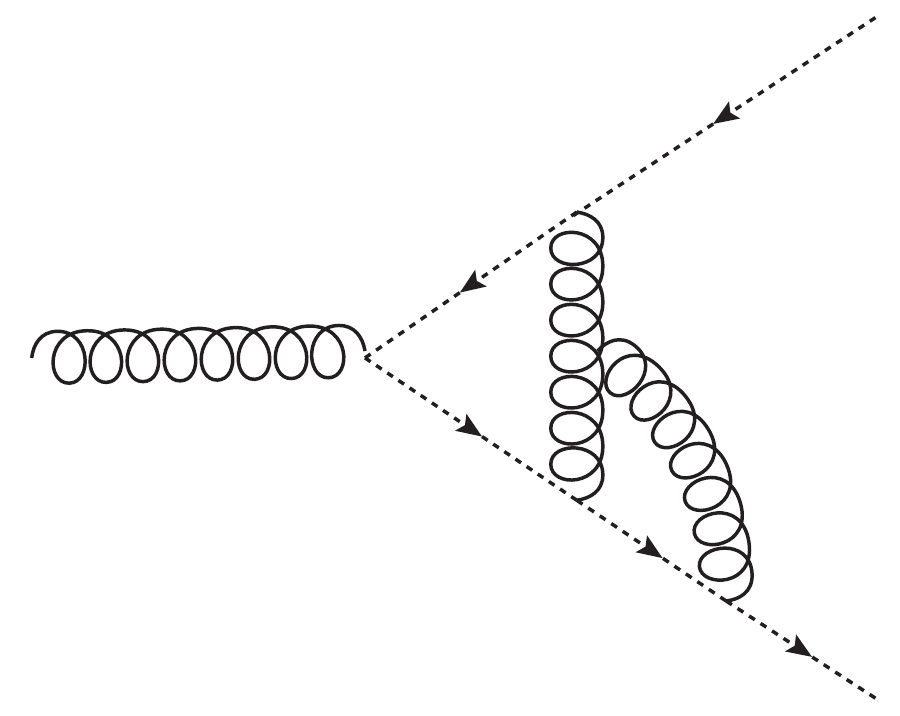}
\includegraphics[width=0.32\linewidth]{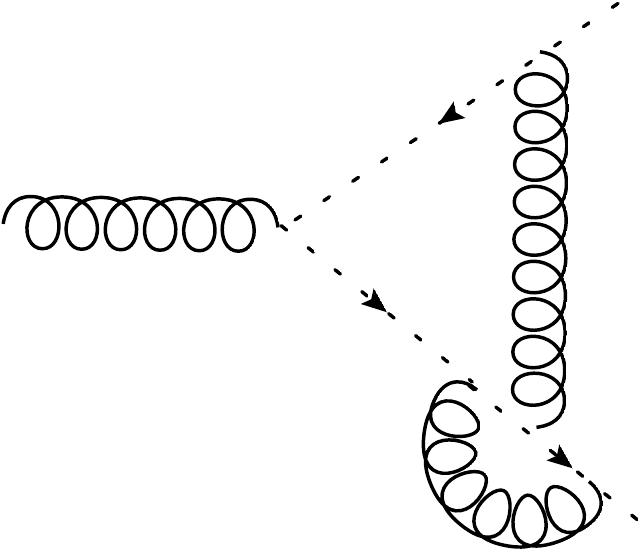}\\
\vspace{0.5cm}

\includegraphics[width=0.37\linewidth]{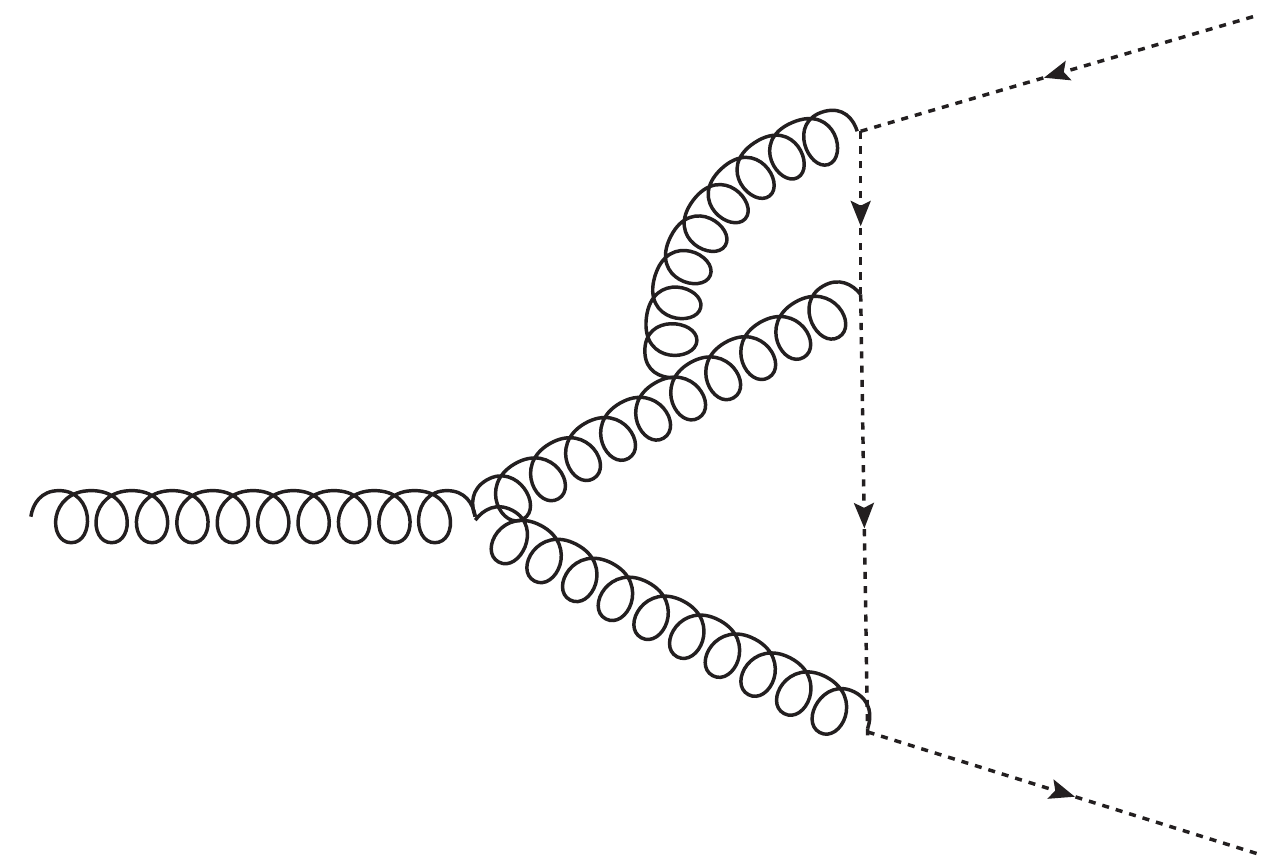}
\includegraphics[width=0.32\linewidth]{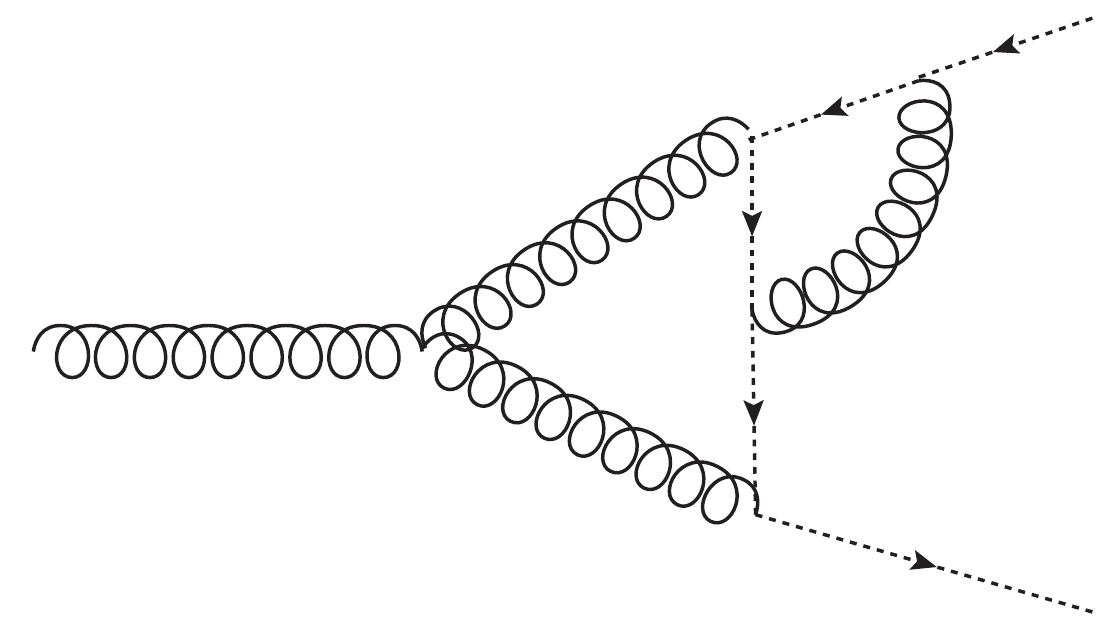}\\
\vspace{0.5cm}

\includegraphics[width=0.37\linewidth]{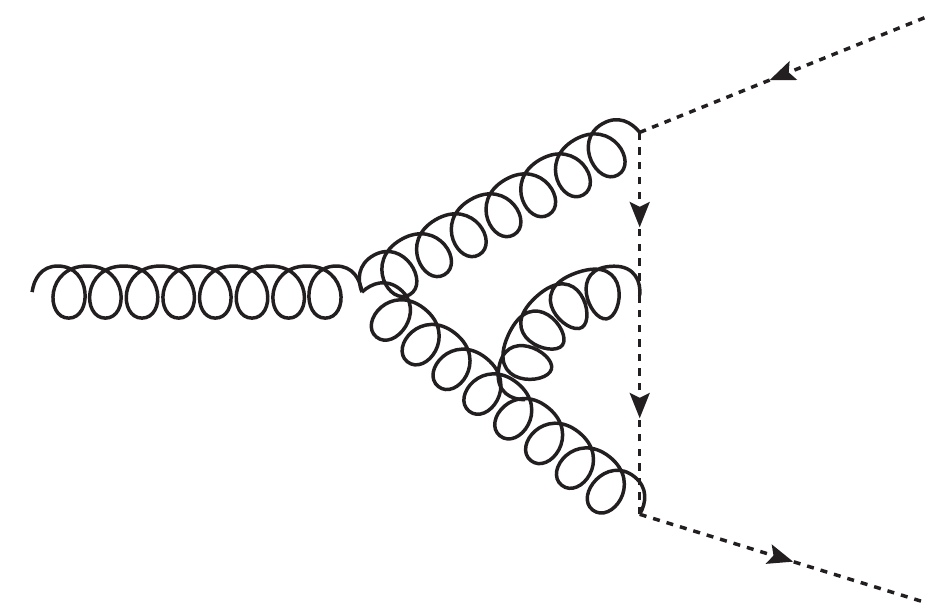}
\includegraphics[width=0.32\linewidth]{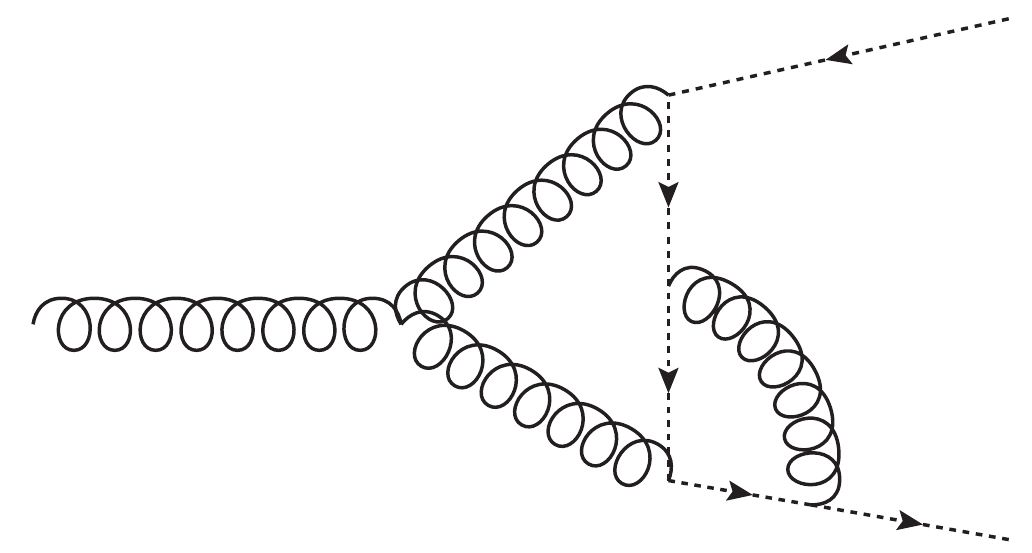}

\caption{Two-loop diagrams corresponding to ghost-gluon vertex corrections inserted in one-loop diagrams.}\label{fig:ghost_vertex_corr}
\end{figure}

\begin{figure}[h]
\includegraphics[width=0.32\linewidth]{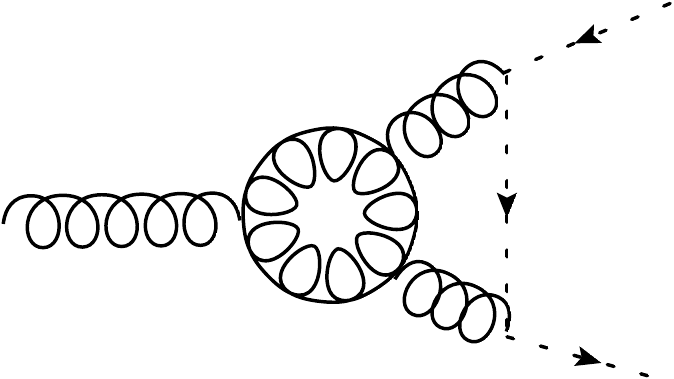}
\includegraphics[width=0.32\linewidth]{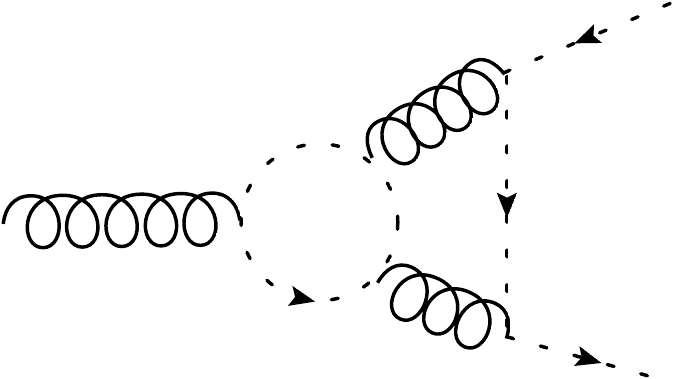}\\
\vspace{0.5cm}

\includegraphics[width=0.37\linewidth]{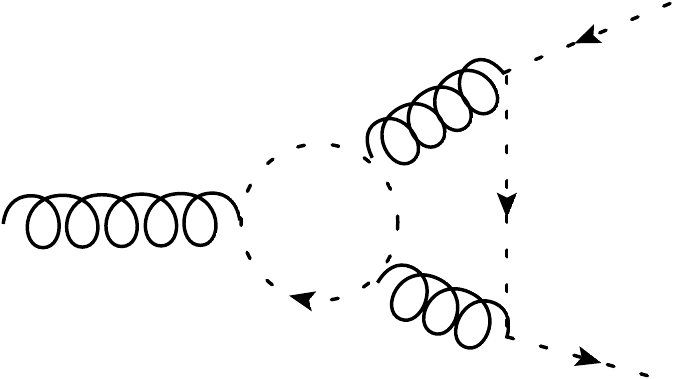}
\includegraphics[width=0.32\linewidth]{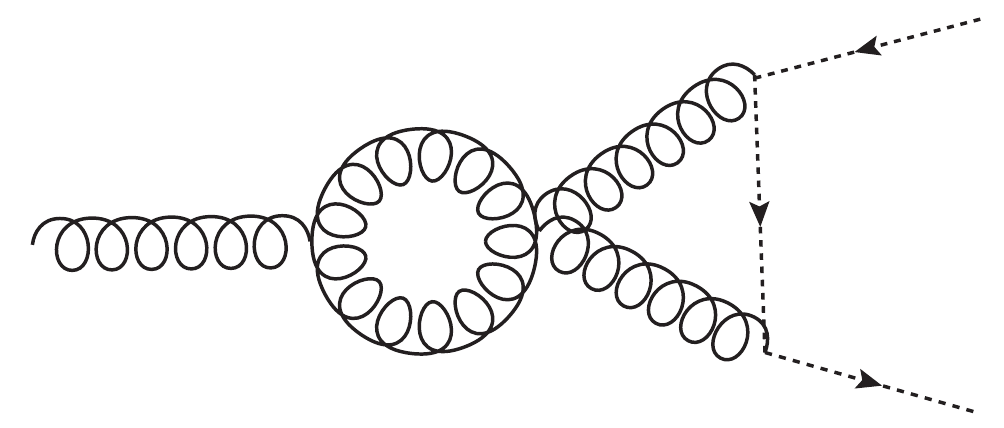}\\
\vspace{0.5cm}

\includegraphics[width=0.37\linewidth]{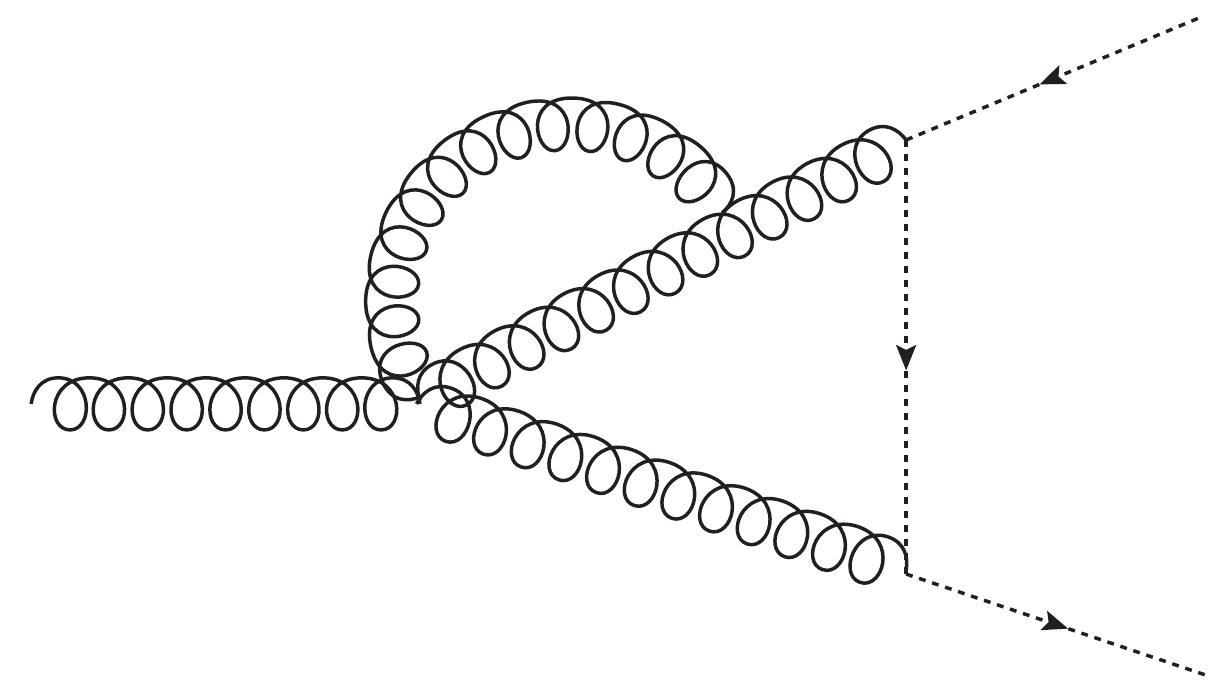}
\includegraphics[width=0.32\linewidth]{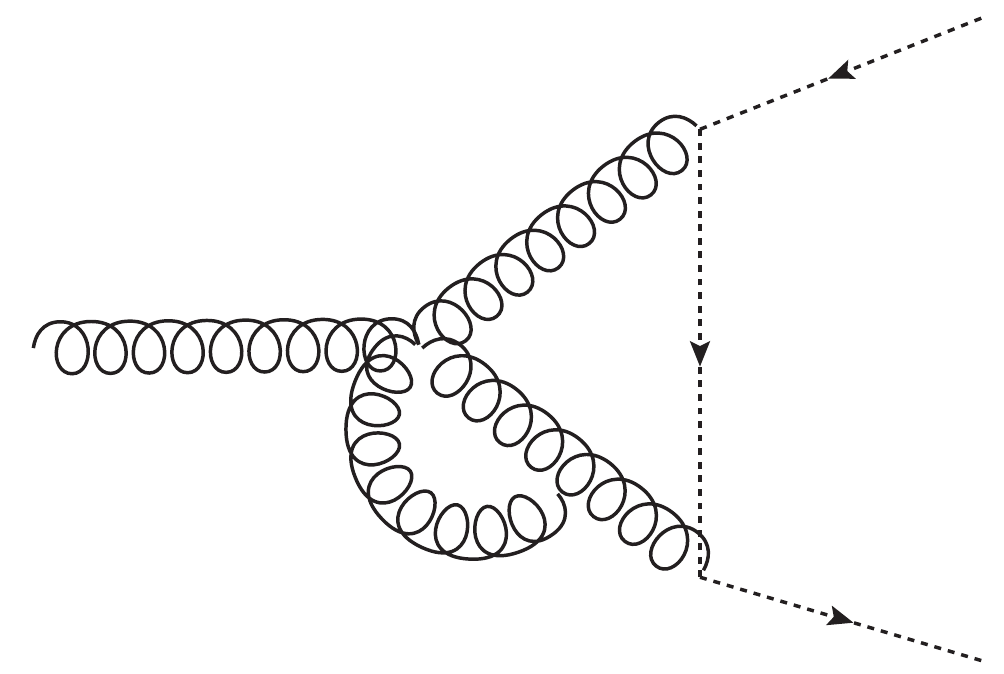}

\caption{Two-loop diagrams corresponding to three-gluon vertex corrections inserted in one-loop diagrams.}\label{fig:gluon_vertex_corr}
\end{figure}

Two-loops diagrams corresponding to ghost and gluon self-energy insertions in one-loop diagrams are shown in
Fig.~\ref{fig:ghost_self_corr} and Fig.~\ref{fig:gluon_self_corr} respectively, whereas two-loops diagrams corresponding to
ghost-gluon and three-gluon vertex corrections inserted in one-loop diagrams appear in Fig.~\ref{fig:ghost_vertex_corr} and
Fig.~\ref{fig:gluon_vertex_corr} respectively.

\end{document}